\documentclass[%
reprint,
nofootinbib,
amsmath,amssymb, 
aps,
prd,
]{revtex4-1}
\usepackage{verbatim}
\usepackage{graphics}
\usepackage{xcolor}
\usepackage{graphicx}
\usepackage[final]{pdfpages}
\usepackage{subfigure}
\usepackage{wrapfig}
\usepackage[font=small,labelfont=bf]{caption}
\usepackage{afterpage}
\usepackage{floatpag}
\usepackage{dcolumn}
\usepackage{bm}
\usepackage{amsmath}
\usepackage{amssymb}
\usepackage{relsize}
\usepackage{mathrsfs}
\usepackage{cases}
\usepackage[retainorgcmds]{IEEEtrantools}
\usepackage{enumitem}
\usepackage[hyperindex,breaklinks]{hyperref}

\usepackage{fullpage}
\usepackage{xr}

\makeatletter
\AtBeginDocument{\let\LS@rot\@undefined}
\makeatother

    \allowdisplaybreaks

\newcommand{\p}{\partial}
\newcommand{\tb}{\textbf}

\newcommand{\tx}{\textit}
\newcommand{\beq}{\begin{equation}}
\newcommand{\eeq}{\end{equation}}
\newcommand{\bdi}{\begin{displaymath}}
\newcommand{\edi}{\end{displaymath}}
\newcommand{\beqn}{\begin{eqnarray}}
\newcommand{\eeqn}{\end{eqnarray}}
\newcommand{\f}{\frac}

\newcommand{\ud}{\mathrm{d}}
\newcommand{\ogw}{\Omega_{\rm{GW}}}

\newcommand{\rgw}{\rho_{\rm{GW}}}
\newcommand{\hini}{h_{k,~\rm{init}}}
\newcommand{\neff}{N_{\rm{eff}}}
\newcommand{\zeq}{z_{\rm{eq}}}
\newcommand{\tre}{T_{\rm{reheat}}}

\newcommand{\mnras}{Mon.Not.Roy.Astron.Soc.}
\newcommand{\apjl}{Astrophy.J. Letter}

\newcommand{\aj}{Astron.J.}
\newcommand{\aap}{Astron.\& Aastrophys.}
\newcommand{\jcap}{J.Cosm.Astropart.Phys.}
\newcommand{\na}{New Astronomy}
\newcommand{\physrep}{Physics Reports}
\begin{document}

\title{Bose-Einstein-condensed scalar field dark matter and the gravitational wave background from inflation: new cosmological constraints and its detectability by LIGO}

\author{Bohua Li}
\email{bohuali@astro.as.utexas.edu}    
\affiliation{
\small Department of Astronomy and Texas Cosmology Center,\\
     \normalsize\it The University of Texas at Austin,
     2515 Speedway C1400, Austin, TX 78712, USA
}     
\author{Paul R. Shapiro}
\email{shapiro@astro.as.utexas.edu}
\affiliation{
\small Department of Astronomy and Texas Cosmology Center,\\
     \normalsize\it The University of Texas at Austin,
     2515 Speedway C1400, Austin, TX 78712, USA
}     
\author{Tanja Rindler-Daller} 
\email{tanja.rindler-daller@univie.ac.at}
\affiliation{
\small Institut f\"ur Astrophysik, Universit\"atssternwarte Wien,
    \normalsize\it University of Vienna, A-1180 Vienna, Austria\\
    \normalsize\it
     \small Department of Physics and Michigan Center for Theoretical Physics, 
    \normalsize\it
    University of Michigan, Ann Arbor, MI 48109, USA
    }

\date{\today}


\begin{abstract}

We consider an alternative to WIMP cold dark
matter (CDM), ultralight bosonic dark matter ($m\gtrsim 10^{-22}$eV$/c^2$) 
described by a complex scalar field (SFDM) with a global $U(1)$ symmetry,
for which the comoving particle number density, or charge density, 
is conserved after particle production during standard reheating. 
We allow for a repulsive self-interaction.
In a $\Lambda$SFDM universe, SFDM starts relativistic, 
evolving from stiff ($w=1$) to radiation-like ($w=1/3$), 
before becoming nonrelativistic at late times ($w=0$). 
Thus, before the familiar radiation-dominated era, 
there is an earlier era of stiff-SFDM-domination.
During both the stiff-SFDM-dominated and radiation-dominated eras, 
the expansion rate is \emph{higher} than in $\Lambda$CDM.
SFDM particle mass $m$ and quartic self-interaction coupling strength $\lambda$, 
are therefore constrained by cosmological observables, 
particularly $\neff$, the effective number of neutrino species during BBN, 
and $\zeq$, the redshift of matter-radiation equality. 
Furthermore, since the stochastic gravitational wave background (SGWB) from inflation 
is amplified during the stiff-SFDM-dominated era, 
it can contribute a radiationlike component large enough to affect these observables, 
by further boosting the expansion rate after the stiff era ends.
Remarkably, this same amplification makes detection of the SGWB possible 
at high frequencies by current laser interferometer experiments, 
e.g., aLIGO/Virgo and LISA. 
For SFDM particle parameters that satisfy these cosmological constraints, 
the amplified SGWB is detectable by LIGO for a broad range of reheat temperatures $\tre$, 
for values of the tensor-to-scalar ratio $r$ currently allowed by CMB polarization measurements. 
For a given $r$ and $\lambda/(mc^2)^2$, the marginally-allowed $\Lambda$SFDM model
for each $\tre$ has the smallest $m$ that satisfies the cosmological constraints, 
and maximizes the present SGWB energy density for that $\tre$.
This SGWB is then maximally \emph{detectable} for values of $\tre$ for which 
modes that reenter the horizon 
when reheating ended have frequencies today that lie within the LIGO sensitive band.
For example, for the family of marginally-allowed models with 
$r = 0.01$ and $\lambda/(mc^2)^2=10^{-18}$ eV$^{-1}$cm$^3$, 
the maximally detectable $\Lambda$SFDM model 
has $\tre\simeq2\times10^4$ GeV and $m\simeq1.6\times10^{-19}$ eV$/c^2$,
for which we predict an aLIGO O1 run detection with signal-to-noise ratio of $\sim10$.  
We show that the null detection of the SGWB recently reported by the aLIGO O1 run
excludes the parameter range $8.75\times10^3\lesssim\tre~(\rm{GeV})\lesssim 1.7\times10^5$ 
for this illustrative family at 95\% confidence, thereby demonstrating that 
GW detection experiments can already place a new kind of cosmological constraint on SFDM.
A wider range of SFDM parameters and reheat temperatures
should be accessible to aLIGO/Virgo O5, 
with the potential to detect this unique signature of the $\Lambda$SFDM model.
For this same illustrative family, for example, 
a $3\sigma$ detection is predicted for $600\lesssim\tre~(\rm{GeV})\lesssim10^7$.

\begin{description}
\item[PACS numbers]
98.80.-k, 95.35.+d, 98.80.Cq, 98.80.Ft, 04.30.-w

\end{description}
\end{abstract}

\maketitle


\section{Introduction}

\subsection{Cold dark matter: WIMPs or something else?}
\label{sec:cdm}

The nature of the dark matter (DM) remains one of the most profound open problems in cosmology.
Observations of the large-scale structure (LSS) of the universe and
the cosmic microwave background (CMB) are consistent with
dark matter which forms structure as if it was created ``cold'', i.e.,
it can be modeled as collisionless particles with nonrelativistic random microscopic motions.
The cold dark matter (CDM) model has been very successful 
in describing structure formation on large scales as hierarchical, 
with the smallest objects forming first and merging over time to form ever-larger objects 
--- ``halos'' in virial equilibrium ---
connected by filaments surrounding largely empty voids in a ``cosmic web of structure''
\cite{2015arXiv150201589P}, \cite{2015PhRvD..92l3516A, 2016arXiv160703155A}, 
\cite{2009ApJ...692.1060V}, 
\cite{2009ApJ...704.1274W, 2010MNRAS.406.1220W}, \cite{2015ApJ...814...26N}.
Candidate particles for DM can be found in many extensions 
to the Standard Model (SM) of particle physics.
Traditionally, the most studied candidate particles for the standard, collisionless CDM
are WIMPs (weakly interacting massive particles), the lightest supersymmetric partner particles
predicted by models of supersymmetry (``SUSY''),
thermal relics whose mass range allows gravitational clustering to form objects down to Earth-mass.

Despite its success on large scales, the standard, collisionless CDM
model has been challenged by observations of galactic and sub-galactic scales. 
First, N-body simulations of collisionless CDM
predict a universal cuspy density profile for DM halos. 
However, measurements of the density profiles of various
dark-matter-dominated systems, e.g., dwarf spheroidal galaxies,
low-surface-brightness galaxies, even some galaxy clusters, 
have suggested shallower profiles, or even cores at their centers (the ``cusp/core problem'') 
\cite{Moore1, 2008AJ....136.2648D, Amorisco2012, 2015AJ....149..180O, 2016MNRAS.460.3610O}. 
Such N-body simulations also predict a large-overabundance of subhalos in the Local Group 
compared with the observed number of satellite galaxies (the ``missing satellites problem'')
\cite{Klypin1999, Moore2}. 
In addition, it has been pointed out that, after abundance matching, 
the most massive subhalos of a Milky-Way-like galaxy predicted by standard CDM simulations 
are too dense to host the brightest satellites of the Milky Way (the ``too big to fail'' problem) 
\cite{2011MNRAS.415L..40B, 2012MNRAS.422.1203B}.

Meanwhile, attempts to detect WIMP DM particles either directly or indirectly
(i.e., as astronomical sources following their decay or annihilation into radiation or other particles)
have thus far been unsuccessful \cite{2016PhRvL.116p1301A, 2016PhRvL.116g1301A, 
2015PhRvL.115w1301A, 2015ApJ...809L...4D}. 
The range of particle models and parameters
which remain viable for WIMP DM has, in fact, been substantially reduced by these nondetections.

These nondetections of WIMPs and the structure formation
discrepancies described above, between theory and observations, for
the standard model of CDM as cold, collisionless particles suggest
that an alternative at the particle level to WIMPs as CDM may be
required. Such an alternative must retain the successes of CDM
with regard to LSS formation and the CMB, as well as the
thermodynamic evolution of the background universe in the standard
Big Bang cosmology.

One such variant of CDM which we have considered before
is that of complex scalar field dark matter (SFDM), 
for which \emph{all} cosmological dark matter is composed of ultralight bosons
\cite{2012MNRAS.422..135R, 2014ASSP...38..163R, 2014MPLA...2930002R} 
(where it is referred to as Bose-Einstein condensate CDM, or BEC-CDM);
\cite{2014PhRvD..89h3536L} (hereafter ``Paper I'').
For additional descriptions of this model and the related literature, we refer the reader to these papers.
With regard to LSS, SFDM provides a natural length scale,
below which structure formation is suppressed, leading to fewer subhalos and generally, 
to a lower density of DM in the central regions of galaxies. 
On larger scales, however, structure formation in SFDM is the same as for cold, collisionless particles.

In Paper I, we considered the cosmological evolution of the homogeneous Big-Bang
background universe in the presence of SFDM and 
showed that the SFDM behaved like a perfect fluid with an equation-of-state (EOS) parameter
$w\equiv p/\rho$ which evolved from stiff ($w=1$) to radiationlike ($w=1/3$) 
to nonrelativistic CDM-like ($w=0$).
The energy density of SFDM during this last CDM-like phase, 
equal to the product of the rest-mass energy density per particle 
and the particle number density, is chosen to match the observed dark matter mass-energy density 
in the Universe today. At early times the stiff EOS
made the SFDM dominate the total energy density of the universe, 
with consequences for the expansion history.
This made it possible for us to use observational constraints
to derive the allowed range of SFDM particle parameters. Here we will revisit this problem
by making two significant advances, as described in the sections below.
First, we will embed the SFDM model more fully in the standard inflationary paradigm,
to create a more holistic $\Lambda$SFDM cosmology.
Second, we will take account of the gravitational-wave (GW) background from inflation
and its amplification in the presence of SFDM, leading to the possibility of its detection
at high frequencies by laser interferometer experiments like the Advanced LIGO/Virgo
experiment (shortened as ``aLIGO/Virgo'').
A preliminary summary of some of this new work was presented in \cite{2015bash.sympE..28L}.

\subsection{Complex SFDM: Bose-Einstein-condensed ultralight particles as cold dark matter}
\label{sec:SFDM1}

The SFDM model considered in Paper I and in
\cite{2012MNRAS.422..135R, 2014ASSP...38..163R, 2014MPLA...2930002R} 
is one type in a family of cold dark matter candidates involving bosonic particles
associated with a scalar field. The best-known example of bosonic
dark matter is the QCD axion, a real (pseudo-)scalar field proposed
to resolve the strong CP problem. Its attractive self-interaction is so weak
that it is usually neglected, leaving only the quadratic mass term
in the potential. The mass of the QCD axion currently allowed by
astronomical observational constraints is $\sim10^{-5}$ eV$/c^2$.
Structure formation in QCD axion DM is like that for cold,
collisionless particles on all scales of astrophysical interest, 
so the small-scale structure problems of CDM described above in
\S\ref{sec:cdm} remain for the QCD axion DM, as well. 
As a generalization, ultralight axions or axion-like particles (ALPs) 
are also predicted by extensions to the Standard Model, 
which could serve as dark matter as long as 
their mass is $> 10^{-33}$ eV$/c^2\sim H_0\cdot\hbar/c^2$ 
($H_0$ is the Hubble constant at the present). 
The self-interaction of these ultralight ALPs, too, is generally assumed to be 
so weak that it can be neglected when comparing model predictions to astrophysical data. 
However, we caution against this neglect, 
since our SFDM results for the case which includes a \emph{repulsive} self-interaction, 
show that even a tiny self-interaction can be dynamically important;
it is not clear why the same should not be true for attractive cases. 
When the mass of the non-interacting axion is above $10^{-18}$ eV$/c^2$, 
dark matter comprised of ALPs is dynamically indistinguishable from collisionless CDM 
on large scales \cite{2015PhRvD..91l3520M}. 
For particle masses smaller than this,
however, their de Broglie wavelength inside galactic halos, which
sets a scale below which structure formation is suppressed, can be
large enough to affect the small scales identified in
\S\ref{sec:cdm} above as problematic for standard CDM.


In fact, other ultralight scalar field particles have been proposed
as DM candidates by various authors, which all mimic standard CDM
above some length scale but deviate on sufficiently small scales,
motivated by the small-scale problems of standard CDM. While the
genesis of ultralight bosonic DM is \tx{a priori} model-dependent,
many of those models share the property of axion DM that 
the DM bosons are considered to be born cold with high occupation number,
such that they can be described by a classical scalar field.
The choice of potentials and particle masses does vary, however.
Non-interacting DM has been considered by, e.g., 
\cite{1994PhRvD..50.3650S}, \cite{1999PhRvD..60j3506P}, 
\cite{2000PhRvL..85.1158H, 2009ApJ...697..850W, 2016arXiv161008297H} (``fuzzy dark matter''),
\cite{2014NatPh..10..496S} (``quantum wave dark matter''),
\cite{2012PhRvD..85j3514M, 2015PhRvD..91j3512H} (``ultralight axions''), 
\cite{2011MNRAS.416...87S, 2012JCAP...10..003M} (``scalar field dark matter''), 
\cite{2016MNRAS.460.4397C}. 
On the other hand, self-interacting DM has been studied in, e.g.,
\cite{2000ApJ...534L.127P} (``fluid dark matter''),
\cite{2000NewA....5..103G, 2012MNRAS.427..839S} (``repulsive dark matter''), 
and \cite{2000PhRvD..62f1301M, MUL2001, 2007JCAP...06..025B,
2011PhRvD..83l3515H, 2012A&A...537A.127C, 2012PhRvD..85b3527K, 2017PhRvD..95f3515S}. 
In the self-interacting\footnote{
The self-interaction term used here should not be confused with 
the kind of self-interacting CDM particles referred to elsewhere in the literature as SIDM, 
suggested by \cite{2000PhRvL..84.3760S}, which we have studied 
in \cite{2005MNRAS.363.1092A} and \cite{2011MNRAS.415.1125K}.
In SIDM, particle self-interaction manifests itself as two-body elastic scattering 
which adds ``collisionality'' to the otherwise collisionless CDM gas,
but does not make a BEC or exhibit any form of macroscopic quantum coherence.
}
DM case (including our SFDM model with a quartic potential
\cite{2012MNRAS.422..135R, 2014ASSP...38..163R, 2014MPLA...2930002R}, 
referred to there as BEC-CDM), 
the suppression of small-scale structure can also result from the
pressure force associated with its repulsive potential, rather than
solely from the ``quantum pressure'' associated with 
large de Broglie wavelength as in the non-interacting case.
When the minimum length scale for structure associated with the repulsive self-interaction 
is greater than that due to quantum pressure, this is referred to as the Thomas-Fermi regime.

Amongst the models mentioned above, there are many which propose that DM bosons are initially in a
Bose-Einstein condensate (BEC), or will form a BEC at some stage in cosmic history. 
In our previous work \cite{2012MNRAS.422..135R, 2014ASSP...38..163R, 2014MPLA...2930002R}, 
we studied the nonlinear behavior of the BEC wave function (or, the order parameter), 
in the context of DM halo structure. 
We applied the Gross-Pitaevskii equation coupled to the Poisson equation
to study the equilibrium structure of BEC-CDM halos, including the effects of
angular momentum and the possible formation of quantum vortices.

The formation of a BEC in QCD axion DM has also been studied in the literature. 
A detailed analysis of the condensation process for the QCD axion has been made by
\cite{2009PhRvL.103k1301S, 2012PhRvD..85f3520E}.
However, controversies remain about the formation of a BEC
and whether it depends on the sign of the self-interaction or
whether the classical field description is sufficient in general
\cite{2013JCAP...12..034D, 2015PhRvD..92j3513G}.
This debate is partly due to the difficulty of forming a BEC for bosons
described by a \tx{real} scalar field (the axion case), 
while the condensation process occurs naturally, even in the early universe,
for bosons described by a \tx{complex} scalar field (the case for our SFDM model) 
with a global $U(1)$ symmetry, associated with a (conserved) Noether charge
\cite{1982PhRvD..25..502H, 1981PhRvD..24..426K, 2001PhRvD..64l3509M}, as described below.

%
%

In the \tx{complex} SFDM model presented here, 
DM appears in the wake of reheating, following inflation. 
An example of such a microphysical implementation can be
found in \cite{2001PhRvD..64l3509M}. 
The idea is that, upon inflaton decay, DM bosons and antibosons are created, 
as are the SM particles. 
We assume that the complex scalar field was born
with a large charge, or comoving charge density, $Q$, 
which is the difference between the comoving number density 
of bosons and antibosons, i.e., $Q\equiv n_+-n_-$.
Owing to the global $U(1)$ symmetry of the complex scalar field, 
$Q$ is a conserved quantity. In thermal equilibrium, while DM bosons and
antibosons are annihilated away (leaving no antibosons behind), the
majority of DM particles will find themselves rapidly occupying
their ground state (the zero-momentum state). In Paper I, 
following \cite{1982PhRvD..25..502H} and \cite{1981PhRvD..24..426K}, 
we pointed out that Bose-Einstein
condensation for DM particles of mass $m$ occurs as long as 
$k_{\rm B}Q/S \gg 1$ initially, where $k_{\rm B}$ is the Boltzmann
constant and $S$ is the comoving entropy density. In a cosmological
setting, both $Q$ and $S$ are conserved. Now, that ground state
which remains is a BEC with charge approximately equal to $Q$. As a
result, the DM can thereafter be described as a classical field,
hence complex scalar field dark matter -- SFDM. 

We note that complex SFDM belongs to the wider family of \emph{asymmetric} DM, 
in which the DM antiparticles annihilate away, along with an equal number of particles, 
leaving only the excess of particles over antiparticles behind. 
In this ``large-charge'' limit, 
the charge density $Q$ (where $Q\simeq n_+$) is then related to the
present-day SFDM energy density, $\rho_{\rm SFDM,0}$, 
by $Qmc^2/\rho_{\rm SFDM,0}\simeq1$. 
This situation is described by
\cite{2002PhRvD..65h3514A} as leading to a ``spintessence'' phase at later times 
(see also \cite{2011PhLB..696..315B}, \cite{2002PhLB..545...17B}).
They also described the other limit in which $Qmc^2/\rho_{\rm SFDM,0}\ll1$, 
the ``small-charge'' limit , i.e., negligible comoving charge density.
This small-charge limit would correspond, instead, 
to the opposite assumption of \emph{symmetric} DM, i.e., 
nearly equal numbers of particles and antiparticles today, so $n_+\simeq n_-$.
Hereafter, since we shall only be interested in the large-charge limit, 
in which the dynamics of the complex scalar field differs 
from that of a real scalar field, as we discuss below in \S\ref{sec:paper1},
the notation ``SFDM'' shall only refer to the \tx{complex} scalar field dark matter 
in the large-charge limit considered here.

\subsection{Cosmic Evolution of $\Lambda$SFDM}
\label{sec:paper1}

We studied the (background) evolution of complex SFDM in detail in Paper I, 
by solving numerically the equation of motion of SFDM in an expanding universe, 
adopting a spatially flat Friedmann-Lema\^itre-Robertson-Walker (FLRW) background metric. 
We called it $\Lambda$SFDM, since all the cosmic components of the $\Lambda$CDM model 
are adopted, except for collisionless CDM, which is replaced by SFDM. 
We assumed that the present cosmic DM abundance is entirely given by 
the current $\rho_{\rm SFDM}$, which also determines 
the (conserved) comoving charge density of SFDM, $Q$, as described in the sections above. 
The evolution of SFDM is determined by the form of the potential in its Langrangian,
as for any other cosmological scalar field.
Let $\psi$ be the complex scalar field describing the condensate of DM bosons.
We adopt the following Lagrangian density (in units of energy density),
\begin{equation} \label{Lag}
\mathcal{L} = \frac{\hbar^2}{2m}g^{\mu \nu}\partial_{\mu}\psi^*\partial_{\nu}\psi - \frac{1}{2}mc^2|\psi|^2 -
\frac{\lambda}{2}|\psi|^4,
\end{equation}
with signature $(+,-,-,-)$. $|\psi|$ denotes the modulus of $\psi$.
$m$ is the DM boson mass, and we choose the energy-independent 2-boson
self-interaction strength to be repulsive or zero, $\lambda \geq 0$.
We will elaborate more on this Lagrangian density in \S\ref{sec:TmnSFDM}.

The range of SFDM parameters of interest is motivated by the small-scale CDM 
structure problems mentioned above.
In its CDM-like phase---when the quadratic term in Eq. (\ref{Lag}) dominates---, 
SFDM can provide two characteristic (Jeans) length scales
below which structure formation is suppressed.
Regardless of self-interaction, the quantum nature of SFDM particles always smoothes fluctuations
below their de-Broglie wavelength.
For example, DM particles with mass
$m \simeq 10^{-22}$ eV$/c^2$ would have a corresponding de-Broglie wave length 
$\lambda_{\rm deB}$ of order a kpc (i.e., typical scale of CDM small-scale structure problems).
Moreover, there arises another length scale, $l_{\rm SI}$, from the (repulsive) self-interaction, 
should it be significant, given by $\lambda/(mc^2)^2$.
In fact, in the Thomas-Fermi regime, 
$l_{\rm SI}$ is the only length scale that is responsible for suppressing
structure growth, because then $l_{\rm SI} \gg \lambda_{\rm deB}$.
For instance, $l_{\rm SI} \simeq 1$ kpc if $\lambda/(mc^2)^2 = 2\times10^{-18}$ eV$^{-1}$ cm$^3$.
Hence, $m$ can be larger than the value of interest suggested by the noninteracting case,
if $\lambda$ is higher as well, and yet the model retains its characteristic length scale,
as long as the ratio $\lambda/(mc^2)^2$ stays constant (see \cite{2012MNRAS.422..135R} for details).
In this paper, the SI is adopted as the system of units, 
in which $[m]=$ eV/$c^2$ and $[\lambda]=$ eV cm$^3$.
We note that fiducial dimensional couplings of order $\lambda \approx 10 ^{-62}$ eV cm$^3$
correspond to dimensionless couplings of order\footnote{
This number is roughly 40 orders of magnitudes below the coupling for 
a $m \sim 10^{-5}$ eV QCD axion. The self-interaction is attractive for the latter, however.}
$\lambda m^2 c/\hbar^3 \approx 10^{-92}$ for $m = 10^{-22}$ eV/$c^2$.
While couplings even this small are enough to resolve the small-scale problems 
for higher mass DM particles, they also render these models qualitatively different\footnote{
This qualitative change in models, when a small coupling is added
(i.e. a quartic term), has been found already earlier in the literature on boson stars, 
which are also described as self-gravitating scalar fields, see \cite{1986PhRvL..57.2485C}.
So, it is not too surprising to rediscover similar consequences for scalar fields as the dark matter.} 
from those with $\lambda \equiv 0$.

In Paper I we found that, for the large-charge regime of interest, 
self-interacting SFDM starts relativistic in the early universe,
with an equation of state (EOS) evolving from stiff
($w\equiv p/\rho \simeq 1$) to radiationlike ($w \simeq 1/3$),
before becoming nonrelativistic at late times ($w \simeq 0$).
In the limit of a vanishing self-interaction ($\lambda\to 0$), 
the intermediate radiationlike phase of SFDM simply vanishes.
In either case, it is the kinetic term in Eq. (\ref{Lag}) that dominates 
the energy density of SFDM at early times, 
with a negligible oscillation whose frequency is less than the expansion rate,
and the EOS of SFDM approaches that of maximally ``stiff'' matter, $w_{\rm SFDM} \simeq 1$. 
This evolutionary phase of a scalar field has sometimes been referred to as ``fast-roll''.
When the fast-rolling scalar field is the dominant component of the universe, 
this period of the expansion history is also referred to as ``kination'' 
\cite{Joyce1997, 1998PhRvD..58b3503F, 2010PhRvD..82h3501D}.

It is important to note that this earliest stiff phase of SFDM 
is a generic feature of scalar field dynamics \cite{1985PhLB..155..232B}.
However, unlike the case of a single real scalar field, 
a complex scalar field in the \emph{large-charge} limit of interest here
does not evolve from the stiff, fast-roll ($w=1$) phase toward a slow-roll attractor 
(i.e., behaving like a cosmological constant, $w=-1$). 
The dynamical possibilities for a complex scalar field are actually richer than this, 
even for simple power-law potentials like $\frac{1}{2}mc^2|\psi|^2$.
This is shown, for example, by \cite{1995PhRvD..51.5698S}. 
For a complex scalar field with a $U(1)$ symmetry, 
the dynamical evolution of the field is different in the large- and small-charge limits, 
respectively. In the \emph{small-charge }limit, the complex field can behave 
as an effective real scalar field, in which case the slow-roll phase described above 
is expected to appear, until the oscillation frequency of the field exceeds the expansion rate. 
After that, the phase angle of the complex scalar field remains almost fixed, 
while the oscillation is in the amplitude alone \cite{2002PhRvD..65h3514A}.
However, in the \emph{large-charge} limit ($Q\simeq\rho_{\rm SFDM,0}/mc^2$, the case of interest here), 
something very different happens. In this case, if the field starts out 
in a fast-roll (stiff) phase, it does not evolve into a slow-roll phase 
before its oscillation frequency exceeds the expansion rate. 
And after that, the field evolves according to the pattern for which 
the oscillation is actually in the phase angle, rather than in the amplitude. 
In the latter case, the behavior of the field when the oscillation frequency 
exceeds the expansion rate is referred to as  ``spintessence'', 
as mentioned in \S\ref{sec:SFDM1}.

During this phase in which the oscillation frequency of the complex scalar field 
exceeds the expansion rate (the spintessence phase), 
the quartic term in Eq. (\ref{Lag}) can dominate the SFDM energy density, 
for large enough $\lambda/(mc^2)^2$.
Then, the EOS of SFDM is that of radiation, namely $w_{\rm SFDM} \simeq 1/3$.
The early universe thus experiences a boost in its expansion rate due to this extra relativistic species
in both the stiff and radiationlike phases of SFDM.
Eventually, the (quadratic) mass term in Eq. (\ref{Lag}) comes to dominate, 
which guarantees that SFDM behaves like CDM in the late universe, 
with or without self-interaction.
More precisely, this term must dominate after
the time of matter-radiation equality at a scale factor of $a_{\rm eq} \simeq 3\times 10^{-4}$,
in order to reproduce a period of ``CDM-like'' matter domination with $w_{\rm SFDM} \simeq 0$, 
the same as that in $\Lambda$CDM during which structure forms.

The transitions between these phases,
determined by SFDM particle mass and self-interaction coupling strength, are therefore
constrained by cosmological observables, particularly $N_{\rm eff}$, 
the effective number of neutrino species during Big Bang nucleosynthesis (BBN),
and $z_{\rm eq}$, the redshift of matter-radiation equality.
There are other models that also change the expansion rate at early times 
relative to the standard model. Some of those do it by making the EOS of the universe
stiffer than radiationlike (i.e., $w>1/3$), while others do it by 
changing the number of relativistic species while leaving the EOS still radiationlike.
For those models with an early era with an EOS stiffer-than-radiation, 
BBN abundance observations primarily place an upper limit on the duration of the stiff era
-- i.e., the stiff era must end before BBN.
For those models with extra contributions to the total energy density with a radiationlike EOS,
instead, the BBN constraint places an upper limit on the number of extra relativistic species,
which is the same during BBN and at later times 
when observables like $z_{\rm eq}$ and the CMB anisotropy 
place additional constraints on $N_{\rm eff}$, independently.
In our model, however, the evolution of the SFDM EOS causes \emph{both} of these effects 
to occur: the stiff-SFDM-dominated early era \emph{and} a relic radiationlike contribution 
\emph{after} the stiff era ends. In this case, the SFDM model must satisfy 
both kinds of constraints, that which limits the expansion rate \emph{during} BBN 
and that which limits the extra radiationlike contributions \emph{after} BBN, as well.
For most other models, those constraints are expressed 
as an allowed range of values of $N_{\rm eff}$, assumed to be a fixed quantity 
which does not evolve during BBN or \emph{between} BBN and $z_{\rm eq}$.
The standard value of $N_{\rm eff,standard}=3.046$ \cite{NEFFS} 
accounts for the presence of the three Standard Model neutrinos.
For the SFDM model, however, we must consider the \emph{evolution} of $N_{\rm eff}$ 
and subject it to different constraints at different epochs. 
SFDM allows $N_{\rm eff}$ to be \emph{higher} at BBN than at $z_{\rm eq}$, in fact, 
which current observations seem to prefer (see \S\ref{sec:constraintBBN}).
In Paper I, we found that $m\geq2.4\times10^{-21}$ eV$/c^2$ and $9.5\times10^{-19}$ eV$^{-1}$
cm$^3\leq\lambda/(mc^2)^2\leq4\times10^{-17}$ eV$^{-1}$ cm$^3$, 
due to cosmological constraints on $N_{\rm eff}$ and $z_{\rm eq}$.

To reiterate, before the familiar radiation-dominated era, there is
an earlier era of stiff-SFDM-domination, and the expansion rate in the early $\Lambda$SFDM
universe is increased compared with that in $\Lambda$CDM.
Interestingly, in our model, dark matter dominates twice in the history of the universe: 
first in its stiff phase before BBN,
and later in its dust-like phase, giving rise to a standard CDM-like matter era.

In this paper, we will expand our previous analysis by embedding $\Lambda$SFDM 
in the standard inflation paradigm and studying the impact of SFDM on 
primordial gravitational waves (GWs) produced during inflation, 
which contribute to $N_{\rm eff}$ as well.


\subsection{SFDM within the standard inflationary cosmology}
\label{sec:intro_inf}

In Paper I, we showed that by setting the conserved charge of the complex scalar field
so as to match the abundance of the DM in the observed universe at the present,
and evolving the field and background universe together over time, 
the field was compelled to dominate the
total energy density at early times. 
We stopped short, however, of asking how this SFDM-dominated phase
was consistent with standard inflationary cosmology 
in which the energy density was dominated initially by the inflaton field. 
Here we merge these two pictures self-consistently by postulating that
the end of inflation was followed by reheating, in which the inflaton decayed primarily into SFDM,
while also producing the other particles of the SM.

More precisely, we envisage the early cosmic evolution as follows.
Standard slow-roll single-field inflation produces nearly
scale-invariant fluctuations of the metric of the universe, which
can be decomposed into scalar, vector and tensor fluctuations. 
The energy scale during inflation is related to 
the ratio of the amplitude of tensor perturbations to the scalar amplitude, 
also called the tensor-to-scalar ratio, $r$. 
This quantity is pursued by CMB polarization experiments \cite{2015arXiv150202114P,
2016PhRvL.116c1302B}, because primordial tensor perturbations induce
quadrupole anisotropies in the CMB temperature, 
which leaves an imprint on the B-mode of CMB polarization 
(the ``recombination bump'' in the BB power spectrum) 
\cite{1997PhRvL..78.2054S, 2016PhRvL.116c1302B}.
The spectrum of (nearly) scale-invariant tensor perturbations can be
parametrized by a power law, determined by the tensor amplitude
$A_t$ (the product of the scalar amplitude $A_s$ and $r$) and the
tensor spectral index $n_t$. These tensor perturbations will become
gravitational waves (GWs) once they reenter the horizon. 
We further assume that inflation is followed by an epoch of reheating with
matter-like EOS, $w = 0$. This is a reasonable, standard choice for a
prolonged period of reheating (see, e.g., \cite{2015IJMPD..2430003A}). 
As already described in \S\ref{sec:SFDM1}, we are interested in scenarios in which 
the DM bosons are born at the end of reheating with a high charge density and
low entropy density, and thus find themselves rapidly occupying their ground state
(the zero-momentum state). 
As soon as SFDM arises, its energy density obeys a stiff EOS ($w=1$). 
Again, this is because, for a scalar field, 
the kinetic term in the Lagrangian, Eq. (\ref{Lag}), goes as $a^{-6}$. 
Since SFDM dominates the cosmic energy budget at early times,
compared to other cosmic components, 
the stiff EOS of the SFDM is also the EOS of the universe at this time.
For simplicity, we adopt an instant transition, i.e. the reheating
temperature $T_{\rm{reheat}}$ at the end of reheating also
corresponds to the point after which there is the ``stiff-SFDM-dominated'' era.

We note that in this work the only source of primordial GWs that we
consider are those predicted by the ``plain-vanilla'' single-field, 
slow-roll inflation model, for which the consistency relation, $n_t=-r/8$, holds. 
Tensor fluctuations from inflation are isotropic
and stochastic in nature. Therefore, they contribute to the
stochastic gravitational wave background (SGWB), 
giving rise to an effective homogeneous energy density of primordial GWs, 
which we will elaborate in more detail in \S\ref{sec:TmnGW}. 
Such a SGWB is described by its energy density spectrum, $\Omega_{\rm GW} (k, a)$,
i.e. the fraction of the critical energy density carried by GWs per logarithmic wavenumber interval 
at any comoving wavenumber $k$ and scale factor $a$. 
The dispersion relation of GWs today is simply given by $f=kc/2\pi$, 
in which $f$ is the (comoving) frequency. 
The differential GW energy density at any frequency $f$
generically decays like radiation $\sim a^{-4}$ once that mode reenters the horizon.

As we will show in this paper, it is the stiff era caused by SFDM
that will amplify the GWs produced during inflation. 
It was first considered by Grishchuk in his seminal paper \cite{1974ZhETF..67..825G} 
that cosmological GWs can be amplified in a universe whose EOS is stiffer than that of radiation, 
i.e., $w > 1/3$, implying that the corresponding $\Omega_{\rm GW} (k, a)$, 
which indicates their contribution to the total energy density of the universe,
will increase over time. 
The results in Grishchuk's paper showed the conditions 
in which gravitons can be massively produced in the early universe, 
from initial tensor-type quantum fluctuations, 
which were later developed and applied to the modern-day inflationary paradigm
(see \cite{1992PhR...215..203M} and references therein for a review).
In contrast, we study the post-inflationary evolution of existing GWs produced during standard inflation.

We will point out in this paper that, for any mode $k$, 
the value of $\Omega_{\rm GW} (k, a)$ measured at some time 
long after that mode reenters the horizon depends on 
the critical energy density of the universe when it is measured, 
and two things which are $k$-dependent: 
the critical energy density at the time of its horizon-crossing 
and the number of e-foldings between the horizon-crossing and the measurement.
We will show that, for any mode $k$ which reenters prior to the end of the stiff phase, 
it is the shortening of the time spent undergoing radiationlike decay due to the stiff phase, 
compared to the $\Lambda$CDM expansion history, 
that is responsible for the amplification of $\Omega_{\rm GW} (k, a)$. 
The expansion history of $\Lambda$SFDM with its stiff era will,
therefore, predict a characteristic GW energy density spectrum
$\Omega_{\rm GW} (k, a)$ at a late time, in which the spectrum shows a blue-tilt, 
$\Omega_{\rm GW} (k, a)\propto k$, for any mode $k$ reentering during the stiff era ($w=1$), 
a peak at $k_{\rm reheat}$ for the mode that reenters at the end of reheating ($w=0$), 
and a decline for higher $k$ as $\Omega_{\rm GW} (k, a)\propto k^{-2}$ (red-tilt). 

We calculate here the present-day GW energy density spectrum, $\Omega_{\rm GW}
(f)\equiv\Omega_{\rm GW}(k=2\pi f/c, a=1)$, as probed by current and
future laser interferometer experiments \cite{2016PhRvL.116m1102A, 2012JCAP...06..027B}. 
These experiments are sensitive to the tensor
deformation of space, or the strain, induced by incoming GWs, to a high accuracy. 
We predict a detectable signal from the SGWB generated by \textit{standard} inflation, 
which is within reach of the sensitivity of the ongoing Advanced LIGO/Virgo experiments, 
for a broad range of $T_{\rm reheat}$ and SFDM parameters. 
This provides a novel science target, given that the expected signal from the standard
cosmological model lies many orders of magnitudes below the
sensitivity limit of those experiments. Meanwhile, pulsar timing
array (PTA) experiments \cite{2016ApJ...821...13A,
2015Sci...349.1522S, 2015MNRAS.453.2576L} also detect strain
signals, but at lower frequency ranges. We remark that the predicted
SGWB signal in the $\Lambda$SFDM model, however, lies well below the
upper bounds reported by current PTA experiments in those frequency ranges.

The SGWB also affects the expansion history
as an extra relativistic degree of freedom by boosting the expansion rate of the background universe,
thereby contributing to $N_{\rm eff}$. In contrast to $\Lambda$CDM with standard inflation,
in which the contribution to the background energy density of the universe from primordial GWs
is negligible (and thus uninteresting),
$\Lambda$SFDM, however, amplifies those primordial GWs so that 
they need to be taken into account in the budget of $N_{\rm eff}$.
Therefore, the SGWB from inflation actually needs to be included in the Friedmann equation
for the average universe in a self-consistent manner.
In other words, we must study the back-reaction of the inflationary SGWB 
on the expansion rate of the average universe,
which in turn affects the evolution of the SGWB, itself, 
an effect which has been neglected in previous literature. 
We stress that we include the fully-coupled evolution of \tx{all} the cosmic components 
in our calculation of the back-reaction.
In light of this effect, $\neff$ thus has two additional sources: the direct contribution from SFDM,
and a new one from the \tx{enhanced} $\ogw$.
This puts additional constraints on the SFDM parameters, $m$ and $\lambda/(mc^2)^2$.
In what follows, we will update the $N_{\rm eff}$ and $z_{\rm eq}$ constraints 
on the SFDM parameters studied in our Paper I, incorporating the new effect from primordial SGWB.

The impact on the primordial SGWB of an early era 
whose EOS is stiffer than radiation (with $1 \geq w \geq 1/3$) 
has been considered in different contexts in previous literature, 
in which such an era was postulated to arise before BBN.
The possibility that inflation ended with the onset of a brief stiff era 
was considered by \cite{1998PhRvD..58h3504G}, 
who calculated the effect on the inflationary SGWB energy density 
of assuming the EOS switched from a constant value of $w$ in the range $1/3<w\leq 1$ 
for the stiff era to $w=1/3$ for the standard radiation-dominated era.
A possible agent considered for the stiff era was quintessential inflation, 
studied in \cite{1999PhRvD..59f3505P}, 
in which the inflaton field transitions from a slow-roll phase to a kinetic-energy-dominated phase. 
Its impact on the SGWB has been considered in 
\cite{1999CQGra..16.2905G, 1999PhRvD..60l3511G}, 
where a blue tilt in the GW energy density spectrum was predicted.
However, unlike the present work, there was no standard reheating epoch 
between the end of inflation and the stiff era in those investigations.
The requirement that the amplification of the SGWB 
relative to the standard radiation components not violate observational constraints 
on the early universe was discussed by \cite{2008PhRvD..78d3531B} 
(based on \cite{2008PhRvD..77f3504B}).  
They expressed this by defining an effective EOS parameter $\hat w$, 
which is a weighted mean of $w$ over cosmic time, for which they calculated an upper limit.
The above works pointed out that the high-frequency extrapolation of the same SGWB 
which contributed to the expansion rate at early times might be detected or constrained 
by GW laser interferometer experiments.
Unlike the present work described, however,  
the back-reaction of the GWs on the expansion rate 
has been neglected in the aforementioned literature. 
\footnote{Regarding back-reaction, \cite{1998PhRvD..58h3504G} 
considered the evolution of a 2-component universe 
consisting of a stiff component with a constant $w$ and a radiationlike SGWB component, 
but treated the latter only as a perturbation. 
\cite{2006PhRvD..73h3505G} also considered the SGWB 
produced in pre-big-bang models \emph{without} inflation 
and its back-reaction on the bouncing solutions for such models.}
Finally, we note that the context in which the stiff era appears in the present work
as an inevitable consequence of the evolution of the complex scalar field in the $\Lambda$SFDM model
has no precedent in earlier work.

This paper is organized as follows. 
In \S\ref{sec:basic}, we present the basic equations 
concerning the composition and expansion history of the $\Lambda$SFDM model, 
and the homogeneous evolution of each component, 
especially SFDM and the SGWB from inflation. 
In \S\ref{sec:Evolution}, we discuss the solutions to these equations, 
providing both analytical insights and numerical treatment, 
especially with regard to the SGWB,
and show the holistic expansion history of $\Lambda$SFDM from inflation through the present. 
We also describe our numerical method for a self-consistent account of the SGWB,
and show the evolution of several example $\Lambda$SFDM models from our numerical calculations,
which delineate the evolutionary phases in $\Lambda$SFDM
and demonstrate a nontrivial contribution from the amplified SGWB from inflation.
In \S\ref{sec:constraint}, we then derive the new constraints on the SFDM particle parameters
required to satisfy the cosmological observables $z_{\rm eq}$ and $N_{\rm eff}$, 
and discuss the impact of the SGWB on these constraints, 
which is dependent on the values of $r$ and $\tre$.
In \S\ref{sec:SGWBspectra}, we present one of our most remarkable results: 
the present-day inflationary SGWB energy density spectrum in the $\Lambda$SFDM model 
is so highly amplified relative to its amplitude in $\Lambda$CDM 
that it may be detectable by the ongoing Advanced LIGO/Virgo (aLIGO/Virgo) experiment. 
We will show that the expected signal-to-noise ratio (SNR) of this unique SGWB signal 
can be significant for a wide range of SFDM parameters and reheat temperatures,
for currently allowed values of $r$.
The SFDM model can thus be tested for parameters in this range. 
In fact, we show that the null detection of the SGWB recently reported by the aLIGO O1 run
excludes part of the parameter range for an illustrative family of $\Lambda$SFDM models, 
thereby demonstrating that GW detection experiments 
can already place a new kind of cosmological constraint on SFDM.
The accessible range will grow over time as aLIGO/Virgo completes its planned observing runs.
Hence, our results provide an additional motivation for LIGO to search for SGWB signals, 
since this has the potential to probe the nature of dark matter, 
reheating physics and inflation parameters.
In \S\ref{sec:discussion}, we briefly discuss several aspects 
in which our results in this paper can be extended
in anticipation of future developments of measurements of BBN light-element abundances, 
and of the space laser interferometer mission LISA.
We summarize our conclusions in \S\ref{sec:conclusion}.
Appendices \ref{ap:GW}--\ref{app:SNR_O1} 
contain some additional materials which we defer from the main text for better readability.


\section{Basic Equations}
\label{sec:basic}

\subsection{The Background universe}
\label{sec:background}

As in Paper I, we will consider the background universe
to be homogeneous and isotropic on large scales, 
as described by the spatially-flat 
Friedmann-Lema\^itre-Robertson-Walker (FLRW) metric tensor.
In this work, we must also consider the perturbations $\delta g_{\mu\nu}$ 
to this unperturbed FLRW metric $\bar g_{\mu\nu}$, 
corresponding to the tensor modes.
In the cosmological ``comoving frame''
\footnote{Rigorously speaking, this reference frame is exactly comoving with cosmic flows 
only if the universe is perfectly homogeneous and isotropic, i.e., no fluctuations.},
it can be written as
\begin{IEEEeqnarray}{rCl}\label{eq:metric}
    \ud s^2 & \equiv & g_{\mu\nu}\ud x^\mu\ud x^\nu = (\bar g_{\mu\nu}+\delta g_{\mu\nu})\ud x^\mu\ud x^\nu\nonumber\\
    & = & c^2\ud t^2-a^2(t)(\delta_{ij}+h_{ij})\ud x^i\ud x^j,
\end{IEEEeqnarray}
where $a(t)$ is the scale factor, 
and $h_{ij}$ is a symmetric tensor which
characterizes tensor perturbations to the metric, $|h_{ij}|\ll 1$ (weak-field limit).
The gauge-invariant $h_{ij}$ satisfies the transverse and traceless
conditions (see, e.g., \cite{2005pfc..book.....M}), \footnote{If
$h_{ij}$ is instead a generic 3-tensor that describes spatial metric perturbations, 
the conditions in Eq. (\ref{eq:tt}) would be regarded
as coordinate conditions, known as the transverse-traceless (TT)
gauge \cite{1973grav.book.....M}.}
\begin{equation} \label{eq:tt}
    \partial_i h^{ij}=0,\quad h_{i}^{~i}=0,
\end{equation}
where indices of $h_{ij}$ are raised and lowered by the spatial background metric $\delta_{ij}$; 
$h^{ij}=\delta^{ik}\delta^{jl}h_{kl}$. 
In this paper, we follow the Einstein summation convention. 
It is understood that there also be generic small perturbations
corresponding to scalar and vector modes as well, 
the growth of which we do not study in this paper.
The metric perturbations associated with tensor modes are special, however, 
in that they also contribute an effective stress-energy tensor $T_{\mu\nu,~\rm GW}$ 
as gravitational waves, 
as we show in \S\ref{sec:TmnGW} and Appendix \ref{ap:GWtmn}.

The evolution of the metric of the background universe is governed by
the Einstein field equations,
\begin{equation}\label{eq:EFE}
    R^\mu_{~\nu}-\frac{1}{2}R=\frac{8\pi G}{c^4}T^\mu_{~\nu},
\end{equation}
where $R^\mu_{~\nu}$ is the Ricci tensor which can be calculated from the metric in Eq. (\ref{eq:metric}).
The time-time component of the stress-energy tensor, $T^0_{~0}$,
defines the energy density.
For the background universe, it is sufficient to solve only the time-time component of
the Einstein field equations, which amounts to the Friedmann equation,
plus the energy conservation equations of each component
that constitutes the total $T_{\mu\nu}$ of the universe (see \cite{Weinberg}).
In many cases, the latter can be derived from the equation of motion of the component.
Therefore, we will evaluate both sides of the time-time component of Eq. (\ref{eq:EFE})
and also find the contribution to the total energy density of the universe from each component.

The expansion of the homogeneous FLRW universe is governed by the Friedmann equation, 
which is derived from the time-time component of the Einstein equations (\ref{eq:EFE}). 
For our model,
\begin{widetext}
\begin{numcases}{\hspace{-0.2in}H^2(t) \equiv \left(\frac{\ud a/\ud t}{a}\right)^2 =}
    H^2_{\rm inf}, & $\hspace{0.in}a<a_{\rm inf},$\label{eq:friedmanninf}\\[0.5em]
    H^2_{\rm inf}\left(\frac{a_{\rm inf}}{a(t)}\right)^3, & $\hspace{0.in}a_{\rm inf}<a<a_{\rm reheat},$ \label{eq:friedmannreheat}\\[0.3em]
    \frac{8\pi G}{3c^2}\left[\rho_r(t)+\rho_b(t)+\rho_\Lambda(t)+\rho_{\rm{SFDM}}(t) + \rho_{\rm{GW}}(t)\right], &  $\hspace{0.in}a>a_{\rm reheat},$ \label{equation:friedmannbu}
\end{numcases}
\end{widetext}
where $a_{\rm inf}$ is the scale factor at the end of inflation when $H(t)=H_{\rm inf}$, 
$a_{\rm reheat}$ is the scale factor when reheating ends at $T=\tre$, 
and we have assumed that $w=0$ during reheating.
In our model, SFDM accounts for all of the cosmological dark matter. 
Apart from SFDM and gravitational waves, all the other cosmic components
are the same than in $\Lambda$CDM, i.e. a radiation component $\rho_r$, baryons $\rho_b$, and a cosmological constant $\rho_{\Lambda}$ (see Eq. [\ref{equation:friedmannbu}]).
The evolution of each component is described in \S\ref{sec:Evolution}.

\subsubsection{Energy density contribution from SFDM}
\label{sec:TmnSFDM}

Let us write down the Lagrangian density of the SFDM again,
\begin{equation*}
	\mathcal{L} = \frac{\hbar^2}{2m}g^{\mu \nu}\partial_{\mu}\psi^*\partial_{\nu}\psi 
	- \frac{1}{2}mc^2|\psi|^2 -\frac{\lambda}{2}|\psi|^4,
\end{equation*}
where the metric $g_{\mu\nu}$ is described in Eq. (\ref{eq:metric})
and the definitions of the particle mass $m$ and self-interaction coupling strength $\lambda$
have been explained in \S\ref{sec:paper1}. The field $\psi$ can be written as
\begin{equation}
    \psi=|\psi|e^{\rm{i}\theta},
\end{equation}
where $|\psi|$ is its modulus and $\theta$ is its phase.

In general, the stress-energy tensor of a field with Lagrangian density $\mathcal{L}$ is given by
\begin{equation}
    T_{\mu\nu}=2\frac{\delta\mathcal{L}}{\delta g^{\mu\nu}}-g_{\mu\nu}\mathcal{L}.
\end{equation}
Hence, the stress-energy tensor of SFDM can be evaluated as
\begin{IEEEeqnarray}{rCl}\label{equation:sfdmtmn}
    T_{\mu\nu,~\rm{SFDM}} & = &\frac{\hbar^2}{2m}(\partial_\mu\psi^*\partial_\nu\psi+\partial_\nu\psi^*\partial_\mu\psi)-\nonumber\\
    & & -g_{\mu\nu}\left(\frac{\hbar^2}{2m}g^{\rho\sigma}\partial_\rho\psi^*\partial_\sigma\psi- \right.\nonumber\\
    & & \left. -\frac{1}{2}mc^2|\psi|^2-\frac{\lambda}{2}|\psi|^4\right).
\end{IEEEeqnarray}
In linear theory with the perturbed FLRW metric (\ref{eq:metric}), we have verified that, 
to the first order, the complex SFDM behaves as a perfect fluid, 
because $T_{\mu\nu,~\rm SFDM}$ can be written in the following form as for a perfect fluid, 
characterized by its energy density $\rho_{\rm SFDM}$, isotropic pressure $p_{\rm SFDM}$ and 
4-velocity $u^\mu\equiv c(\ud x^\mu/\ud s)$, with no anisotropic stress,
\begin{equation} \label{equation:fluidt}
    T_{\mu\nu,~\rm SFDM}=(\rho_{\rm SFDM}+p_{\rm SFDM})u_\mu u_\nu/c^2-g_{\mu\nu}p_{\rm SFDM}.
\end{equation}

For the homogeneous and isotropic background universe,  $u^0=c$ and $u^i=0$, 
and thus, $T_{\mu\nu,~\rm SFDM}$ becomes diagonal. Its time-time component 
is recognized as the spatially-averaged energy density of SFDM, 
\begin{IEEEeqnarray}{rCl} \label{equation:densitybu}
    \rho_{\rm SFDM} & \equiv & T^0_{~0,~\rm{SFDM}} = \frac{\hbar^2}{2mc^2}|\partial_t\psi|^2+\nonumber\\
    & & + \frac{1}{2}mc^2|\psi|^2+\frac{1}{2}\lambda|\psi|^4\nonumber\\
    & = & \frac{\hbar^2}{2mc^2}\left[\dot{|\psi|}^2+|\psi|^2\dot{\theta}^2\right]\nonumber\\
    & & + \frac{1}{2}mc^2|\psi|^2+\frac{1}{2}\lambda|\psi|^4,
\end{IEEEeqnarray}
where an ``overdot'' ($\dot{~}$) indicates the derivative with respect to the cosmic time $\ud/\ud t$,
throughout this paper. In the equation above and thereafter, we assume that
the complex function $\psi$ always means the spatially-averaged value of the BEC wave function,
which adequately accounts for the SFDM contribution to the background universe.
The space-space component of $T_{\mu\nu,~\rm SFDM}$ is recognized as 
the spatially-averaged pressure, 
\begin{IEEEeqnarray}{rCl}\label{equation:pressurebu}
	p_{\rm SFDM} & \equiv & -T^i_{~i,~\rm SFDM}=\frac{\hbar^2}{2mc^2}\left[\dot{|\psi|}^2+|\psi|^2\dot{\theta}^2\right]\nonumber\\
	& & -\frac{1}{2}mc^2|\psi|^2-\frac{1}{2}\lambda|\psi|^4. 
\end{IEEEeqnarray}
Hereafter in this paper, $\rho_{\rm SFDM}$ and $p_{\rm SFDM}$ will always refer to 
the homogeneous part of the energy density and pressure of SFDM, which are only functions of time. 

Eqs. (\ref{equation:densitybu}) and (\ref{equation:pressurebu}) 
can also be rearranged into a useful form, 
in which $\rho_{\rm SFDM}$ and $p_{\rm SFDM}$ are related to $|\psi|^2$, 
\begin{IEEEeqnarray}{rCl}
     \rho_{\rm SFDM} & = & \frac{\hbar^2}{2mc^2}\left(\frac{(\ud|\psi|^2/\ud t)^2}{4|\psi|^2}
	+\frac{(a^3|\psi|^2\dot{\theta})^2}{a^6|\psi|^2}\right)\nonumber\\
	& & +\frac{1}{2}mc^2|\psi|^2+\frac{1}{2}\lambda|\psi|^4,\label{eq:rhopsi2}\\
     p_{\rm SFDM} & = & \frac{\hbar^2}{2mc^2}\left(\frac{(\ud|\psi|^2/\ud t)^2}{4|\psi|^2}
	+\frac{(a^3|\psi|^2\dot{\theta})^2}{a^6|\psi|^2}\right)\nonumber\\
	& & -\frac{1}{2}mc^2|\psi|^2-\frac{1}{2}\lambda|\psi|^4\label{eq:ppsi2}.
\end{IEEEeqnarray}
We note that in the numerator of the 2nd term above, $a^3|\psi|^2\dot{\theta}$ is a conserved quantity 
as it is proportional to the comoving charge density (see Appendix B in Paper I). 
In fact, 
\beq\label{eq:conservedQ}
	a^3|\psi|^2\dot{\theta}\equiv\frac{mc^2}{\hbar}Q=\rho_{\rm SFDM,0}/\hbar, 
\eeq
where $\rho_{\rm SFDM,0}$ is the present-day dark matter energy density. 
The last equality in the equation above expresses the fact that
our SFDM today can be treated as nonrelativistic particles. 
On the other hand, it is shown in Paper I that $\dot\theta\cong mc^2/\hbar$ 
when SFDM is nonrelativistic as ``dust-like''. 
Therefore, the number density of SFDM at present, equivalent to the comoving charge density, 
is given by
\beq\label{eq:chargedensity}
	|\psi|^2\Big |_{a=1}=Q=\rho_{\rm SFDM,0}/mc^2.
\eeq

\subsubsection{Energy density contribution from gravitational waves}
\label{sec:TmnGW}

As pointed out in \cite{1973grav.book.....M}, gravitational waves,
squeezing and stretching the local metric
perpendicular to their direction of propagation through space-time,
must carry energy. In fact, an (effective) stress-energy tensor of GWs,
$T_{\mu\nu,~\rm{GW}}$, can be defined for small tensor perturbations
to the background metric, which is slowly-varying on scales larger than the wavelength,
as shown in Appendix \ref{ap:GWtmn}.

The effective energy density associated with tensor perturbations $h_{ij}$ can be written as follows: 
\beq\label{eq:rhoGW1}
	\rho_{\rm GW}\equiv T^0_{~0,~\rm{GW}}=\frac{c^2}{64\pi G}\langle\p_th_{ij}\p_th^{ij}+\f{c^2}{a^2}\nabla h_{ij}\cdot\nabla h^{ij}\rangle,
\eeq
\cite{2008PhRvD..77f3504B} where the brackets $\langle\cdot\rangle$ 
denote the spatial average over several wavelengths. 
In particular, this will describe the effect of the SGWB from inflation of interest here. 
Since primordial fluctuations (including the tensor sector) produced by most inflation models 
are predicted to be Gaussian, the SGWB can therefore be fully characterized by its power spectrum.
As a result, the spatial average $\langle\cdot\rangle$ defined above is equal to the ensemble average. 
Furthermore, we assume that this ensemble average of tensor fluctuations 
is unpolarized and isotropic on large scales, 
according to the standard paradigm of inflation and reheating, as mentioned in \S\ref{sec:intro_inf}. 
This guarantees that the SGWB produced by inflation is homogeneous on large scales.
Hence, applying the Fourier decomposition to $h_{ij}$ 
(see Appendix \ref{ap:GWfourier}, Eq. [\ref{eq:fourier}]), 
we can write down the (dimensionless) power spectrum of the SGWB, $\Delta^2_h(k,t)$, 
or the tensor power spectrum, in terms of its mode functions $h^P_{\mathbf{k}}$, as follows:
\beq\label{eq:tensorps}
	\hspace{-0.1in}
	k^3\langle h^P_{\mathbf{k}}(t)(h^{P'}_{\mathbf{k'}}(t))^*\rangle\equiv 2\pi^2\Delta^2_h(k,t)\delta^{(3)}_D(\mathbf{k}-\mathbf{k'})\f{\delta_{PP'}}{4},
\eeq
where $k=|\mathbf{k}|$ is the comoving wavenumber, 
$P=+,\times$ stands for the two linear polarization states of $h_{ij}$, 
$\delta^{(3)}_D$ is the Dirac delta function and $\delta_{PP'}$ denotes the Kronecker delta. 
As required, $\Delta^2_h(k,t)$ does not depend on the direction of the comoving wave vector 
$\mathbf{k}$ nor the polarization state $P$, but only on the magnitude $k=|\mathbf{k}|$, 
capturing all statistical properties of the stochastic metric perturbation $h_{ij}$.\footnote{
Strictly speaking, to ensure that the tensor power spectrum 
is only a function of the wavenumber $k$ at any time $t$, 
we also need to investigate the evolution of $h_{ij}$ via its equation of motion, 
which is explained in \S\ref{sec:tensoreq}. 
}
Inserting Eq. (\ref{eq:tensorps}) and Eq. (\ref{eq:fourier}) into Eq. (\ref{eq:rhoGW1}) yields
\begin{widetext}
\beq\label{eq:rhoGWfourier}
	\rho_{\rm{GW}}(t) = \frac{c^2}{64\pi G}\int^{\infty}_0\ud \ln k\left(\left|\f{\dot h^P_{\mathbf{k}}(t)}{h^P_{\mathbf{k}}(t)}\right|^2\Delta^2_h(k,t)+\frac{k^2c^2}{a^2(t)}\Delta^2_h(k,t)\right),
\eeq
\end{widetext}
where the term $|\dot h^P_{\mathbf{k}}(t)/h^P_{\mathbf{k}}(t)|^2$ has been extracted out of 
the ensemble average $\langle\cdot\rangle$, because it is deterministic, 
governed by the equation of motion of $h^P_{\mathbf{k}}(t)$, 
which we will show in \S\ref{sec:tensoreq}. 
There we will also explain why $|\dot h^P_{\mathbf{k}}(t)/h^P_{\mathbf{k}}(t)|^2$ 
does not depend on $P$, i.e., $P$ can be either $+$ or $\times$. 
As expected, Eq. (\ref{eq:rhoGWfourier}) shows that $\rho_{\rm{GW}}(t)$ is homogeneous in space, 
since it does not depend on position $\mathbf{x}$.

It is useful to define the differential SGWB energy density per logarithmic $k$ as
\vspace{-0.5in}
\beq\label{eq:rhoGWdiff}
	\f{\ud \rho_{\rm{GW}}}{\ud\ln k}(k,t) = \frac{c^2}{64\pi G}\left(\left|\f{\dot h^P_{\mathbf{k}}(t)}{h^P_{\mathbf{k}}(t)}\right|^2+\frac{k^2c^2}{a^2(t)}\right)\Delta^2_h(k,t).
\eeq

\subsection{Equation of motion: scalar field dark matter}
\label{sec:KGeq}

The equation of motion for SFDM is the Klein-Gordon equation. For a homogeneous scalar field, 
it is written as 
\beq \label{eq:KG}
\f{\hbar^2}{2mc^2}\ddot{\psi} + 3\f{\hbar^2}{2mc^2}\f{\dot{a}}{a}\dot{\psi} + \f{1}{2}mc^2\psi + \lambda |\psi|^2\psi = 0,
\eeq
 in terms of the BEC wave function $\psi(t)$.
It can be transformed into an equivalent form, namely, the energy conservation equation, 
in terms of the energy density $\rho_{\rm{SFDM}}$ and the corresponding pressure,
$p_{\rm{SFDM}}$, as follows:
\vspace{-0.1in}
\beq \label{eq:EC}
\dot{\rho}_{\rm{SFDM}} + 3\f{\dot{a}}{a}(\rho_{\rm{SFDM}} + p_{\rm{SFDM}}) = 0.
\eeq

The Klein-Gordon equation (\ref{eq:KG}) can be rearranged into the following form, 
\begin{widetext}
\begin{equation}\label{eq:KGpsi2}
	\frac{\hbar^2}{2mc^2}\left(\frac{\ud^2|\psi|^2}{\ud t^2}-\frac{(\ud|\psi|^2/\ud t)^2}{2|\psi|^2}\right)
	+\frac{\hbar^2}{2mc^2}\frac{3\dot{a}}{a}\frac{\ud |\psi|^2}{\ud t}
	-\frac{\hbar^2}{mc^2}\frac{(\rho_{\rm SFDM,0}/\hbar)^2}{a^6|\psi|^2}
	+mc^2|\psi|^2+2\lambda|\psi|^4=0,
\end{equation}
\end{widetext}
where we have made use of Eq. (\ref{eq:conservedQ}), replacing $a^3|\psi|^2\dot{\theta}$ by
$\rho_{\rm SFDM,0}/\hbar$.
In Eq. (\ref{eq:KGpsi2}), the dependent variable is essentially $|\psi|^2$, rather than $\psi$. 
We will see later in \S\ref{sec:nummeth} that it is this equation that we solve numerically
to obtain the early phase of the evolution of SFDM.

\subsection{Equation of motion: tensor perturbations}
\label{sec:tensoreq}

In the absence of anisotropic stresses,\footnote{
Actually, the presence of free-streaming relativistic neutrinos has been shown by 
 \cite{2004PhRvD..69b3503W} to contribute anisotropic stress 
 which modifies Eq. (\ref{eq:tensorC}). The correction which results can be treated by 
 a post-facto multiplicative factor which does not depend on wavenumber $k$, 
 as described in \cite{2005PhRvD..72h8302D}.
}
the Einstein equation for tensor perturbations $h_{ij}$ in a spatially flat FLRW universe with
scale factor $a$ reads
 \beq \label{eq:tensorC}
 \p^2_t{h}_{ij}(\mathbf{x},t) + 3\f{\dot{a}(t)}{a(t)}\p_t{h}_{ij}(\mathbf{x},t) - \f{c^2}{a^2(t)}\nabla^2h_{ij}(\mathbf{x},t) = 0.
 \eeq
 The equation above is essentially a cosmological wave equation, 
 its corresponding solutions are thus gravitational waves. 
In fact, the wave nature can be more directly manifested 
by rewriting the equation of motion above, Eq. (\ref{eq:tensorC}), 
in terms of Fourier mode functions $h^P_{\mathbf{k}}(t)$ 
(and their conjugate $(h^P_{\mathbf{k}}(t))^*$, see Eq. [\ref{eq:fourier}] for their definition),
 \beq \label{eq:tensork}
 \ddot{h}^P_{\mathbf{k}}(t) +
 3\f{\dot{a}(t)}{a(t)}\dot{h}^P_{\mathbf{k}}(t) +
 \f{k^2c^2}{a^2(t)}h^P_{\mathbf{k}}(t) = 0.
 \eeq
The equation above manifestly shows that the equation of motion for tensor perturbations 
only involves the magnitude $k=|\mathbf{k}|$ of the wave vector, 
not its direction nor the polarization state $P=+$ or $\times$. 
Therefore, as long as the initial condition for tensor perturbations is isotropic and unpolarized, 
so will they always be at any time throughout their evolution.
 This completes our justification to treat $\Delta^2_h(k,t)$ only as a function of $k$ at any time $t$.
 With no loss of generality, we can thereby assume 
 $h^+_{\mathbf{k}}(t)=h^\times_{\mathbf{k}}(t)\equiv h_k(t)$ 
 and henceforth treat $h_k(t)$ only.
 
 If we neglect the cosmological expansion (i.e., set $\dot a =0$) in Eq. (\ref{eq:tensork}), 
 then its solutions are simply 
 traveling plane waves with the dispersion relation $\omega_k=kc$. 
 Therefore, on time scales much less than a Hubble time, 
 tensor modes are plane waves propagating at the speed of light, 
 just like the GWs detected recently by the Advanced LIGO experiment, 
 sourced by binary black hole merger events \cite{2016PhRvL.116f1102A}. 
 GWs are also known as gravitational radiation, 
 or radiative degrees of freedom \cite{2005NJPh....7..204F}. 
 
It is convenient to express Eq. (\ref{eq:tensork}) with respect to the conformal time (length), 
$\tau = c\int dt/a(t)$, leading to an equation for $h_k(\tau)$,
  \beq \label{tensoreq}
  h_k''(\tau) + 2\f{a'(\tau)}{a(\tau)}h_k'(\tau) + k^2h_k(\tau) = 0,
 \eeq
 where the prime denotes the derivative with respect to conformal time $'\equiv d/d\tau$. 
 We will discuss the evolution of gravitational waves, and some analytical solutions, 
 in \S\ref{sec:EvolGW}.

\section{Evolution in the $\Lambda$SFDM universe}
\label{sec:Evolution}

In Paper I, we considered a universe with the same cosmic inventory
as the basic $\Lambda$CDM model except that CDM is replaced by SFDM,
the $\Lambda$SFDM model, since the late-time evolution of the $\Lambda$SFDM universe
is indistinguishable from that of standard $\Lambda$CDM after $z_{\rm eq}$, 
except for small-scale structure. We used the set of cosmological parameters
from the Planck 2013 data release \cite{Planck2013}. 
In this work, we will add to $\Lambda$SFDM the contribution due to $\rho_{\rm GW}$, 
as it is currently constrained by upper bounds. 
Also, we use the updated 2015 Planck data to solve for the evolution of this homogeneous
background universe \cite{2015arXiv150201589P}. 
A summary of the parameters we use can be found in Table \ref{table:tab1}.
The fractional energy densities are defined via $\Omega_i(t) \equiv \rho_i(t)/\rho_{\rm{crit}}(t)$ 
with the critical energy density of the universe at time $t$,
 \begin{equation}
    \rho_{\rm{crit}}(t)=\frac{3H^2(t)c^2}{8\pi G}.
 \end{equation}
Hereafter in this paper, unless otherwise noted as fluctuations, 
all physical quantities in space will refer to 
their spatially homogeneous, isotropic part, i.e., only functions of time.

First, we discuss the evolution of each of the cosmic components separately in 
\S\ref{sec:SFDMbackground}--\S\ref{sec:EvolOthers}, highlighting certain heuristic aspects.
We then put them altogether in \S\ref{sec:EvolALL} 
to derive the expansion history of the entire background $\Lambda$SFDM universe. 
In Paper I, we took the point of view that, 
since the cosmological parameters are known at the present 
(e.g., from CMB measurements), our solutions of the coupled Klein-Gordon 
and Friedmann equations must match this late-time universe.
In particular, the observed dark matter energy density at late times,
when the SFDM is nonrelativistic (dust-like), 
sets the value of the conserved comoving charge density $Q$, 
which in turn sets the amplitude of the field $|\psi|$ at the present 
(see Eq. [\ref{eq:chargedensity}]).
This field value combines with the observed Hubble constant 
and energy densities of the other components in Table \ref{table:tab1} 
to make the boundary conditions for the coupled evolution equations. 
In Paper I, these B.C.'s were satisfied by integrating backward in time from the present.
The results were checked against a forward time-integration.
Here, however, unlike in Paper I, the evolution is affected by $\Omega_{\rm GW}$, too, 
which must be included self-consistently in a forward time-integration 
which starts from the end of inflation, whose energy scale is set by our choice of $r$.
As a result, a more elaborate scheme is required than in Paper I.
In this paper, we only use backward integration to produce a guess 
for the initial conditions of the forward integration, and then converge on the final solution 
by an iterative scheme involving both forward and backward integrations.
We present the details of the numerical method in \S\ref{sec:nummeth} below.


\begin{table}[h]
\begin{center}
{\renewcommand{\arraystretch}{1.5}
\renewcommand{\tabcolsep}{.3cm}
\begin{tabular}[t]{cc|cc}
$h$ & 0.6781 & $\Omega_mh^2$ & 0.141\\
\hline
$\Omega_bh^2$ & 0.02226 & $\Omega_rh^2$ & $4.184\times 10^{-5}$\\
\hline
$\Omega_ch^2$ &  0.1186 & $z_{\rm{eq}}$ & 3365\\
\hline
$T_{\rm{CMB}}$/K & 2.7255 & $\Omega_\Lambda$ & 0.694\\
\hline
$10^9A_s$ & 2.139 & $r_{0.05}$ & $<$ 0.07\quad (95\%)\\
\end{tabular}
}\caption{Cosmological parameters. 
All values except $r_{0.05}$ are quoted from the Planck 2015 results: 
central values of the 68\% confidence intervals for the base $\Lambda\text{CDM}$
model with TT+LowP+Lensing data, see Table 4 in \cite{2015arXiv150201589P}. 
The upper bound of $r_{0.05}$ at the pivot scale $k_*=0.05$ Mpc$^{-1}$ at $95\%$ confidence
is quoted from the latest result of the BICEP2/Keck Array CMB polarization experiment
\cite{2016PhRvL.116c1302B}.
} 
\label{table:tab1}
\end{center}
\end{table}

\subsection{Evolution of SFDM}
\label{sec:SFDMbackground}

In our model, DM is entirely made up of SFDM, i.e. $\Omega_ch^2$ in Table \ref{table:tab1}
will be taken to refer to the present-day SFDM energy density, instead of CDM.
The discussion in this subsection follows largely the one in Paper I, 
but since it is of central importance to our model,
we want to repeat some of it here for the sake of the reader.

One basic behavior of a scalar field is that it oscillates over time,
characterized by its changes in phase $\theta$. The oscillation angular frequency
is defined as $\omega\equiv\dot{\theta}$. SFDM behaves differently, depending on whether $\omega$
predominates over the expansion rate $H$ or not (oscillation vs. roll).
As a result, SFDM passes through certain limit cases as it evolves, in which its EOS is simply
barotropic, as we have shown in Paper I.
At early times the expansion is much faster than the field oscillation ($\omega/H\ll 1$).
Eventually, however, the expansion rate declines faster than 
the oscillation frequency and the inequality reverses ($\omega/H\gg 1$).

\subsubsection{Scalar field oscillation faster than Hubble expansion
($\omega/H\gg1$)} \label{sec:SFDMbackground1}

Once the expansion rate drops below the (angular) oscillation frequency of the field, 
the oscillation frequency can be derived as (see Paper I)
\begin{equation} \label{equation:dispersion}
    \omega=\frac{mc^2}{\hbar}\sqrt{1+\frac{2\lambda}{mc^2}|\psi|^2}.
\end{equation}
In this regime,
the exact calculation of the cosmological time evolution of the scalar field is numerically prohibitive,
since the necessary time step is too small ($\propto1/\omega$).
Instead, it has been customary in the literature to follow 
the evolution of the time-averaged values of $\rho$ and $p$ 
over several oscillation cycles of the field. In this subsection \S\ref{sec:SFDMbackground}, 
we omit the subscript ``SFDM'' in $\rho$ and $p$ for brevity. Multiplying the field equation
(\ref{eq:KG}) by $\psi^*$ and then averaging over
a time interval that is much longer than the field oscillation
period, but much shorter than the Hubble time, results in 
(see also \cite{1983PhRvD..28.1243T, 1999PhRvD..60j3506P})
\begin{equation}\label{equation:kineticvsother}
    \frac{\hbar^2}{2mc^2}\langle|\ud\psi/\ud t|^2\rangle\cong\frac{1}{2}mc^2\langle|\psi|^2\rangle+\lambda\langle|\psi|^4\rangle.
\end{equation}
Combining this relation with the expressions for energy density (\ref{equation:densitybu}) 
and pressure (\ref{equation:pressurebu}) yields,
\begin{IEEEeqnarray}{rCl}
    \langle\rho\rangle & = & mc^2\langle|\psi|^2\rangle+\frac{3}{2}\lambda\langle|\psi|^4\rangle\nonumber\\
    & \approx & mc^2\langle|\psi|^2\rangle+\frac{3}{2}\lambda\langle|\psi|^2\rangle^2,\label{equation:averagerho}\\
    \langle p\rangle & = & \frac{1}{2}\lambda\langle|\psi|^4\rangle\approx\frac{1}{2}\lambda\langle|\psi|^2\rangle^2.
\end{IEEEeqnarray}
In this regime, we take $\rho=\langle\rho\rangle$ and $p=\langle p\rangle$.
The equation of state is then approximately
\begin{equation} \label{equation:eos1}
    p=\frac{m^2c^4}{18\lambda}\left(\sqrt{1+\frac{6\lambda\rho}{m^2c^4}}-1\right)^2,
\end{equation}
 or equivalently,
\begin{equation}
  w \equiv \frac{p}{\rho} = \frac{1}{3}\left[\frac{1}{1+\frac{2mc^2}{3\lambda
  \langle |\psi|^2\rangle}}\right]
\end{equation}
(see also \cite{1986PhRvL..57.2485C, MUL2001}). 
This is referred to as the fast-oscillation approximation in Paper I. 
We call this regime the \textit{fast-oscillation regime}, and, henceforth, 
drop the $\langle\rangle$'s around $\rho$ and $p$ in what follows. 
It encompasses two evolutionary phases of SFDM, as follows:

\begin{enumerate}

\item[(1)] CDM-like (or ``dust''-like) phase: non-relativistic ($w=0$)
\label{sec:newtonlimit}

As the universe expands, the dark matter energy density
will continuously decrease to the point when the
rest-mass energy density dominates the total SFDM energy density, i.e.,
$\frac{3}{2}\lambda\langle|\psi|^2\rangle^2\ll
mc^2\langle|\psi|^2\rangle$. In this limit, equation
(\ref{equation:eos1}) reduces to
\begin{equation}\label{equation:polytropiceos}
    p\approx\frac{\lambda}{2m^2c^4} \rho^2\approx 0,
\end{equation}
thus SFDM behaves like non-relativistic dust. Its self-interaction
is weak, so that on large scales SFDM is virtually collisionless.
Therefore, it evolves like CDM, following the familiar relation,
\begin{equation}
    \rho\propto a^{-3}.
\end{equation}
Then, the field amplitude decays as $|\psi| \propto a^{-3/2}$
and the scale factor goes as $a \sim t^{2/3}$.

\item[(2)] Radiationlike phase: relativistic ($w=1/3$)

At some point early enough, SFDM will be so dense
that the quartic term in the energy density (\ref{equation:averagerho}), the
self-interaction energy, dominates, i.e.,
$\frac{3}{2}\lambda\langle|\psi|^2\rangle^2\gg
mc^2\langle|\psi|^2\rangle$. In this limit, equation
(\ref{equation:eos1}) reduces to
\begin{equation}
  p\approx \frac{1}{3}\rho\approx\frac{1}{2}\lambda\langle|\psi|^2\rangle^2,
\end{equation}
and SFDM behaves like radiation. The time evolution is
accordingly
\begin{equation}
    \rho\propto a^{-4},
\end{equation}\\
while the field amplitude decays as $|\psi| \propto a^{-2}$ with the
scale factor $a \sim t^{1/2}$.

It is important to note that SFDM without self-interaction, i.e.,
when $\lambda=0$, does \textit{not} undergo this radiationlike phase.

\end{enumerate}

\subsubsection{Scalar field oscillation slower than Hubble expansion ($\omega/H\ll1$)} 
\label{sec:SFDMbackground2}

At earlier times, the Hubble parameter 
\emph{exceeded} the oscillation frequency. 
In this early regime,
the fast oscillation approximation above is not valid, 
so there is no closed-form expression for the EOS.
In this \emph{slow-oscillation regime}, 
one has to solve the rearranged Klein-Gordon equation (\ref{eq:KGpsi2}) exactly, 
coupled with the Friedmann equations (\ref{equation:friedmannbu}). 
Nonetheless, one can still find a heuristic qualitative description, as follows:

\begin{enumerate}

\item[(1)] Stiff phase: relativistic limit ($w=1$)
\label{sec:stiffphase}

At sufficiently early times, the expansion rate is much greater than
the oscillation frequency, $\omega/H\ll1$. The energy density and
pressure are both dominated by the kinetic term of
(\ref{equation:densitybu}) and (\ref{equation:pressurebu}).
Therefore,
\begin{equation} \label{stiff}
    p\approx\rho\approx\frac{\hbar^2}{2mc^2}|\partial_t\psi|^2.
\end{equation}
This stiff EOS implies that the sound speed almost reaches the speed
of light,  the maximal possible value (this is formally analoguous to the
incompressible fluid in Newtonian gas dynamics, where the sound
speed is infinity). In this case,
 \begin{equation}\label{eq:rhostiff}
    \rho\propto a^{-6},
\end{equation}
and it can be shown that $\partial_t \psi \propto a^{-3}$, and hence $\psi \propto \log a$,
where $a \sim t^{1/3}$. 
An important implication immediately follows from relation (\ref{eq:rhostiff}) that, 
as we go back in time approaching the Big Bang ($a\to0$), the energy density of SFDM 
should dominate the total energy density of the universe, because it increases faster than 
that of radiation and of any other component. Therefore, 
we predict an early era of stiff-SFDM-domination in a $\Lambda$SFDM universe, 
which will be demonstrated in \S\ref{sec:EvolALL}.

\end{enumerate}

\subsection{Tensor fluctuations from inflation and the SGWB}
\label{sec:EvolGW}

In this subsection, we describe the evolution and implementation of our calculation of the SGWB. 
To anticipate our full numerical treatment presented in \S\ref{sec:EvolALL}, 
in which we solve the coupled equations for the SGWB, the SFDM, 
the standard cosmic components and the expansion rate of the background universe, 
it will be instructive to show some analytical results first, 
for the simpler case of constant $w$ (the EOS parameter of the universe).
For this purpose, we must derive the energy density contributed by the SGWB, 
for which we will first need to derive the evolution of the tensor metric perturbations, 
by solving their equation of motion presented in \S\ref{sec:tensoreq} 
along with the initial condition posed in \S\ref{sec:hinit} below.
As we shall see, there are two limits in which this evolution is simplified 
for a given mode of comoving wavenumber $k$, in terms of its wavelength ($\propto k^{-1}$) 
relative to the horizon. 
It will be sufficient to represent the evolution at all times by stitching these two limits together, 
in what is known as the \emph{thin-horizon approximation}. 
With this solution, we will have both the spectrum of the primordial tensor perturbations 
and of their associated energy density as a function of time.

\subsubsection{Primordial amplitude}
\label{sec:hinit}

The equation of motion for the tensor modes $h_k(\tau)$ in Eq. (\ref{tensoreq}) 
requires an initial condition. For our purpose, the initial amplitude of $h_k(\tau)$ 
is given by the primordial tensor amplitude produced by inflation. 
During slow-roll inflation, in which the Hubble constant $H(a)$ is slowly varing, 
fluctuations are exponentially stretched in space, so that for many modes, 
their proper wavelengths, $2\pi a/k$, will become larger than the Hubble radius $c/H(a)$ (or the horizon).
In other words, these modes exit the horizon during inflation. 
Once a mode is far outside the horizon, the amplitude (of its growing mode) is conserved (``frozen'') 
throughout its superhorizon evolution \cite{Weinberg}, even after inflation ends.
Therefore, we will begin our integration of Eq. (\ref{tensoreq}) for a given mode $k$ 
when it is far outside the horizon (i.e., $kc\ll aH(a)$) 
and its initial amplitude, $h_{k,~\rm init}$, is given by this superhorizon value. 
Modes of interest are all far outside the horizon by the end of inflation at $a=a_{\rm inf}$, 
so $h_k(a_{\rm inf})=h_{k,~\rm init}$ for these modes.

These modes will later reenter the horizon at different cosmic times according to their wavelength, 
 while the EOS of the background universe evolves through different cosmic eras. 
 On the other hand, modes reentering during different eras
 do not know of each other, which means that each mode inherits the memory 
 of its own superhorizon amplitude $h_{k,~\rm init}$ with which it started out, 
 at its respective reentry point $a_k$.\footnote{There are also modes at the low-$k$ end, 
 whose comoving wavelengths are even larger than the present-day horizon size. 
Hence, they will never reenter the horizon 
 as the universe has already been in the $\Lambda$-dominated era.
 We do not study these modes in this paper.
 }
Hereafter in this paper, unless otherwise noted, we will use $a_k$ and $H_k\equiv H({a_k})$ 
to indicate those quantities at the horizon reentry for mode $k$, $kc=a_kH_k$.
\footnote{
It is customary to describe a tensor mode of comoving wavenumber $k$ as 
``reentering the horizon'' when $k=aH/c$. We follow that convention here.
However, this actually corresponds to the time 
when the comoving wavelength of the mode equals the comoving Hubble radius $c/a(t)H(t)$, 
not the particle horizon $c\int\ud t/a(t)$.
When the effective EOS of the universe is $w=-1$, for example, 
the comoving Hubble radius shrinks, while the particle horizon always grows.
}

%

Note that this initial amplitude $h_{k,~\rm init}$ is not unique, 
 because of the stochastic nature of the primordial tensor fluctuations produced by inflation.
However, this does not prevent us from evaluating $\rho_{\rm GW}(\tau)$, 
the mean energy density of the inflationary SGWB, in Eq. (\ref{eq:rhoGWfourier}), 
because the stochasticity in $h_k$ is fully accounted for 
by the tensor power spectrum $\Delta^2_{h}(k, \tau)$ defined in Eq. (\ref{eq:tensorps}). 
In fact, we only need to know the primordial tensor power spectrum, 
$\Delta^2_{h,~\rm init}(k)\equiv\Delta^2_{h}(k, a_{\rm inf})$,
evaluated at $a_{\rm inf}$ for all modes of interest. 
The evolution of $\Delta^2_{h}(k, \tau)$, or equivalently, $h_k(\tau)$, 
at any time later is deterministic, separable from its stochastic initial condition. 
We can define the tensor transfer function $T_h(k,\tau)$, 
which encodes this deterministic evolution, as 
\beq \label{trans}
T_h(k,\tau) \equiv \left|\f{h_k(\tau)}{h_{k,~\rm{init}}}\right|^2=\f{\Delta^2_h(k,\tau)}{\Delta^2_{h,~\rm{init}}(k)}, 
\eeq 
the solution of which we will show in \S\ref{sec:GWanalytics}.

The primordial power spectrum of tensor fluctuations generated during inflation, 
$\Delta^2_{h,~\rm init}(k)$, is predicted to be nearly scale-invariant, 
if inflation is driven by a single slow-rolling scalar field.
It can be parametrized by a power law,
\begin{equation}\label{eq:scaleinvtensor}
    \Delta^2_{h,~\rm init}(k)=A_t(k/k_*)^{n_t}\equiv rA_s(k/k_*)^{n_t},
\end{equation}
where $A_t$ ($A_s$) is the tensor (scalar) amplitude, and the pivot
scale $k_*=0.05$ Mpc$^{-1}$, following the 2015 Planck data
convention \cite{2015arXiv150201589P}. 
The value of $A_s$ and the latest upper bound of $r=r_{0.05}$ is given in Table \ref{table:tab1}.
The tensor spectral index $n_t$ is related to the tensor-to-scalar ratio $r$ by
\begin{equation}
	n_t=-r/8,
\end{equation}
which is known as the consistency relation. In this paper, we will presume that this relation is valid.

\subsubsection{Analytical solutions for tensor metric perturbations in the subhorizon limit}
\label{sec:GWanalytics}

Closed-form solutions of Eq. (\ref{tensoreq}) for $h_k(\tau)$ exist,
if $a$ and the conformal time $\tau$ are related via a powerlaw,
\beq\label{eq:aoftau}
 \f{a}{a_0} = \left(\f{\tau}{\tau_0}\right)^{\alpha},
 \eeq
where the exponent $\alpha$ depends on the EOS parameter $w$ of the universe, according to
\beq\label{eq:eosalpha}
\alpha = \f{2}{1+3w}.
\eeq
In our case, however, $w$ changes with time, so we cannot adopt Eq. (\ref{eq:aoftau}) in general. 
Fortunately, when a mode is well outside the horizon, $h_k$ is independent of time 
and of the change in $w$. Furthermore, as long as there are eras of the background expansion 
history in which $w$ is relatively constant, i.e., in each of these eras Eq. (\ref{eq:aoftau}) 
can be applied with a respective constant $\alpha$ over a range of $\tau$, 
we can insert this relation into Eq. (\ref{tensoreq}), to obtain an analytical solution 
for the evolution of $h_k(\tau)$ during these eras.



Particularly, if a mode $k$ reenters the horizon in such an era with a constant $\alpha$, 
and later becomes deep within the horizon (i.e., $k \gg aH/c$, or $k\tau \gg \alpha$) 
while still in the same era, 
one can show that in this subhorizon limit the solution for $h_k(\tau)$, in the eras of interest to us, 
respectively reads as
\begin{widetext}
\begin{itemize}
 \item reheating and matter-dominated era: $w=0, \alpha = 2$
 \beq
h_{k,~\rm{m}}(\tau) \simeq \hini \Gamma \left(\f{5}{2}\right) \f{4}{\sqrt{\pi}} \f{\cos(k\tau-\pi)}{(k\tau)^2}
~~\mbox{ for } k\tau \gg 2,
\eeq
\item stiff-SFDM-dominated era: $w=1, \alpha = \f{1}{2}$
  \beq
 h_{k,~\rm{stiff}}(\tau) \simeq \hini \sqrt{\f{2}{\pi}} \f{\cos(k\tau-\pi/4)}{(k\tau)^{1/2}}~~\mbox{ for } k\tau \gg \f{1}{2},
 \eeq
 \item radiation-dominated era: $w=\f{1}{3}, \alpha = 1$
  \beq
   h_{k,~\rm{rad}}(\tau) \simeq \hini \Gamma \left(\f{3}{2}\right) \f{4}{\sqrt{\pi}} \f{\cos(k\tau-\pi/2)}{k\tau}
  ~~\mbox{ for } k\tau \gg 1.
   \eeq
\end{itemize}
\end{widetext}
Thus, the initial (superhorizon) amplitude from inflation,
$h_{k,~\rm{init}}$, suffers decay upon horizon reentry, according to those expressions. 
In terms of the tensor transfer function $T_h(k,\tau)$ defined in Eq. (\ref{trans}), 
we can see that, in the respective eras considered above, 
\beq\label{transmat} 
T_h^{\rm{m}}(k,\tau) = \Gamma^2
\left(\f{5}{2}\right) \f{16}{\pi} \f{\cos^2(k\tau-\pi)}{(k\tau)^4},
\eeq 
\beq \label{transstiff} 
T_h^{\rm{stiff}}(k,\tau) = \f{2}{\pi}
\f{\cos^2(k\tau-\pi/4)}{k\tau}, 
\eeq and 
\beq \label{transrad}
T_h^{\rm{rad}}(k,\tau) = \Gamma^2 \left(\f{3}{2}\right) \f{16}{\pi}
\f{\cos^2(k\tau-\pi/2)}{(k\tau)^2}. 
\eeq 
Indeed, it can be shown that the tensor transfer function in the subhorizon limit 
for general $\alpha$ reads as 
\bdi 
T_h(k,\tau) =
\f{\Gamma^2(\alpha_k+\f{1}{2})}{\pi}\left(\f{2}{k\tau}\right)^{2\alpha_k}
\cos^2(k\tau-\alpha_k\pi/2) \edi 
\beq \label{transalpha} \simeq
\f{1}{2}\left(\f{a_k}{a}\right)^2 
\f{\Gamma^2(\alpha_k+\f{1}{2})}{\pi}\left(\f{2}{\alpha_k}\right)^{2\alpha_k}
\eeq 
(compare also to \cite{2008PhRvD..77f3504B}), where $a_k$ is
the scale factor at which the mode $k$ reenters the horizon, $a_k=kc/H_k$, 
and we averaged over $\cos^2(..)$ to arrive at the second line. 
The era-dependent parameter $\alpha_k = 2/(1+3w(a_k))$
should be evaluated at horizon reentry for each mode $k$ as well. 
We note that in the second line of Eq. (\ref{transalpha}), the explicit time variable is $a$ 
rather than $\tau$. Therefore, this expression for the tensor transfer function 
$T_h(k,a)$ as a function of $a$, can be applied at any later time 
in the subhorizon limit for a given $k$, 
regardless of any later change in the EOS parameter $w$ of the background universe.

The factor $\frac{1}{2}(a_k/a)^2$ in Eq. (\ref{transalpha}) 
will simply lead to the well-known behavior that for a given $k$, $\ud\rho_{\rm GW}/\ud\ln k$ 
(see Eq. [\ref{eq:rhoGWdiff}]) will decay like radiation ($\propto a^{-4}$) 
after the mode reenters the horizon 
(called ``redshift-suppression'' factor $C_1$ in \cite{2008PhRvD..77f3504B}), 
while the remaining factors make sure that
the correct subhorizon limit is retrieved when matching the solution
at horizon crossing to the superhorizon limit 
(called ``horizon-crossing'' factor $C_2$ in \cite{2008PhRvD..77f3504B}).

There is an additional multiplicative factor which takes account of 
the effects of anisotropy due to neutrino free streaming 
(as described in \cite{2004PhRvD..69b3503W, 2005PhRvD..72h8302D}). 
When relativistic neutrinos are important during the radiation-dominated era, 
they can damp the tensor fluctuations $h_{k}(\tau)$ by a multiplicative factor $A\sim0.8$.
This multiplicative factor is \tx{not} included in the analytical solutions above 
(e.g., Eq. [\ref{transalpha}] for the tensor transfer function), 
but will be included later in our numerical solutions
(this effect was called ``anisotropy factor'' $C_3$ in \cite{2008PhRvD..77f3504B}).

\subsubsection{Evaluating $\Omega_{\rm GW}$}
\label{sec:eval_ogw}

The energy density fraction of the SGWB, $\ogw(a)\equiv8\pi G\rho_{\rm GW}(a)/3H^2(a)c^2$,
is calculated by integrating Eq. (\ref{eq:rhoGWdiff}) over all modes of interest, 
divided by $\rho_{\rm crit}(a)$, 
\begin{IEEEeqnarray}{rCl}\label{eq:Ogw}
    \hspace{-0.2in}
    \ogw (k,a) & \equiv & \frac{\ud \ogw(a)}{\ud \ln k}=\f{1}{\rho_{\rm crit}(a)}\frac{\ud \rho_{\rm GW}(a)}{\ud \ln k}\nonumber\\
    & = & \frac{\Delta^2_h(k,a)c^2}{24a^2H^2(a)}\left(\left|\f{h'_{k}(a(\tau))}{h_{k}(a(\tau))}\right|^2+k^2\right).
\end{IEEEeqnarray} 
This form is written in a way that makes apparent the
contribution from superhorizon evolution, i.e., the second term in Eq. (\ref{eq:Ogw}). 
In the subhorizon limit, the two terms are equal, as $|h'_{k}(a(\tau))|^2\cong k^2|h_{k}(a(\tau))|^2$. 
This can be shown by neglecting the Hubble friction term ($\propto a'/a$)
in the wave equation (\ref{tensoreq}). 
In the superhorizon limit, on the other hand, only the second term remains, 
since $h'_k(a(\tau))\cong 0$. There remains uncertainty in whether 
superhorizon modes physically contribute an average stress-energy 
that can affect the background metric of the universe. 
However, this contribution, should it exist, is negligible compared to subhorizon modes anyway, 
as we have confirmed in this work.

\begin{enumerate}
 \item  \tb{Subhorizon limit:}\\
 In the subhorizon limit $k\gg aH/c$, the energy density spectrum of GWs,
 $\Omega_{\rm GW}(k,a)$, can be calculated by solving the linear evolution equation (\ref{tensoreq}).
 For modes which reenter the horizon when the universe has a fixed EOS, 
 $\Omega_{\rm GW}(k,a)$, defined above in Eq. (\ref{eq:Ogw}), 
 is related to the tensor transfer function defined in Eq. (\ref{trans}), as follows:
\begin{IEEEeqnarray}{rCl}\label{eq:gwsubhor}
    \ogw(k,a) & = & \frac{\Delta_h^2(k,a)}{12}\left(\f{kc}{aH}\right)^2\nonumber\\
     & = & \frac{\Delta_{h,~\rm{init}}^2(k)}{12}\left(\f{kc}{aH}\right)^2 T_h(k,a).
\end{IEEEeqnarray}
The expressions for $\ogw(k, a(\tau))$ which correspond to the above analytical solutions 
in Eqs. (\ref{transmat}-\ref{transrad}), after averaging over $\cos^2(..)$, are given by
 \beq
  \ogw^{\rm{m}}(k,\tau) \simeq \f{\Delta_{h,~\rm{init}}^2(k)}{24} \cdot \f{9}{4}\f{1}{(k\tau)^2},
 \eeq
 \beq
 \ogw^{\rm{stiff}}(k,\tau) \simeq \f{\Delta_{h,~\rm{init}}^2(k)}{24} \cdot \f{8}{\pi} k\tau,
 \eeq
 \beq
  \ogw^{\rm{rad}}(k,\tau) \simeq \f{\Delta_{h,~\rm{init}}^2(k)}{24}.
 \eeq
 For $\Delta^2_{h,~\rm{init}}(k)\simeq k^0$, this yields the $k$-dependence of $\ogw(k,\tau)$ 
 for a matter-, stiff-SFDM-, or radiation-dominated universe as follows:
  $\ogw^{\rm{m}}(k,\tau) \propto k^{-2}$, $\ogw^{\rm{stiff}}(k,\tau) \propto k$, and
 $\ogw^{\rm{rad}}(k,\tau) \propto k^{0}$, respectively. 
 This dependence on $k$ will be reflected in our prediction of 
 the SGWB energy density spectrum at the present in \S\ref{sec:SGWBspectra}.\\

We can now illustrate the effect of the amplification of
the (differential) GW energy density of a certain mode with wavenumber $k$ which
reenters the horizon during the stiff phase, compared to that if the mode reenters the horizon
during the radiation-dominated era, as in a standard $\Lambda$CDM universe.
In fact, combining Eq. (\ref{eq:scaleinvtensor}), Eq. (\ref{transalpha}) and Eq. (\ref{eq:gwsubhor}) yields
\begin{equation}\label{eq:diffgwsubhor}
     \ogw(k,a)\simeq\frac{rA_s}{24}\left(\frac{ck}{aH}\right)^2\left(\frac{a_k}{a}\right)^2,
\end{equation}
where we have, for simplicity, neglected the dependence on $n_t$ in Eq. (\ref{eq:scaleinvtensor})
and ignored the factor $\f{\Gamma^2(\alpha+\f{1}{2})}{\pi}\left(\f{2}{\alpha}\right)^{2\alpha}$ 
in Eq. (\ref{transalpha}).
In this equation above, the scale factor $a$ is at a late time when 
the expansion histories of the two scenarios, $\Lambda$SFDM vs. $\Lambda$CDM, converge, 
so that the Hubble parameter $H=H(a)$ is the same for both. 
Therefore, the only uncommon factor in Eq. (\ref{eq:diffgwsubhor}) is $a_k$ for the two scenarios.
In a $\Lambda$SFDM universe, suppose now that the stiff era ends and the universe becomes
radiation-dominated at $a_{\rm rad}$. The Hubble constant $H_{\rm rad}$ at that time 
must be approximately the same as that in the $\Lambda$CDM scenario, 
since the evolution of the two universes
from that point on up to the present must be the same.
From the evolution of the homogenous background universe we have
\begin{equation}
    \left(\frac{H_{k,\rm{stiff}}}{H_{\rm rad}}\right)^2=\left(\frac{a_{k,\rm{stiff}}}{a_{\rm rad}}\right)^{-6},
\end{equation}
and
\begin{equation}
    \left(\frac{H_{k,\rm{rad}}}{H_{\rm rad}}\right)^2=\left(\frac{a_{k,\rm{rad}}}{a_{\rm rad}}\right)^{-4},
\end{equation}
where $a_{k,i}$ ($i=\rm{stiff, rad}$) is the scale factor at which the mode $k$ reenters the horizon,
for each scenario, and $H_{k,i}$ is the corresponding Hubble constant. Therefore,
\begin{equation}
    \frac{H_{k,\rm{stiff}}}{H_{k,\rm{rad}}}=\frac{a_{k,\rm{stiff}}^{-3}a_{\rm rad}}{a_{k,\rm{rad}}^{-2}}.
\end{equation}
Taking into account the fact that $ck=a_{k,\rm{stiff}}H_{k,\rm{stiff}}=a_{k,\rm{rad}}H_{k,\rm{rad}}$,
we rearrange the equation above and obtain
\begin{equation}
    a_{k,\rm{rad}}=\left(\frac{a_{k,\rm{stiff}}}{a_{\rm rad}}\right)a_{k,\rm{stiff}}.
\end{equation}
Since $a_{k,\rm{stiff}}/a_{\rm rad}<1$, from the equation above $a_{k,\rm{stiff}}>a_{k,\rm{rad}}$.
The mode reenters the horizon later (i.e. at a larger scale factor) during the stiff phase
than it would during a radiation-dominated universe. Thus, according to Eq. (\ref{eq:diffgwsubhor})
we conclude that a mode that reenters the horizon during the stiff era will contribute a
higher GW energy density at late times than it would in the standard scenario,
when that mode reenters in the radiation-dominated era.

To view this effect from another perspective, there are two competing factors
which combine to make the contribution of a given mode to the GW energy density
of the universe bigger in the presence of the stiff phase, as follows.
Whatever the initial GW energy density upon horizon reentry at $a_k$ is, thereafter
it dilutes like radiation, in proportion to $(a_k/a)^4$. Since $a_{k,\rm{stiff}}>a_{k,\rm{rad}}$, there is less
dilution to a given late time for the $\Lambda$SFDM case with a stiff phase.
On the other hand, the superhorizon tensor amplitude is the same in both cases, since we consider the same inflationary model.
Therefore, the GW contribution of a mode expressed as a fraction of the critical density
at horizon reentry (see Eq. [\ref{eq:diffgwsubhor}]) is also the same. Since this critical density is proportional to
$H_k^2=c^2k^2/a_k^2$, it is, however, smaller in $\Lambda$SFDM than in $\Lambda$CDM.
This effect makes the contribution to $\rgw$ at horizon reentry smaller for $\Lambda$SFDM
than for $\Lambda$CDM. To elucidate both of the effects, we can rewrite Eq. (\ref{eq:diffgwsubhor})
in the following way:
\begin{IEEEeqnarray}{rCl}\label{eq:diffgwsubhorizon2}
    \ogw(k,a) & = & \frac{rA_s}{24}\frac{H_k^2}{H^2}\left(\frac{a_k}{a}\right)^4\nonumber\\
    & = & \frac{\ud\rgw}{\ud\ln k}\bigg |_{a=a_k}\left(\frac{a_k}{a}\right)^4\frac{1}{\rho_{\rm{crit}}(a)}.
\end{IEEEeqnarray}
While a tensor mode reenters the horizon
with a lower energy density when it reenters during the stiff phase of a $\Lambda$SFDM universe,
since $H_{k,\rm{stiff}}<H_{k,\rm{rad}}$, however, according to Eq.(\ref{eq:diffgwsubhorizon2}), 
it reenters at a later scale factor.
Hence, its radiationlike energy density does not thereafter dilute so much as in $\Lambda$CDM,
since $(a_{k,\rm{stiff}}/a)^4>(a_{k,\rm{rad}}/a)^4$. Overall, the latter effect wins, and, therefore, 
there is a boost in the GW energy density for a mode that reenters the horizon during 
the stiff-SFDM-dominated era (predicted in \S\ref{sec:SFDMbackground2}), relative to what it would have been in $\Lambda$CDM.

As we will see in \S\ref{sec:EvolALL}, due to this amplification effect, at a later time, 
the total $\rho_{\rm GW}(a)$, integrated over all $k$ but dominated by high-frequency modes 
which have reentered the horizon by the end of the stiff-SFDM-dominated era, 
will evolve nearly as radiation ($\propto a^{-4}$). 
Consequently, $\rho_{\rm GW}(a)$ will emerge as a significant contribution 
to the critical energy density of the $\Lambda$SFDM universe, 
as soon as the radiation-dominated era begins.

\item \tb{Superhorizon limit:}\\
According to Eq. (\ref{eq:Ogw}), the superhorizon ($k\ll aH/c$) GW energy density spectrum 
can be written as
\begin{equation}\label{eq:suphorGW}
    \ogw(k, a)  =\frac{\Delta_{h,~\rm{init}}^2(k)}{24}\left(\f{kc}{aH}\right)^2.
\end{equation}
Eq. (\ref{eq:suphorGW}) can be applied at all times during the superhorizon evolution of each mode $k$. 
Since $kc=a_kH_k$, this equation tells us that every mode reenters the horizon
with almost the same fractional energy density ($\approx\Delta_{h,~\rm{init}}^2(k)/24$).

\item \tb{Thin-horizon approximation:}\\
In the {\it thin-horizon approximation}, the horizon crossing of mode $k$ 
is assumed to occur suddenly at $a=a_k$, and immediately follows the asymptotic behavior 
of the subhorizon evolution.
We confirm that the assumption of thin-horizon crossing 
is a very good approximation for all eras of interest to us in the expansion history.
As an example, we show in Appendix \ref{app:thinhor} 
the exact solution for $h_k(\tau)$ and $\Omega_{\rm{GW}}(k, \tau)$, 
along with the asymptotic solutions for the latter in the sub- and superhorizon regime, 
for modes which reenter the horizon during reheating with a matter-like EOS ($w=0$). 
One can see that the asymptotic solutions of $\Omega_{\rm{GW}}(k, \tau)$ 
not only perfectly trace the exact solution, in their regime of validity, 
but also that the range in $k\tau$ around horizon crossing is rather narrow, 
validating the thin-horizon approximation.

\item \tb{Total GW energy density:}\\
We apply the {\it thin-horizon approximation}, so that for each mode $k$, 
the superhorizon evolution of $\ogw(k, a)$ is given by Eq. (\ref{eq:suphorGW}) for all $a<a_k$, 
and the subhorizon evolution is given by Eq. (\ref{eq:gwsubhor}) 
combined with Eq. (\ref{transalpha}) for all $a>a_k$ (or equivalently, $\tau>\tau_k=\alpha/k$).
We can then integrate the fraction of total SGWB energy density 
over all wavenumbers at any given time, 
\vspace{-0.2in}
\begin{widetext}
\begin{IEEEeqnarray}{rCl}\label{eq:gwspectrum}
    \ogw(a) & = & \int^{k_{\rm hor}}_0 \ogw(k,a)\ud \ln k +
     \int^{k_{\rm inf}}_{k_{\rm hor}}\ogw(k,a)\ud \ln k\nonumber\\
    & = & \frac{rA_s}{24a^2H^2}\int^{k_{\rm hor}}_0 c^2k^2\left(\frac{k}{k_*}\right)^{n_t}\ud \ln k +
    \frac{rA_s}{12a^2H^2}\int_{k_{\rm hor}}^{k_{\rm inf}}c^2k^2\left(\frac{k}{k_*}\right)^{n_t}T_h(k,a)\ud\ln k\nonumber\\
    & = & \frac{rA_s}{24(2+n_t)}\left(\frac{k_{\rm hor}}{k_*}\right)^{n_t} +
    \frac{rA_s}{12a^2H^2}\int_{k_{\rm hor}}^{k_{\rm inf}}a_k^2H_k^2\left(\frac{k}{k_*}\right)^{n_t}T_h(k,a)\ud\ln k,
\end{IEEEeqnarray}
\end{widetext}
where $k_{\rm inf}$ is the wavenumber of the mode that just exits the horizon 
and then immediately reenters, when inflation ends at $a_{\rm inf}$, 
and we have used the relation $k_{\rm hor}=aH/c$ for the mode that fills the horizon 
at scale factor $a$, i.e., $a_{k_{\rm hor}}=a$.
The integral in the above equation can be divided into two parts by $k_{\rm reheat}$, 
the wavenumber of the mode that fills the horizon at the end of reheating, 
when $T=T_{\rm reheat}$ and $a=a_{\rm reheat}$,
\begin{widetext}
\vspace{-0.1in}
\beq\label{eq:OGWsubhor}
     \int_{k_{\rm hor}}^{k_{\rm inf}}a_k^2H_k^2\left(\frac{k}{k_*}\right)^{n_t}T_h(k,a)\ud\ln k
    =  \int_{k_{\rm reheat}}^{k_{\rm inf}}a_k^2H_k^2\left(\frac{k}{k_*}\right)^{n_t}T_h(k,a)\ud\ln k
     +\int_{k_{\rm hor}}^{k_{\rm reheat}}a_k^2H_k^2\left(\frac{k}{k_*}\right)^{n_t}T_h(k,a)\ud\ln k.
\eeq
In the equation above, the contribution from reheating, assuming an EOS with $w=0$, 
can be integrated analytically. The result is
\begin{equation}\label{eq:GWreheat}
    \int_{k_{\rm reheat}}^{k_{\rm inf}}a_k^2H_k^2\left(\frac{k}{k_*}\right)^{n_t}T_h(k,a)\ud\ln k
    =\frac{1}{2(2-n_t)}\f{a_{\rm inf}^4H_{\rm inf}^2}{a^2}\left(\frac{a_{\rm reheat}}{a_{\rm inf}}\left(\frac{k_{\rm reheat}}{k_*}\right)^{n_t}-\left(\frac{k_{\rm inf}}{k_*}\right)^{n_t}\right).
\end{equation}
\end{widetext}

\end{enumerate}


\subsection{Other cosmic components}
\label{sec:EvolOthers}

Apart from SFDM and GWs, the other components are the same than in $\Lambda$CDM. 
In Table \ref{table:tab1}, $\Omega_rh^2$, 
calculated from the CMB temperature today $T_{\rm CMB}$, 
accounts for the ordinary radiation component,
i.e. photons and neutrinos. For simplicity, the neutrinos are
considered as \textit{massless} (i.e. SM neutrinos), 
such that the total matter density fraction today is
$\Omega_m=\Omega_b+\Omega_c$, where
$\Omega_b$ stands for the baryon density fraction at the present. 
The energy density of baryons (the ``ordinary matter'') always decays like non-relativistic ``dust'',
$\rho_b(a)\propto a^{-3}$.
While the radiation component decays asymptotically like $\rho_r(a)\propto a^{-4}$, 
photons do get extra heat during various processes in the early evolution. 
These effects are usually described via a quantity called $g_*$ (or $g$ factor), 
which reflects the change (decrease) of relativistic species over time.
It amounts to calculating the thermal history exactly, i.e. the photon temperature $T$ 
as a function of $a$ during such periods.
As in Paper I, we will again take into account the most notable of these changes, 
namely the time of electron-positron annihilation that occurs around $0.5$ MeV. 
This effect will be reflected in our solutions as a little dip in the density fraction of radiation at that time.
Finally, we assume a cosmological constant, $\rho_{\Lambda} = \rm{const.}$, 
whose present-day density fraction is given by $\Omega_\Lambda=1-\Omega_m-\Omega_r$.

\subsection{``Putting it together'': homogeneous $\Lambda$SFDM universe}
\label{sec:EvolALL}

In this section, we couple the evolution of all cosmic components
to obtain the expansion history of the homogeneous $\Lambda$SFDM universe. 
We will also introduce several cosmological observables, 
which we later use to constrain the $\Lambda$SFDM model. 
Inserting Eq. (\ref{eq:rhopsi2}) and the relations mentioned in \S\ref{sec:EvolOthers} 
into the post-reheating Friedmann equation (\ref{equation:friedmannbu}) yields
\begin{widetext}
\vspace{0.1in}
\begin{IEEEeqnarray}{rCl}\label{equation:friedmann2}
    H^2(a) & = & H_0^2\left(\frac{\Omega_r(a)}{a^4}+\frac{\Omega_b}{a^3}+\Omega_\Lambda\right)+H^2(a)\Omega_{\rm{GW}}(a)+\frac{8\pi G}{3c^2}\rho_{\rm{SFDM}}\nonumber\\
    & = & H_0^2\left(\frac{\Omega_r(a)}{a^4}+\frac{\Omega_b}{a^3}+\Omega_\Lambda\right)+H^2(a)\Omega_{\rm{GW}}(a)\nonumber\\
    & &+\frac{8\pi G}{3c^2}\left[\frac{\hbar^2}{2mc^2}\left(\frac{(\ud|\psi|^2/\ud t)^2}{4|\psi|^2}
	+\frac{(\rho_{\rm SFDM,0}/\hbar)^2}{a^6|\psi|^2}\right)
	+\frac{1}{2}mc^2|\psi|^2+\frac{1}{2}\lambda|\psi|^4\right],\IEEEeqnarraynumspace
\end{IEEEeqnarray}
\end{widetext}
where $\Omega_b$ and $\Omega_\Lambda$ are given in Table \ref{table:tab1}, 
the parameter $\Omega_r(a)$ is different before and after 
the electron-positron annihilation\footnote{
After the $e^-e^+$ annihilation, $\Omega_r(a)$ is equal to 
the present-day radiation energy density fraction given in Table \ref{table:tab1}. 
It is slightly smaller before the $e^-e^+$ annihilation 
because photons get heated as $e^-e^+$ pairs annihilate into photons in thermal equilibrium. 
We take this into account in our evolution of the thermal history of the universe.
}, 
and $\Omega_{\rm GW}(a)$ is evaluated by Eqs. (\ref{eq:gwspectrum}) -- (\ref{eq:GWreheat}).
Unlike the standard $\Lambda$CDM universe in which 
the Friedmann equation can be solved separately from the equations of motion
for each component, it is necessary in the case of SFDM, to solve Eq. (\ref{equation:friedmann2}) 
fully coupled to the Klein-Gordon equation (\ref{eq:KGpsi2}), 
the equation of motion for the SFDM.
Therefore, a numerical integration is required, which we will describe in \S\ref{sec:nummeth}.

To start the description of the evolution of the homogeneous universe, 
we first remind the reader that, 
$\Lambda$SFDM is embedded in the standard inflationary paradigm 
in a way similar to $\Lambda$CDM, 
that a $\Lambda$SFDM universe commences 
in a period of cosmic inflation which ends in reheating, as explained in \S\ref{sec:intro_inf}. 
In the single-field slow-roll inflation picture, the energy scale of inflation, or equivalently, 
the Hubble constant at the end of inflation, $H_{\rm inf}$,
can be determined by the value of the tensor-to-scalar ratio $r$, 
\begin{equation}\label{eq:hubbleinf}
	H_{\rm inf}=\frac{\pi M_{\rm pl}}{\hbar}\sqrt{rA_s}, 
\end{equation}
where $M_{\rm pl}$ is the reduced Planck mass, 
$M_{\rm pl}\equiv\sqrt{\frac{\hbar c}{8\pi G}}$.

When inflation ends, the inflaton oscillates and decays, 
which results in particle production and reheating ($w=0$). 
The end of reheating is considered as the emergence of SFDM 
as well as the SM particles, produced during reheating. 
Unlike in $\Lambda$CDM, in our $\Lambda$SFDM model 
reheating dumps most of the energy of the inflaton into SFDM, 
which quickly forms a Bose-Einstein condensate, as argued in \S\ref{sec:SFDM1}. 
Meanwhile, a subdominant amount of energy is dumped into the SM particles, 
which was a radiation component at $T=\tre$.
In $\Lambda$SFDM, this is the moment when the cosmic expansion history 
starts to be distinguishable from $\Lambda$CDM.

\begin{figure*}[t]
\begin{minipage}{\linewidth}
     \centering
     \includegraphics[width=1\linewidth]{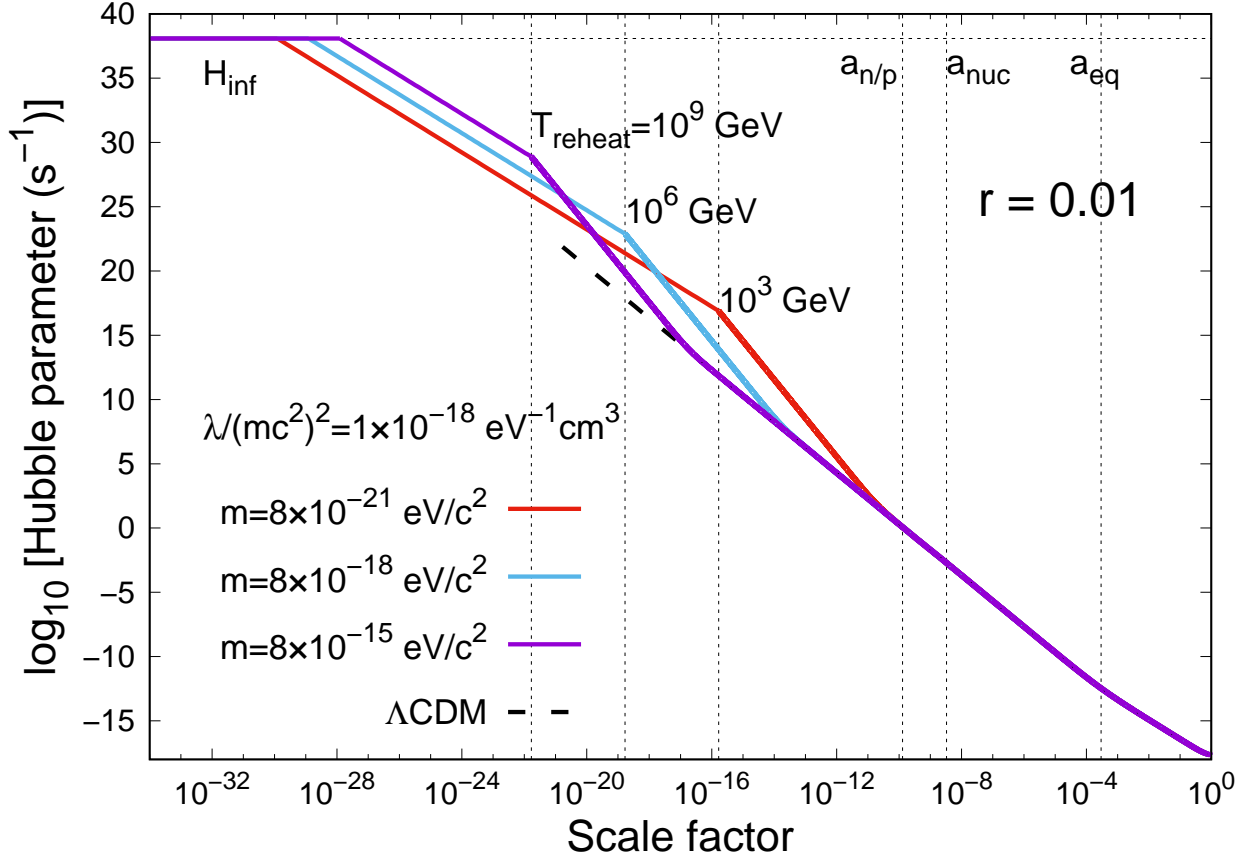}
     \end{minipage}
  \caption{Expansion history of 3 example $\Lambda$SFDM models in the standard inflation paradigm 
  including an epoch of standard reheating ($w=0$). 
  }
  \label{fig:hubble_all}
\end{figure*}

While the Hubble constant when inflation ends 
is fixed in $\Lambda$SFDM by Eq. (\ref{eq:hubbleinf}), 
the value of $H$ when reheating ends is set 
by the value of $a=a_{\rm reheat}$ when $T=T_{\rm reheat}$,
which cannot be determined on its own without solving the holistic evolution 
that follows to match the observed universe at present in the presence of SFDM.
This will be apparent if we preview the generic behavior of the expansion history 
in the full solutions we will calculate later in this section.
Fig. \ref{fig:hubble_all} (based upon the calculation detailed later) 
 shows a plot of the Hubble parameter for several $\Lambda$SFDM models 
 with different parameters, as a function of scale factor,
in which the varying EOS of the background universe is reflected in different slopes.
Following the end of inflation at $a=a_{\rm inf}$, $H\propto a^{-3/2}$ during reheating 
until $a=a_{\rm reheat}$. At this point, 
the $\Lambda$SFDM universe is dominated by stiff SFDM, rather than radiation. 
We have described the relativistic nature of SFDM at early times in \S\ref{sec:SFDMbackground}, 
that BEC SFDM starts as stiff matter ($w=1$), and then transitions into a radiationlike 
($w=1/3$) component, before a final transition into dustlike CDM ($w=0$). 
Therefore, we expect to see that, as the energy density of the dominant stiff SFDM decreases as 
$\rho_{\rm SFDM}\propto a^{-6}$ (faster than radiation), 
the initially stiff-SFDM-dominated universe ($H\propto a^{-3}$) will later experience a transition in its EOS 
to radiation-dominated, when SFDM and other relativistic species 
combine to make the critical energy density of the universe $\rho_{\rm crit}\propto a^{-4}$, 
so $H\propto a^{-2}$,
until the SFDM transitions to CDM-like and once again dominates $\rho_{\rm crit}$, 
then $H\propto a^{-3/2}$.

It can be inferred from above that, during the stiff and radiationlike phase of SFDM, 
the expansion rate of the background $\Lambda$SFDM universe in its early stage 
is increased, compared to that in $\Lambda$CDM (see Fig. \ref{fig:hubble_all}). 
Hence, in the $\Lambda$SFDM model, SFDM will contribute to 
the effective number of relativistic species, 
also known as effective number of neutrino species, $\neff$. 
In $\Lambda$CDM, where there are only three SM neutrinos, 
$N_{\rm eff}=N_{\rm eff,standard} = 3.046$.
In $\Lambda$SFDM, an increased expansion rate
can be translated into an increased $\neff$, or vice versa. 
Thus, measurements of the value of $\neff$ at a certain time 
will constrain the expansion rate of the $\Lambda$SFDM universe at that time. 

In fact, BBN is such an epoch during which the value of $\neff$ can be measured, 
by determining primordial light element (He, D, etc.) abundances from observations. 
Standard BBN proceeds in a period between the freeze-out of the neutron-proton ratio
when the photon temperature $T\simeq T_{\rm{n/p}}\equiv1.293$ MeV 
(the difference between the neutron and the proton mass) 
and the epoch of nuclei production when $T\simeq T_{\rm{nuc}}\approx0.07$ MeV. 
We denote the respective scale factors as $a_{\rm{n/p}}$ and $a_{\rm{nuc}}$.
A detailed analysis on how the value of $\neff$ during BBN constrains 
the expansion rate of $\Lambda$SFDM, and thereby the SFDM particle parameters, 
will be carried out in \S\ref{sec:constraintBBN}.

%
%

Later in the expansion history of $\Lambda$SFDM, the universe undergoes another transition 
from radiation-dominated (RD) to matter-dominated (MD). 
The division of these two eras is described by 
the epoch of matter-radiation equality, the redshift at which is denoted as $\zeq$. 
Note that in $\Lambda$SFDM, matter-radiation equality refers to the
equality between the energy density of the matter component (SFDM plus baryons)
and the radiation component (including GWs). 
After $\zeq$, the overdensities in the matter-dominated universe start to grow 
in proportion to the scale factor, which become seeds for forming cosmic structures. 
Since we consider SFDM as a variant of CDM, which retains the cosmic structure 
on large enough scales as predicted by standard CDM (see \S\ref{sec:cdm}), 
we should expect that the expansion history of the background $\Lambda$SFDM universe 
be nearly identical to that in $\Lambda$CDM, after the same $\zeq$. 
Besides LSS, $\zeq$ is a cosmological observable well determined by 
CMB anisotropy measurements independently.
Therefore, $\Lambda$SFDM must respect the value of $\zeq$ measured by the CMB. 
In other words, $\zeq$ puts constraints on $\Lambda$SFDM, too, 
which we will discuss in \S\ref{sec:ConstraintEQ}.


The combination of these constraints will allow us to determine allowed ranges of 
SFDM particle parameters. Allowed regions will correspond to those SFDM models which
comply to all the current measurements of the background evolution. 
The results will be summarized in \S\ref{sec:PPS}.

\begin{figure*}[t]
\begin{minipage}{\linewidth}
     \centering
     \includegraphics[width=1\linewidth]{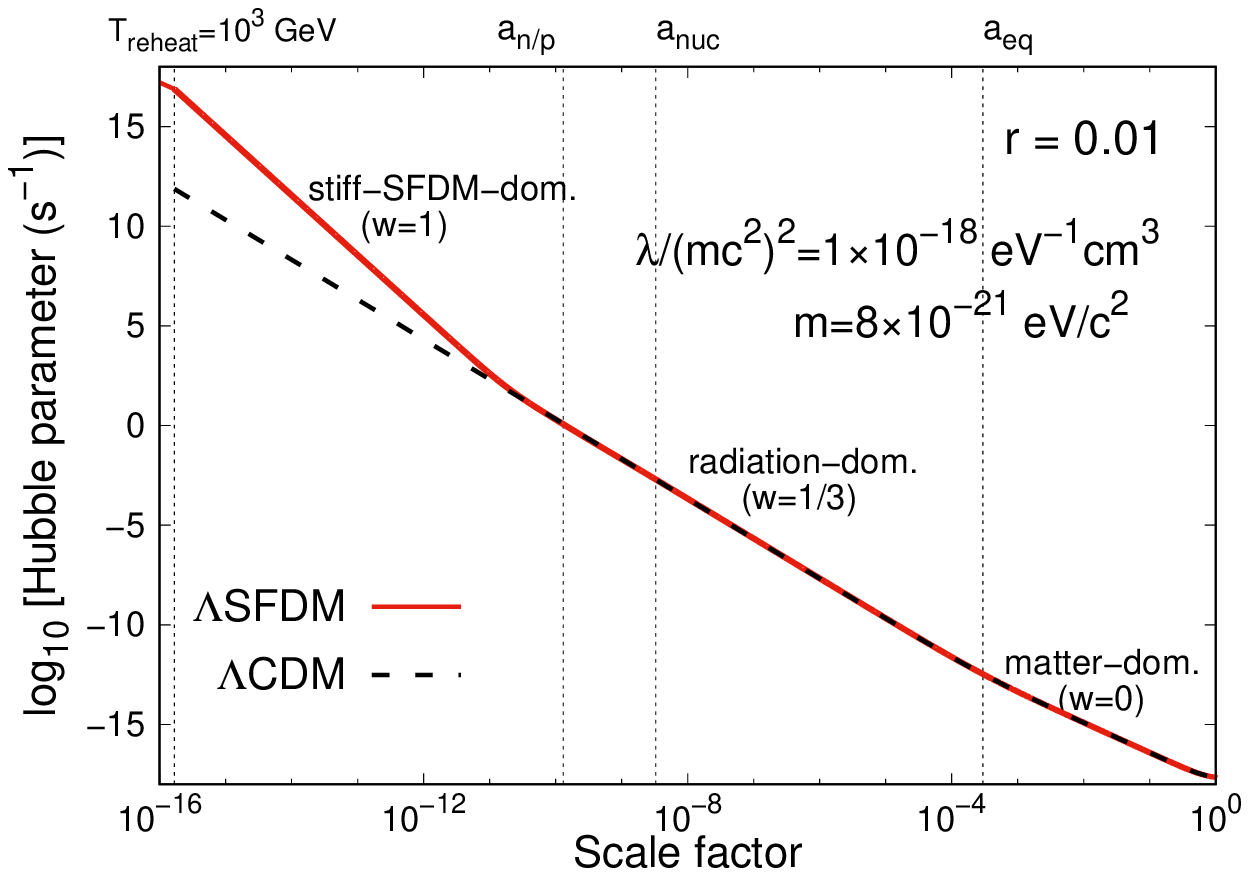}
     \end{minipage}
  \caption{Expansion history of a $\Lambda$SFDM model, of which the particle parameters 
  are $\lambda/(mc^2)^2 = 1 \times 10^{-18}~\rm{eV}^{-1}~\rm{cm}^3$ and 
  $m = 8\times10^{-21}~\rm{eV}/c^2$. This is one of the example models in Fig. \ref{fig:hubble_all}.
  }
  \label{fig:hubble}
\end{figure*}

\subsubsection{Numerical method}
\label{sec:nummeth}

In Paper I, we presented many details of how the evolution of SFDM is numerically calculated, 
so we refer the reader to that paper for more technical details.
We emphasize that, as in Paper I, there are basically two different calculational regimes, as follows. 
When $\omega/H \gg 1$, the fast-oscillation approximation applies, 
as described in \S\ref{sec:SFDMbackground1}. 
As long as the oscillation is much faster than the rate at which the scale factor changes, 
the exact SFDM energy density and pressure should be well approximated by
the corresponding time-averaged quantities, and we confirmed in Paper I that this is indeed the case. 

At earlier times, $\omega/H$ decreases and the fast-oscillation approximation becomes invalid. 
Then, we have to work in the slow-oscillation regime as described in \S\ref{sec:SFDMbackground2}, 
and the evolution of SFDM has to be calculated exactly, with no reference to an averaging procedure. 

The presence of $\rho_{\rm GW}$, which is dependent on the expansion history and, in turn, 
affects that history, requires us to generalize the method of Paper I. 
In addition, we have improved the accuracy of our numerical solutions.

\begin{enumerate}

\item[(1)] Nondimensionalized equations

We have rewritten the coupled Klein-Gordon and Friedmann equations 
in a nondimensionalized form which takes advantage of the characteristic scales 
of the dimensional quantities expected during the early, slow-oscillation regime, 
to improve the accuracy of our numerical solutions.

In the early-time slow-oscillation regime, we solve the Klein-Gordon equation (\ref{eq:KGpsi2}) 
directly in terms of the field amplitude square $|\psi|^2$ as the dependent variable, 
coupled with the Friedmann equation (\ref{equation:friedmann2}). 
The hydrodynamical variables $\rho_{\rm SFDM}$ and $p_{\rm SFDM}$ 
are then related to $|\psi|^2$ by Eqs. (\ref{eq:rhopsi2}) and (\ref{eq:ppsi2}).

In particular, we have nondimensionalized this set of ordinary differential equations (ODEs) 
by expressing variables in terms of their values at the matching point at $a=a_M$, 
between the slow-oscillation regime and the late-time fast-oscillation regime. 
We define the dimensionless dependent variable for our numerical integration as follows:
\beq
	y \equiv \frac{|\psi|^2}{|\psi_M|^2},
\eeq
where $\psi_M$ is the value of the scalar field at the matching point.
The independent variable, cosmic time $t$ is nondimensionalized as:
\beq
	x \equiv \omega_Mt,
\eeq
where $\omega_M$ is the oscillation frequency of the scalar field at the matching point.
Likewise the dimensionless Hubble parameter is defined as: 
\beq
	\mathscr{H} \equiv H/\omega_M.
\eeq
We note that according to the definition of the Hubble parameter, 
\begin{equation} \label{eq:ndhubble}
	\dot a=\mathscr{H}a,
\end{equation}
where the upper dot denotes the derivative with respect to the dimensionless time variable $x$, 
{\it throughout this subsection}.

Given these variables, the dimensionless equivalent of the Klein-Gordon equation 
(\ref{eq:KGpsi2}) can be written as
\begin{equation}\label{eq:ndkge}
	\ddot y=-3\mathscr{H}\dot y+\frac{\dot{y}^2}{2y}+\frac{2F_1}{a^6y}-2F_2y-4F_3y^2,
\end{equation}
in which dimensionless constants $F_1$, $F_2$ and $F_3$ are defined as
\begin{IEEEeqnarray}{rCl}
	F_1 & \equiv & \frac{(\rho_{\rm SFDM,0}/\hbar)^2}{\omega_M^2|\psi_M|^4},\\
	F_2 & \equiv & \frac{(mc^2)^2}{(\hbar\omega_M)^2}\\
	F_3 & \equiv & \frac{\lambda mc^2|\psi_M|^2}{(\hbar\omega_M)^2}.
\end{IEEEeqnarray}

For the dimensionless version of the Friedmann equation, 
combining Eq. (\ref{equation:friedmann2}) with the expressions of 
the dimensionless variables and constants above yields
\begin{IEEEeqnarray}{rCl}\label{eq:ndfe}
	\mathscr{H}^2 & = & \mathscr{H}_0^2\left(\frac{\Omega_r(a)}{a^4}+\frac{\Omega_{b}}{a^3}
	+\Omega_{\Lambda}\right)
	+\mathscr{H}^2\Omega_{GW}(a)\nonumber\\
	& & +\frac{\dot{y}^2}{24y}+\frac{F_1}{6a^6y}+\frac{F_2y}{6}+\frac{F_3y^2}{6},
\end{IEEEeqnarray}
where $\mathscr{H}_0=H_0/\omega_P$ apparently. 

The ODEs (\ref{eq:ndhubble}), (\ref{eq:ndkge}) and (\ref{eq:ndfe}) will be coupled 
to solve the holistic evolution of $\Lambda$SFDM, provided we are able to evaluate $\ogw(a)$ 
at any scale factor self-consistently.

\item[(2)] Integration and iteration scheme

We use a publicly-available ODE solver, {\it DVODE} \cite{DVODE_f90}, 
which can solve stiff systems in double precision, for all our numerical integrations. 
In Paper I, we integrated the evolution backward in time, 
using cosmological parameters at the present as the initial condition, 
given by the Planck 2013 results \cite{Planck2013} . 
This was necessary in Paper I because otherwise we would have needed to know 
the initial value of the scalar field and its time derivative, in the early universe,
as well as the (conserved) comoving charge density $Q$, in order to integrate forward in time, 
but only $Q$ is known in advance (see Eq. [\ref{eq:chargedensity}]).

However, it is difficult for the backward calculation to take into account 
the SGWB produced by inflation self-consistently. 
Therefore, in this paper, we must evolve the ODEs forward in time, 
and iterate. We use a backward integration to make a first guess for the starting values 
to use in the next forward integration, and subsequently iterate 
by a sequence of backward-forward integrations designed to converge. 
Convergence in this case means that the end result of a \emph{forward} integration 
reaches the values of the present-day cosmological parameters in Table \ref{table:tab1} 
at $a=1$ with sufficient accuracy, as described below.

For each forward integration, we need to guess the starting values to use 
for the scalar field and its time derivative at $a=a_{\rm reheat}$.
For this, we depend upon a backward integration 
from the known values of the cosmological parameters at $a=1$.
Unfortunately, the contribution to the total energy density from $\rho_{\rm GW}$
depends upon the accumulation of tensor modes over time 
as they reenter the horizon, which can only be determined self-consistently 
by a \emph{forward} integration. Hence, backward integrations, too, 
must incorporate some guess, for the evolution of $\rho_{\rm GW}(a)$.

For the very first iteration, we integrate backward, neglecting $\rho_{\rm GW}$.
A forward integration is then performed from $a=a_{\rm reheat}$ to $a=1$ 
and the outcome compared with the cosmological parameters in Table \ref{table:tab1}
used to start the backward integration. 
In particular, we see how close the ending of $\rho_{\rm SFDM}$ is from $\rho_{\rm SFDM, 0}$.
For $\Delta\equiv\rho_{\rm SFDM}/\rho_{\rm SFDM, 0}-1$, 
if $\Delta\leq0.001$, the iteration is deemed to have converged.
If, however, $\Delta>0.001$, then we guess the evolution of $\rho_{\rm GW}(a)$
based upon that first forward integration and insert it in a new backward integration, 
to find better starting values for the next forward integration.
There is a simplification that makes a good $\rho_{\rm GW}(a)$ guess possible, 
based upon the generic behavior of solutions that are cosmologically allowed.
While $\rho_{\rm GW}(a)$ increases over time 
during reheating and the stiff-SFDM-dominated era, as more and more modes 
reenter the horizon, this increase peaks when the stiff era ends.
Thereafter, for most cases of interest, with a substantial stiff era, 
$\rho_{\rm GW}(a)$ evolves like radiation, i.e., 
$\rho_{\rm GW}(a)\simeq\rho_{\rm GW}(a=1)/a^4$.
As a result, we can use this assumed behavior, 
along with the final value of $\rho_{\rm GW}$ at $a=1$ from the last forward integration.
This can be extrapolated safely back to $a_{\rm reheat}$ in the following backward integration, 
since $\rho_{\rm GW}$ does not affect the expansion history at earlier times 
when the energy density of the universe is dominated by the SFDM in the stiff phase.
In cases in which the stiff era is too limited in duration 
to boost $\rho_{\rm GW}$ significantly above the value in $\Lambda$CDM, 
$\rho_{\rm GW}$ is so small that there is no back-reaction on the expansion rate, 
so this radiationlike extrapolation from $a=1$ backward in time is fine, as well,
since it makes no difference.

In general, each new forward integration in this iteration scheme yields 
a new, improved $\rho_{\rm GW}(a)$ guess to use in the next backward integration.
These iterations are continued until the threshold for convergence is achieved 
($\Delta\leq0.001$) for a forward integration.
For example, in the case in which successive iterations cause an increase in $\Delta$, 
we discard the current iteration and examine carefully the last iteration, 
to improve the $\rho_{\rm GW}(a)$ guess for the next iteration, 
by a bisection of the guesses in two previous iterations.
There are details for safely converging, which we leave aside.

For the backward integration, we follow the same numerical method as in Paper I. 
We apply the {\it fast-oscillation approximation} from the present
up to the matching point at $a=a_M$, where that approximation is still valid. 
We refer to the solution obtained in this regime as the ``late-time solution''. 
Then, starting from the matching point, we calculate the {\it exact} evolution (without any approximation), 
all the way back to the point at $a_{\rm reheat}$, i.e. the point at which SFDM comes into existence. 
We refer to this part as the ``early-time solution''.

When we integrate forward in time, starting from $a_{\rm reheat}$ with the initial condition 
provided by the backward integration, we obtain the early-time solution first.
We have to solve the coupled ODEs (\ref{eq:ndhubble}), (\ref{eq:ndkge}) and (\ref{eq:ndfe}) exactly, 
since we are in the slow-oscillation regime.
This recipe is carried out up to the matching point, after which we can apply the 
fast-oscillation approximation again. 
Then we combine Eqs. (\ref{equation:friedmann2}), (\ref{eq:EC}) and (\ref{equation:eos1}) 
to calculate the late-time solution.

The contribution from the SGWB is accounted for {\it self-consistently} in the forward integration, 
by the following treatment of $\ogw(a)$ 
which appears in the dimensionless Friedmann equation (\ref{eq:ndfe}). 
As shown in Eq. (\ref{eq:gwspectrum}), $\ogw(a)$ is integrated over all wavenumbers $k$. 
At each time step, we add to the integral the contribution from the mode that reenters the horizon
at the current time step. In fact, using Eqs. (\ref{eq:gwspectrum}) -- (\ref{eq:GWreheat}),
\begin{widetext}
\begin{IEEEeqnarray}{rCl}\label{eq:GWspectranum}
    \ogw(a) & = & \left(\int^{k_{\rm inf}}_{k_{\rm reheat}}\ogw(k,a)\ud\ln k
    + \int^{k_{\rm reheat}}_{k_{\rm hor}+\Delta k}\ogw(k,a)\ud\ln k +\ogw(k,a)\Delta\ln k\right)
    + \int^{k_{\rm hor}}_0\ogw(k,a)\ud \ln k\nonumber\\
    & = & \frac{rA_s}{24(2-n_t)}
    \f{a_{\rm inf}^4H_{\rm inf}^2}{a^4H^2}\left(\frac{a_{\rm reheat}}{a_{\rm inf}}\left(\frac{k_{\rm reheat}}{k_*}\right)^{n_t}-\left(\frac{k_{\rm inf}}{k_*}\right)^{n_t}\right)+\f{rA_s}{12a^2H^2}\int_{k_{\rm hor}+\Delta k}^{k_{\rm reheat}}a_k^2H_k^2T_h(k,a)\left(\frac{k}{k_*}\right)^{n_t}\ud\ln k\nonumber\\
    & & +~\frac{rA_s}{12}T_h(k_{\rm hor},a)\left(\frac{k_{\rm hor}}{k_*}\right)^{n_t}\Delta\ln k
    +\frac{rA_s}{24(2+n_t)}\left(\frac{k_{\rm hor}}{k_*}\right)^{n_t}, 
\end{IEEEeqnarray}\
\end{widetext}
where $k_{\rm hor}=aH/c$ is the wavenumber of the mode that fills the horizon 
at the current time step and $\Delta k$ is the difference 
between such a wavenumber at the current time step and the previous one, 
$\Delta\ln k\equiv \Delta k/k$. 
The equation above demonstrates how we account for $\ogw(a)$ 
in the coupled ODEs, for both the early-time and late-time solution. 

The tensor transfer function $T_h(k,a)$ that we use in Eq. (\ref{eq:GWspectranum}) 
is the one in Eq. (\ref{transalpha}) in which $\alpha_k$ is evaluated 
with the corresponding $w(a_k)=p(a_k)/\rho(a_k)$ of the background universe, 
multiplied by a factor $A^2$ mentioned at the end of \S\ref{sec:GWanalytics}, 
which accounts for the damping of tensor modes from free-streaming neutrinos, 
which is nontrivial during the radiation-dominated era.
It was first pointed out by \cite{2004PhRvD..69b3503W} that, 
since a free-streaming relativistic component contributes an anisotropic stress-energy tensor $\pi_{ij}$
on the right-hand side of the tensor wave equation (\ref{tensoreq}), 
the growth of $h_k$ will be damped, once it reenters the horizon, 
compared with the solution without anisotropic inertia (see \S\ref{sec:GWanalytics}). 
This effect amounts to a multiplicative factor $A$ as a function of the fraction of 
the free streaming species, calculated by \cite{2005PhRvD..72h8302D}.
In cosmology, the only important case is the free streaming neutrinos 
during the radiation-dominated era, when their fraction $\Omega_\nu(a)$ is not negligible.
Therefore, $A=A(\Omega_\nu(a))$ should be applied to modes which reenter the horizon 
during the radiation-dominated era. In this paper, we will only quote the result of 
 $A(\Omega_\nu(a))$ from \cite{2005PhRvD..72h8302D}, 
 and incorporate it into Eq. (\ref{eq:GWspectranum}), which yields
 \begin{widetext}
 \begin{IEEEeqnarray}{rCl}\label{eq:GWspectrafinal}
	\ogw(a) & = &  \frac{rA_s}{24(2-n_t)}
    \f{a_{\rm inf}^4H_{\rm inf}^2}{a^4H^2}\left(\frac{a_{\rm reheat}}{a_{\rm inf}}\left(\frac{k_{\rm reheat}}{k_*}\right)^{n_t}-\left(\frac{k_{\rm inf}}{k_*}\right)^{n_t}\right)\nonumber\\
    & & +~\f{rA_s}{12a^2H^2}\int_{k_{\rm hor}+\Delta k}^{k_{\rm reheat}}a_k^2H_k^2T_h(k,a)\left(\frac{k}{k_*}\right)^{n_t}A^2(\Omega_\nu(a))\ud\ln k\nonumber\\
    & & +~\frac{rA_s}{12}T_h(k_{\rm hor},a)\left(\frac{k_{\rm hor}}{k_*}\right)^{n_t}A^2(\Omega_\nu(a))\Delta\ln k
    +\frac{rA_s}{24(2+n_t)}\left(\frac{k_{\rm hor}}{k_*}\right)^{n_t}. 
\end{IEEEeqnarray}
\end{widetext}
This equation (\ref{eq:GWspectrafinal}) is the final version of $\ogw(a)$ 
which we insert into the dimensionless Friedmann equation (\ref{eq:ndfe}) for our numerical calculation.
We are hereby able to treat the back reaction of GWs 
unto the expansion history of the background $\Lambda$SFDM universe, 
an effect that has not been self-consistently taken into account in previous literature. 
In this paper, we provide the first example 
of a holistic numerical evolution of the homogenous universe, 
which correctly accounts for the back reaction from GWs, 
while including all contributions to the total energy density of the universe.

%
%

\end{enumerate}

\subsubsection{Results: example $\Lambda$SFDM models}
\label{sec:fidSFDM}

We will now show the evolutionary aspects of $\Lambda$SFDM by presenting results 
for some example models obtained from our numerical calculation in detail.\footnote{
The fiducial model in Paper I was
$m = 3 \times 10^{-21}~\rm{eV}/c^2$,
$\lambda/(mc^2)^2 = 2 \times 10^{-18}~\rm{eV}^{-1} \rm{cm}^3$, 
see Fig. 1,2,3 in \cite{2014PhRvD..89h3536L}.}
As we sill see in \S\ref{sec:constraint}, 
these models are chosen to fulfill the constraints from the observables 
described there and in \S\ref{sec:EvolALL}, 
while still being in the range of parameters of interest for solving the
small-scale structure problems of CDM.

Again, we refer the reader to Fig. \ref{fig:hubble_all}, the evolution of the Hubble parameter 
of several example $\Lambda$SFDM models with different parameters, as a function of scale factor. 
As in Paper I, we find it convenient to work with the ratio $\lambda/(mc^2)^2$, 
rather than $\lambda$, because many observables constrain the former, rather than the latter. 
For all these example models, the value of $\lambda/(mc^2)^2$ is chosen to be
\beq \label{eq:lgtom2}
\lambda/(mc^2)^2 = 1 \times 10^{-18}~\rm{eV}^{-1}~\rm{cm}^3.
\eeq
The value of $\lambda/(mc^2)^2$ corresponds, for example, 
to the minimum size of a virialized halo in SFDM models with significant self-interaction, 
in the Thomas-Fermi regime (see \S\ref{sec:paper1}), 
i.e. choosing a fixed value for $\lambda/(mc^2)^2$ amounts to 
fixing the minimum clustering scale below which structure formation is suppressed.
Since observations suggest a scale of order kpc, we adopt the above value, 
corresponding to a scale of 0.8 kpc (which is smaller than that of the fiducial model in Paper I).

Also, the value of the tensor-to-scalar ratio is fixed, $r = 0.01$, 
for all three models in Fig \ref{fig:hubble_all}. 
It satisfies the latest upper bound given in Table \ref{table:tab1}, $r<0.07$, 
from CMB polarization experiments.
The other input parameters, the SFDM particle mass $m$ 
and the reheat temperature $\tre$, are varied among the three models, 
as illustrated by the plot labels. 
We have chosen three values for the reheat temperature, 
$\tre=10^3,~10^6$ and $10^9$ GeV, which span a wide range 
of possible $\tre$ in $\Lambda$SFDM (the energy density at $\tre$ 
should not exceed the inflationary energy scale). 
We vary the SFDM particle mass $m$ accordingly with these choices of $\tre$, 
so as to satisfy the constraints described in \S\ref{sec:constraint}. 

As shown in Fig \ref{fig:hubble_all}, the Hubble parameter of the universe drops from the initial plateau, $H_{\rm inf}$, when inflation ends, at different scale factors $a_{\rm inf}$ 
for different example models. The duration of the prolonged $w=0$ reheating, 
in which $H(a)\propto a^{-3/2}$, is also different among these models. 
In accordance with the definition of $\tre$, 
the higher it is, the shorter the duration of reheating.
The end of reheating marks the emergence of BEC SFDM 
and all the SM particles. 
To describe the homogeneous evolution of the $\Lambda$SFDM universe hereafter, 
we will focus on one of the example models, in which $\tre=10^3$ GeV, and 
\beq \label{eq:mass}
m = 8\times10^{-21}~\rm{eV}/c^2.
\eeq

\begin{figure*}[t]
\begin{minipage}{\linewidth}
     \centering
     \hspace*{-0.2in}
     \includegraphics[width=0.5\linewidth]{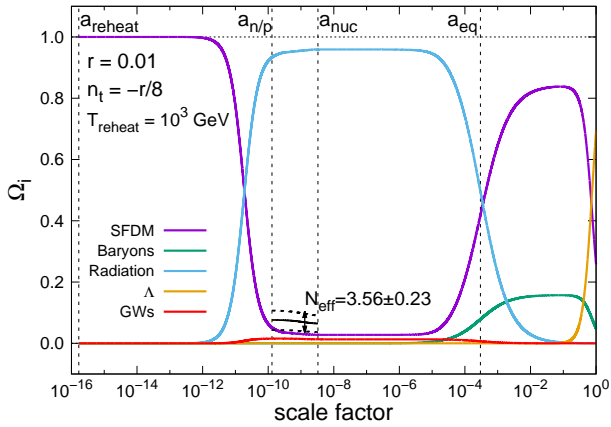}
     \includegraphics[width=0.5\linewidth]{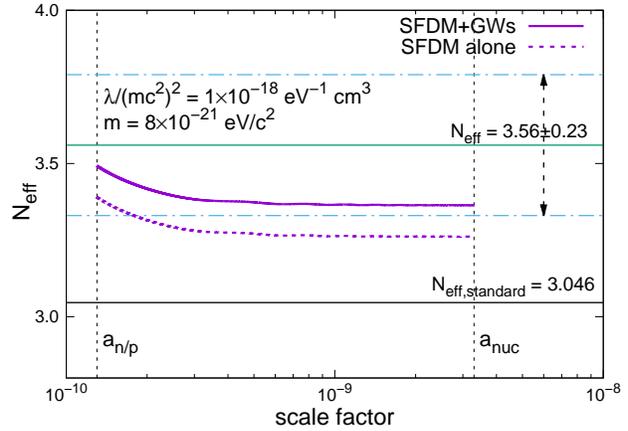}
     \end{minipage}

  \caption{\tx{left-hand plot:} Post-inflationary evolution of energy density fractions 
  of all cosmic components, $\Omega_i$ \emph{vs.} $a$, for an example $\Lambda$SFDM model.
  \tx{right-hand plot:} Evolution of $N_{\rm eff,BBN}$ \emph{vs.} $a$, 
  for the same model during BBN. The solid purple curve accounts for 
  all contributions to $N_{\rm eff,BBN}$, including SFDM and the SGWB from inflation.
  }
  \label{fig:fiducial_model}
\end{figure*}

\begin{figure*}[t]
\begin{minipage}{\linewidth}
     \centering
     \hspace*{-0.2in}
     \includegraphics[width=0.5\linewidth]{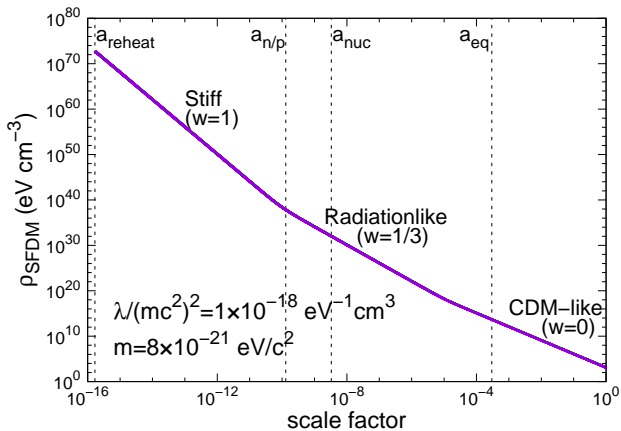}
     \includegraphics[width=0.5\linewidth]{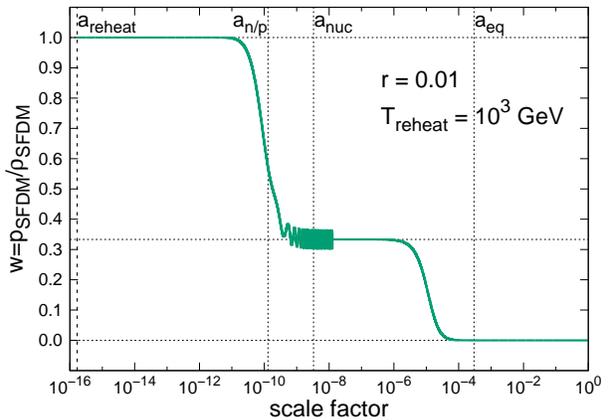}
     \end{minipage}

  \caption{\tx{left-hand plot:} Evolution of the energy density of SFDM.
  \tx{right-hand plot:} Evolution of equation of state of SFDM. 
  Oscillations are shown explicitly for $a<a_M$, which are manifest in the EOS plot, 
  but only the smooth average is plotted for $a>a_M$.
  }
  \label{fig:fiducial_SFDM}
\end{figure*}

For this model, the evolution of the Hubble parameter as a function of scale factor 
is plotted in Fig. \ref{fig:hubble}, and the evolution of the energy density fractions 
of all its components can be found in the left-hand plot of Fig \ref{fig:fiducial_model}. 
We can see that SFDM dominates in the universe twice: 
first, from the time of the onset of the stiff phase 
---which follows the epoch of reheating at $a_{\rm{reheat}}$,
to shortly before the time of neutron-proton freeze-out $a_{n/p}$, 
and then after the time of matter-radiation equality at $a_{\rm{eq}}$ to shortly before the present,
which is $\Lambda$-dominated. At present, $\Omega_i$ of all the components, 
as well as the Hubble constant $H_0$, 
match the cosmological parameters measured by Planck in Table \ref{table:tab1}.

The intermediate radiation-dominated era of $\Lambda$SFDM 
also has a different expansion history from that of $\Lambda$CDM.
There are two extra radiation components besides the standard radiation (photons plus neutrinos),
namely radiationlike SFDM and primordial GWs amplified by the stiff era.
As shown in the left-hand plot of Fig. \ref{fig:fiducial_model}, 
$\Omega_{\rm SFDM}$ is constant during its radiationlike phase, as a ``plateau'' 
(see Paper I for a more detailed description). 
In the same era, this model allows for another plateau contributed by
the energy density fraction of the SGWB from inflation, $\ogw$. 
As predicted in \S\ref{sec:eval_ogw}, it is possible that $\rho_{\rm GW}$ can emerge 
as a significant contribution to the total energy density of the universe during the RD era 
(indicated by the plateau of $\ogw$ in the left-hand plot of Fig. \ref{fig:fiducial_model}), 
resulting from the amplification of subhorizon GWs during the stiff-SFDM-dominated era.
For all the example models shown here, the boost effect is significant, 
due to the considerable number of e-foldings during the stiff era. 
For tensor modes that reenter the horizon after the stiff-SFDM-domination ends, 
of lower frequencies than those that reentered before, 
their energy density is not boosted relative to that of the background universe, 
so they add little to the total energy density of the SGWB, 
or its fraction $\ogw(a)$ given by Eq. (\ref{eq:gwspectrum}), 
throughout their subhorizon evolution.
Hence, for $\Lambda$SFDM models like these, $\ogw(a)$ will always be dominated by modes 
which have reentered the horizon by the end of the stiff-SFDM-dominated era. 
From that moment on, the relative contributions to the total $\rho_{\rm GW}(a)$ are fixed
for all modes that contribute significantly, and, as subhorizon modes, 
they evolve thereafter like radiation, $\ud\rho_{\rm GW}/\ud\ln k\propto a^{-4}$, 
thus, so must $\rho_{\rm GW}(a)\propto a^{-4}$ approximately.
Therefore, $\ogw(a)$, only beginning to emerge at the end of the stiff-SFDM-dominated era, 
soon stops growing and becomes a plateau when the stiff-to-radiation transition finishes.

The evolution of the SFDM, itself, is shown in Fig. \ref{fig:fiducial_SFDM},
from our numerical calculation. 
The respective phases of stiff, radiationlike, and CDM-like evolution 
are indicated in the left-hand plot.
They follow the behavior derived heuristically in \S\ref{sec:SFDMbackground}.
The right-hand plot shows the evolution of the EOS parameter of the SFDM
$w = p_{\rm{SFDM}}/\rho_{\rm{SFDM}}$, respectively. 
The wiggles in this figure reflect the oscillatory nature of the scalar field $\psi$, 
which generally appear in exact solutions of all types of DM modeled by a scalar field 
(see, e.g., \cite{2014ASSP...38..107S}). 
This oscillation feature stops at $a=a_M$ when we change the calculational method, 
between the slow- and the fast-oscillation regime 
(see and \S\ref{sec:SFDMbackground} and \S\ref{sec:nummeth}).
Note that there are no wiggles in the left-hand plot of Fig \ref{fig:fiducial_SFDM}, 
indicating that the oscillations are not manifest in $\rho_{\rm SFDM}$, only in $p_{\rm SFDM}$.
This guarantees that the expansion history of the background universe, 
which only depends on the mean energy density of SFDM, is not affected by these oscillations.

For fixed $r$ and $\tre$, the transition of the SFDM EOS between the radiationlike ($w=1/3$) 
and CDM-like (matter-like, $w=0$) phase is determined solely by the parameter $\lambda/(mc^2)^2$. 
The larger $\lambda/(mc^2)^2$ is, the later the transition. 
In contrast, the transition between the stiff ($w=1$) and radiationlike phase is determined by 
both SFDM particle parameters, $m$ and $\lambda/(mc^2)^2$. 
In other words, while the beginning of the stiff phase is set by $\tre$, 
its end is determined by both $m$ and $\lambda/(mc^2)^2$. 
For fixed $\lambda/(mc^2)^2$, the larger the mass $m$, the earlier the stiff phase ends. 
For fixed $m$, the larger $\lambda/(mc^2)^2$ is, the earlier the stiff phase ends, as well.
In the limit of small $\lambda/(mc^2)^2$, the end of the stiff phase is determined 
primarily by $m$ alone. 
Hence, the duration of each phase can be tuned by SFDM particle parameters. 
It was also shown in Paper I how changing these parameters affects the evolution of SFDM. 

We highlight again that SFDM in its early stiff phase dominates 
the energy density of the background universe,
which gives rise to several interesting implications on cosmological observables 
as mentioned above in \S\ref{sec:EvolALL}. 
For example, both SFDM and GWs contribute to $\neff$ during BBN 
(from $a_{\rm n/p}$ to $a_{\rm nuc}$),
increasing the expansion rate of the background universe. 
The evolution of $\neff$ during BBN is illustrated in the right-hand plot of Fig \ref{fig:fiducial_model}. 
For the example model, the contribution from the SGWB from inflation is noticeable. 
When SFDM transitions from radiationlike to CDM-like, it will no longer contribute to $\neff$. 
However, the SGWB contribution will always remain, which could affect other 
cosmological observables at later times, such as $\zeq$. 
Therefore, such observables will be capable of constraining $\Lambda$SFDM parameters, 
via the relic SGWB from inflation. 
We will carry out the analyses and show results from these constraints in the next section.

\section{Results: new constraints on SFDM particle parameters from cosmological observables}
\label{sec:constraint}

\subsection{Constraint from matter-radiation equality $z_{\rm{eq}}$} 
\label{sec:ConstraintEQ}

As briefly mentioned in \S\ref{sec:EvolALL}, 
a $\Lambda$SFDM model has to preserve the redshift of matter-radiation equality, 
$z_{\rm eq}$, according to the measurement from CMB.
The constraint on the value of $z_{\rm eq}$ from the Planck 2015 results reads
\beq\label{eq:zeqPlanck}
	z_{\rm eq}=3365\pm 44,\quad (68\%{\rm ~confidence~limit}).
\eeq 
This requires that SFDM should be well into its CDM-like phase 
(i.e. be fully non-relativistic) at $z_{\rm eq}$. 
As a result, it sets a constraint on the transition point 
between the radiation-like and CDM-like phases of SFDM, 
which is a function of $\lambda/(mc^2)^2$, as described in \S\ref{sec:fidSFDM} and in Paper I. 
As long as SFDM has completed this transition well before $z_{\rm eq}$, 
one can derive from the definition of $z_{\rm eq}$ that
\begin{equation}\label{eq:zeqconstraint}
    1+z_{\rm{eq}}\equiv\frac{1}{a_{\rm{eq}}}=\frac{\Omega_bh^2+\Omega_ch^2}{\Omega_rh^2
    +\Omega_{\rm GW}h^2
    },
\end{equation}
where $a_{\rm{eq}}$ is the scale factor at matter-radiation equality, 
and $\Omega_ih^2$ ($i=b,c,r$) is given in Table \ref{table:tab1}. 
In particular, $\rho_{\rm SFDM}$ after the transition evolves as matter all along until today 
and matches the present-day value determined by $\Omega_ch^2$.
We first ignore the term $\Omega_{\rm GW}h^2$ in Eq. (\ref{eq:zeqconstraint}) for a moment, 
as in Paper I. 
Then the value of $z_{\rm{eq}}$ calculated by Eq. (\ref{eq:zeqconstraint}) must 
exactly agree with the constraint in Eq. (\ref{eq:zeqPlanck}). 
Therefore, without GWs, the only aspect through which SFDM is 
constrained is its radiation- to CDM-like transition point, governed by $\lambda/(mc^2)^2$.
We have shown this constraint on $\lambda/(mc^2)^2$ in Paper I. 
Here we update it with the latest constraint on $z_{\rm eq}$ in Eq. (\ref{eq:zeqPlanck}), 
but use the same threshold value of $w= p/\rho = 0.001$
(neglecting the subscript SFDM here), a tiny deviation from zero,
to indicate the point after which SFDM can be considered as fully non-relativistic 
(i.e., $w<0.001$ for $a>a_{w={0.001}}$). 
The requirement of $a_{w={0.001}}\leq a_{\rm eq}$ can be translated 
into the following constraint on $\lambda/(mc^2)^2$:
\begin{equation} \label{equation:constraint1}
    \frac{\lambda}{(mc^2)^2} \leq 4.3 \times 10^{-17}~\rm{eV}^{-1}~\rm{cm}^3.
\end{equation}
The choice of this threshold $w= 0.001$ is artificial. 
If we relaxed it to higher values of $w$, 
the corresponding constraint on $\lambda/(mc^2)^2$ would become less
tight, allowing a broader range of values. 
A more precise threshold would require a recalculation of the CMB power spectrum 
for different SFDM particle parameters,
to solve for the best-fitting $\Lambda$SFDM parameters, 
which is well-beyond the scope of this paper.

Now we add the contribution from the amplified inflationary SGWB.
From Eq. (\ref{eq:zeqconstraint}), we see that not only should SFDM
be fully non-relativistic by $z_{\rm{eq}}$, 
but the amount of $\ogw$, amplified by the stiff era, is also subject to the constraint. 
Since for fixed $r$ and $\tre$, $\ogw(a)$ is determined by SFDM particle parameters, 
 $m$ and $\lambda/(mc^2)^2$, \emph{both} these parameters will be constrained further. 
By matter-radiation equality, $\ogw(a)$ has already evolved through the ``plateau'' 
described in \S\ref{sec:fidSFDM}, the height of which is determined 
by the duration of the stiff-SFDM-dominated era.
 The later the stiff era ends, the more modes get amplified 
 and thus the higher the plateau of $\ogw(a)$ is, 
 which will result in a later $z_{\rm{eq}}$ as inferred from Eq. (\ref{eq:zeqconstraint}).
 Therefore, to keep it in agreement with the measured value of $z_{\rm{eq}}$, 
 it is required that the stiff phase of SFDM ends early enough. 
We adopt the $-1\sigma$ confidence limit in Eq. (\ref{eq:zeqPlanck}) 
 as the minimum allowed value for $z_{\rm{eq}}$. 
 Thus, for fixed $r$ and $\tre$, there will be a lower limit on the mass $m$ 
 for each allowed value of $\lambda/(mc^2)^2$.
With the inclusion of GWs, the allowed range of ($\lambda/(mc^2)^2,~m$) 
due to the $z_{\rm{eq}}$ constraint will be more stringent 
than the half-plane given by Eq. (\ref{equation:constraint1}) for the case without GWs. 
This is illustrated in the SFDM particle parameter space, 
shown in Figs. \ref{fig:pssingle} and \ref{fig:psmultiple}.
A detailed description of the allowed ranges from the $z_{\rm{eq}}$ constraint, 
parametrized by $r$ and $\tre$, will be given in \S\ref{sec:PPS}.


\subsection{Constraint from $N_{\rm{eff}}$ during Big Bang nucleosynthesis}
\label{sec:constraintBBN}

The effective number of neutrino species, $N_{\rm{eff}}$, 
is introduced in \S\ref{sec:EvolALL} as a measure of relativistic degrees of freedom of the universe.
It affects the expansion rate in the early universe, 
at all times before the matter-dominated era, 
which encompasses the important epoch of Big Bang nucleosynthesis.
The abundances of primordial light elements produced by BBN 
are very sensitive to the expansion rate then.
As a result, measurements of these abundances through astronomical observations 
of metal-poor systems set a constraint on $N_{\rm{eff,BBN}}$ during BBN 
\cite{2014MNRAS.445..778I, 2015PhRvD..91h3505N, 2016ApJ...830..148C}.
BBN is not an instantaneous event; it undergoes two important stages 
which we explained in \S\ref{sec:EvolALL}, 
first, neutron-to-proton freeze-out occurs at $a_{\rm n/p}$ 
and then, light nuclei production occurs at $a_{\rm nuc}$ \cite{Dodelson}, 
where $a_{\rm nuc}/a_{\rm n/p}\simeq T_{\rm n/p}/T_{\rm nuc}\approx20$. 
Therefore, BBN actually cares about the evolution of $N_{\rm{eff,BBN}}(a)$ 
throughout this window ($a_{\rm n/p},~a_{\rm nuc}$).
Nevertheless, it is often the case that 
only a single value of $N_{\rm{eff,BBN}}$ is reported from observations,
in which the expansion history is modeled by a constant $N_{\rm{eff,BBN}}$ at all times,
since it is the simplest model to fit.
In this paper, we use the following measurement result \cite{2015PhRvD..91h3505N} 
to constrain the SFDM model,
\begin{equation} \label{equation:neff}
    N_{\rm{eff,BBN}}=3.56 \pm 0.23, \quad (68\%{\rm ~confidence~limit}).
\end{equation}
We comment that this value is not required to be consistent with the $N_{\rm{eff,CMB}}$ 
measured by CMB anisotropies, because $N_{\rm{eff}}(a)$ can in principle evolve over time 
as in our $\Lambda$SFDM model (see Fig. \ref{fig:fiducial_model}), 
and $N_{\rm{eff,CMB}}$ is only affected by its values later at around recombination.
In other words, $N_{\rm{eff,BBN}}$ and $N_{\rm{eff,CMB}}$ indicate 
relativistic degrees of freedom at \emph{different} epochs of the expansion history.
As a matter of fact, current measurements mildly suggest that 
$N_{\rm{eff,BBN}}$ be greater than $N_{\rm{eff,CMB}}$ by $\sim1\sigma$
\cite{2014MNRAS.445..778I, 2015PhRvD..91h3505N, 2016ApJ...830..148C, 2015arXiv150201589P}.

In $\Lambda$CDM, where there are only three SM neutrino species all the time, 
$N_{\rm{eff,BBN}}(a)=N_{\rm{eff, standard}}=3.046$. 
In contrast, in $\Lambda$SFDM, SFDM has an EOS which evolves over time, 
affecting the expansion rate during BBN if SFDM is relativistic then, 
and will hence contribute to 
$N_{\rm{eff,BBN}}(a)\equiv N_{\rm{eff, standard}}+\Delta N_{\rm{eff,BBN}}(a)$ 
as an extra relativistic component, 
as we pointed out in \S\ref{sec:EvolALL} and in Paper I.
In addition, the inflationary SGWB which we have included self-consistently, 
amplified by the earlier stiff-SFDM-dominated era, 
also adds to $\Delta N_{\rm{eff,BBN}}(a)$, so it must be taken into account as well.
In fact, in a $\Lambda$SFDM model with the SGWB, 
we infer $N_{\rm{eff,BBN}}(a)$ between $a_{\rm{n/p}}$ and $a_{\rm{nuc}}$, 
from the energy density fractions of relativistic SFDM, $\Omega_{\rm{SFDM}}$, 
and the GWs, $\Omega_{\rm{GW}}$. 
Both are sources to $\Delta N_{\rm{eff,BBN}}(a)$,
\begin{equation}\label{equation:defneff}
    \frac{\Delta N_{\rm{eff,BBN}}(a)}{N_{\rm{eff, standard}}}= \frac{\Omega_{\rm{SFDM}}(a) 
    + \Omega_{\rm{GW}}(a)
    }{\Omega_\nu(a)},
\end{equation}
where $\Omega_\nu(a)$ denotes the energy density fraction of the SM neutrinos. 
The evolution of $N_{\rm{eff,BBN}}(a)$ for one example $\Lambda$SFDM model 
has been shown in the right-hand plot of Fig. \ref{fig:fiducial_model}.

We compare the $N_{\rm{eff,BBN}}(a)$ obtained this way 
to the measured value given by Eq. (\ref{equation:neff}), 
and impose on it a conservative threshold that throughout BBN 
(from $a_{\rm{n/p}}$ to $a_{\rm{nuc}}$), 
it shall be within the 1$\sigma$ confidence interval of the measured value.
In Eq. (\ref{equation:defneff}), both values of $\Omega_{\rm{SFDM}}(a)$ and $\Omega_{\rm{GW}}(a)$ 
are controlled by the properties of SFDM, i.e., its particle parameters $m$ and $\lambda/(mc^2)^2$, 
once the values of $r$ and $\tre$ are fixed, as described in \S\ref{sec:fidSFDM}. 
Therefore, the constraint on $N_{\rm{eff,BBN}}(a)$ will again translate 
as a constraint on the SFDM particle parameter pair ($\lambda/(mc^2)^2,~m$).

Eq. (\ref{equation:defneff}) shows that, for fixed $r$ and $\tre$, 
if the stiff phase of SFDM ends too late into the BBN epoch, 
the considerable amount of $\Omega_{\rm{SFDM}}(a)$ can lead to too large an $N_{\rm{eff,BBN}}(a)$ 
which violates its measured value given by Eq. (\ref{equation:neff}).
In addition, the later the stiff-to-radiationlike transition of SFDM is, 
the larger the amplified $\ogw(a)$ is, increasing the value of $N_{\rm{eff,BBN}}(a)$ as well.
Therefore, any change in the stiff-to-radiationlike transition point 
affects both $\Omega_{\rm{SFDM}}(a)$ and $\ogw(a)$ in the same direction.
In order for this transition to finish early enough that 
the sum of $\Omega_{\rm{SFDM}}(a)$ and $\ogw(a)$ should observe
the $+1\sigma$ confidence limit of $N_{\rm{eff,BBN}}$, 
there must be a lower bound on $m$, for any allowed value of $\lambda/(mc^2)^2$.

The radiationlike ``plateau'' of SFDM (see \S\ref{sec:EvolALL}), 
as well as its stiff-to-radiationlike transition, is subject to the BBN constraint.
If the plateau overlaps BBN, i.e., SFDM is well into its radiationlike phase by $a_{\rm nuc}$,
then $\Omega_{\rm{SFDM}}(\rm{plateau})$ during the plateau, 
as a function of $\lambda/(mc^2)^2$, must comply with 
the constraint on $N_{\rm{eff,BBN}}(a)$ according to Eq. (\ref{equation:defneff}).
In particular, for large enough $m$, the stiff phase of SFDM ends so early 
that not only the radiationlike phase of SFDM would enclose BBN, 
but also the amplification of the inflationary SGWB be insignificant, which leads to $\ogw(a)\simeq0$.
In this limit, the constraint from BBN amounts to 
a constraint on the value of $\Omega_{\rm{SFDM}}(\rm{plateau})$, 
and hence on $\lambda/(mc^2)^2$ alone.
The BBN constraint can thereby be analyzed the same way as in Paper I, for the case without GWs. 
We will not repeat that analysis here but just write down the result as follows: 
\begin{equation} \label{equation:constraint2}
	\hspace{-0.1in}
    2.3 \times 10^{-18}~\rm{eV}^{-1}~\rm{cm}^3\leq\frac{\lambda}{(mc^2)^2} \leq 4.1 \times 10^{-17}~\rm{eV}^{-1}~\rm{cm}^3,
\end{equation} 
for $\Lambda$SFDM models in which the SGWB from inflation is negligible,
and the radiationlike phase of SFDM overlaps BBN.
The lower and upper bounds on $\lambda/(mc^2)^2$ in the equation above
correspond to the $-1\sigma$ and $+1\sigma$ confidence limits of 
the measured value of $N_{\rm{eff,BBN}}$ given by Eq. (\ref{equation:neff}), respectively. 
The difference between Eq. (\ref{equation:constraint2}) and the equivalent bounds in Paper I 
only reflects our update on the measured value of $N_{\rm{eff,BBN}}$.

If $\lambda/(mc^2)^2$ is less than the lower bound in Eq. (\ref{equation:constraint2}), 
the SFDM plateau alone cannot make up a $\Delta N_{\rm{eff,BBN}}(a)$ 
which meets the $-1\sigma$ confidence limit of its measured value.
Therefore, for any of these smaller values of $\lambda/(mc^2)^2$, 
there must be an upper bound on $m$, 
which sets a constraint on how early the stiff phase can end, 
so that the sum of $\Omega_{\rm{SFDM}}(a)$ and $\Omega_{\rm{GW}}(a)$ 
can be substantial enough to reach the $-1\sigma$ limit of $N_{\rm{eff,BBN}}$.

\begin{figure*}[t]
\begin{minipage}{\linewidth}
     \centering
     \hspace*{-0.5in}
     \includegraphics[scale=0.9]{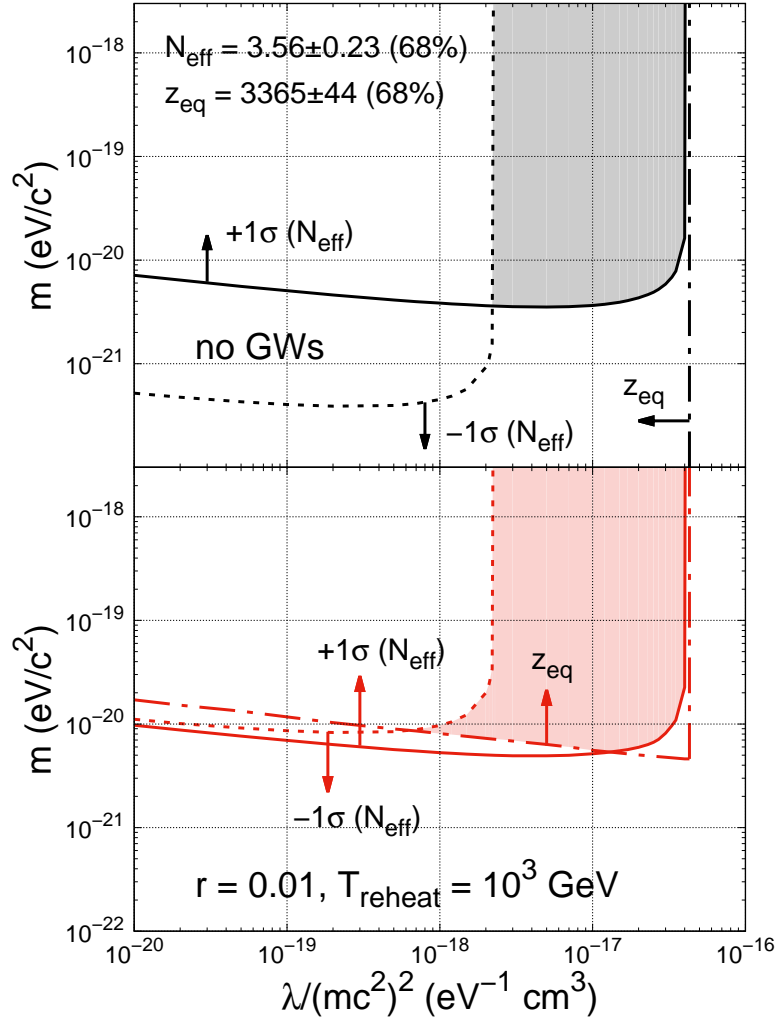}
\end{minipage}
  \caption{Cosmological constraints expressed in the SFDM parameter space 
  for values $(\lambda/(mc^2)^2,m)$. 
  {\it Upper plot:} for the case which does not include GWs, as in Paper I.
  {\it Lower plot:} for the case which self-consistently includes GWs, 
  in which $r=0.01$ and $\tre=10^3$ GeV.
  In both plots, the solid curve corresponds to the constraint 
  from the $+1\sigma$ confidence limit of $N_{\rm eff,BBN}$ at $a_{\rm n/p}$, 
  the dashed curve corresponds to the constraint 
  from the $-1\sigma$ confidence limit of $N_{\rm eff,BBN}$ at $a_{\rm nuc}$,
  and the dash-dotted curve indicates the constraint from $z_{\rm eq}$.
  The arrows indicate the directions in which the SFDM particle parameters 
  satisfy the respective cosmological constraints.
 In each plot, the shaded region denotes 
 the overall allowed range of the SFDM particle parameters, 
 for the respective case.
  }
  \label{fig:pssingle}
\end{figure*}

\begin{figure*}[t]
\begin{minipage}{\linewidth}
     \centering
     \hspace*{-0.2in}
     \includegraphics[width=0.82\linewidth]{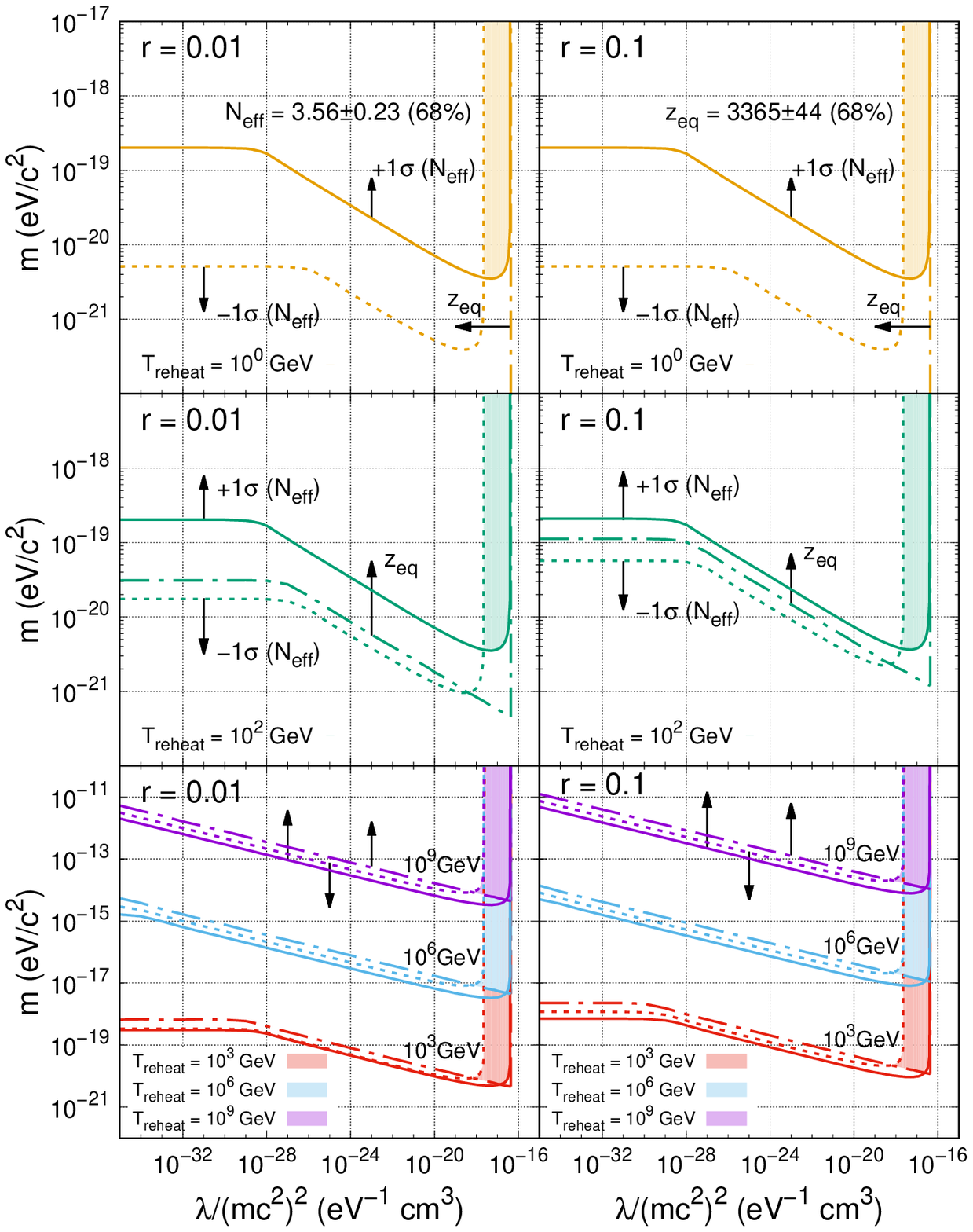}
     \end{minipage}
  \caption{Cosmological constraints expressed in the SFDM parameter space 
  for values $(\lambda/(mc^2)^2,m)$, for multiple choices of $\tre$ and $r$. 
  In each panel, the shaded region indicates the values of the SFDM particle parameters 
  which are allowed by the cosmological constraints,
  and the arrows indicate the directions of these constraints, the same as in Fig. \ref{fig:pssingle}. 
  In the bottom two panels, the allowed regions of the SFDM particle parameters, 
  for multiple choices of $\tre$, are plotted together; 
  all of them actually extend and overlap in the direction of larger-mass, exceeding the plot range,
  same as in the upper and middle panels.
  }
  \label{fig:psmultiple}
\end{figure*}

These constraints from $N_{\rm eff,BBN}$ on the allowed ranges of ($\lambda/(mc^2)^2,~m$) 
can also be illustrated in the SFDM particle parameter space plots, 
Figs. \ref{fig:pssingle} and \ref{fig:psmultiple}, for a wide range of $r$ and $\tre$.
In the upper plot of Fig. \ref{fig:pssingle}, we show the result for the case without GWs 
(i.e., setting $\ogw(a)=0$ in Eq. [\ref{equation:defneff}]), 
which can be compared to our previous result on the corresponding allowed region in Paper I. 
The bounds given by Eq. (\ref{equation:constraint2}) 
are also reflected in Figs. \ref{fig:pssingle} and \ref{fig:psmultiple}, 
as described in \S\ref{sec:PPS}.
There we will discuss in more details the allowed region due to the $N_{\rm eff,BBN}$ constraint
in the SFDM particle parameter space and its dependence on the values of $r$ and $\tre$.


\subsection{Results: allowed SFDM particle parameter space}
\label{sec:PPS}

Combining the constraints from the two cosmological observables described above, 
we can confine the allowed values of the SFDM particle parameters, $(\lambda/(mc^2)^2,m)$, 
in the two-dimensional parameter space, for various choices of $r$ and $\tre$ 
(see Figs. \ref{fig:pssingle} and \ref{fig:psmultiple} for the parameter space plots). 

In both figures, the constraints from $z_{\rm eq}$ and $N_{\rm eff,BBN}$ 
are expressed by curves of critical parameter values that marginally satisfy the respective constraints.
Specifically, in each plot, the constraint from $z_{\rm{eq}}$ is indicated by the dash-dotted curve, 
and the region above this curve is allowed 
by the $-1\sigma$ confidence limit of the measured value of $z_{\rm{eq}}$, 
given by Eq. (\ref{eq:zeqPlanck}). 
The solid curve refers to the constraint 
from the $+1\sigma$ confidence limit of $N_{\rm eff,BBN}$ at $a_{\rm n/p}$,
and the dashed curve to the constraint 
from the $-1\sigma$ confidence limit of $N_{\rm eff,BBN}$ at $a_{\rm nuc}$, 
given by BBN measurements (see Eq. [\ref{equation:neff}]).
The region below the solid curve and above the dashed curve is consistent with
the $1\sigma$ confidence interval of the measured value of $N_{\rm eff,BBN}$ throughout BBN
(see the right-hand plot of Fig. \ref{fig:fiducial_model} for reference).
The arrows in each plot indicate the directions in which the values of the SFDM particle parameters 
can satisfy the respective constraints,
which result in the shaded region that denotes 
the overall allowed range of the SFDM particle parameters, 
satisfying all cosmological constraints.

Fig. \ref{fig:pssingle} is a blow-up of Fig. \ref{fig:psmultiple}.
It shows the comparison between the case which does not include GWs 
in the evolution of $\Lambda$SFDM (the upper plot) 
so the values of $r$ and $\tre$ are not important, as studied in Paper I, 
and the case in which the SGWB from inflation is self-consistently included (the lower plot), 
which we study in this paper. 
In the upper plot, the constraint from $z_{\rm{eq}}$ 
is given by the upper bound on $\lambda/(mc^2)^2$ in Eq. (\ref{equation:constraint1}),
indicated by the vertical dash-dotted line: the half-plane on its left side is allowed. 
In the lower plot, the corresponding critical curve takes the same vertical line 
but pivots at a minimum value of $m$ and provides a lower bound on $m$ 
for every value of $\lambda/(mc^2)^2$ below its upper bound.
This change is due to the inclusion of the inflationary SGWB, which contributes a radiation component 
at $z_{\rm{eq}}$, as explained in \S\ref{sec:ConstraintEQ}.
In both the upper and lower plots of Fig. \ref{fig:pssingle}, 
it is easily seen that for large enough $m$, 
the parameter values ($\lambda/(mc^2)^2$, $m$) allowed by the $N_{\rm{eff,BBN}}$ constraint
indeed correspond to models in which the radiationlike phase of SFDM overlaps BBN 
and the effect from the SGWB is negligible,
so that the value of $\lambda/(mc^2)^2$ must be bounded 
between the asymptotic vertical solid and dashed lines given by Eq. (\ref{equation:constraint2}), 
as explained in \S\ref{sec:constraintBBN}.
In this limit, the allowed region in the lower plot 
becomes indistinguishable from the one in the upper plot, 
since the SGWB makes no difference to the background evolution of the universe.



Multiple cases are plotted in Fig. \ref{fig:psmultiple}, with different choices for $r$ and $\tre$.
In every panel, the overall allowed region for the SFDM particle parameters
is given by combining all the cosmological constraints, leaving the shaded area. 
In the above and middle panels, i.e., the four cases with $r=0.01$ or either $0.1$, 
and $\tre=1$ GeV or either $100$ GeV, the shaded regions are nearly 
indistinguishable from one case to another.
This reflects the fact that if the reheat temperature is too low, 
the stiff era is then too short to boost the inflationary SGWB to a considerable degree. 
In this situation, the allowed range of SFDM particle parameters simply reduces to 
that in the ``no GWs'' case as shown in Fig. \ref{fig:pssingle}.
Reheat temperatures $\tre\gtrsim10^3$ GeV start to make differences to the allowed regions,  
as shown in the bottom two panels of Fig. \ref{fig:psmultiple},
where the allowed regions for $\tre=10^3,~10^6$ and $10^9$ GeV are plotted together and overlap.
In these cases, they are significantly affected by the SGWB from inflation.
The larger the energy density of the SGWB amplified by the stiff era, 
resulting from an increase in either the value of $r$ or $\tre$,
the more stringent the constraints on the SFDM particle parameters, as one should expect. 
In fact, for fixed $\tre$, the allowed region contracts slightly when the value of $r$ increases 
from $r=0.01$ to $0.1$.
Its dependence on the value of $r$ is found to be relatively weak.
On the other hand, however, for fixed $r$, the allowed region shrinks significantly 
every time $\tre$ increases by a factor of 1000.
We find that, for given values of $r$, for $\tre\gtrsim10^3$ GeV, 
the minimum value of the SFDM particle mass, $m_{\rm min}$, 
among the models which satisfy all the cosmological constraints, 
is proportional to $\tre$. 
For $r\gtrsim0.01$, the dependence of $m_{\rm min}$ on both $r$ and $\tre$ 
can be empirically expressed as
\begin{widetext}
\begin{numcases}{\hspace{-0.2in}m_{\rm min}\simeq(5\times10^{-21}~{\rm eV}/c^2)\times}
    \f{\tre}{10^3~{\rm GeV}}\sqrt{\f{r}{0.01}}, & $\hspace{0.3in}~\tre\gtrsim10^3 \rm{~GeV},$\label{eq:mminlaw}\\[1em]
    1, &  $\hspace{0.3in} ~\tre<10^3 \rm{~GeV}.$ \label{eq:mmin2}
\end{numcases}
\end{widetext}

\section{Results: present-day SGWB energy density spectrum and its detectability by LIGO}
\label{sec:SGWBspectra}

In the $\Lambda$SFDM model, the integrated inflationary SGWB energy density 
predicted in \S\ref{sec:Evolution} 
contributes only a small fraction of the total energy density today, $\ogw(a=1)\sim10^{-8}-10^{-7}$.
As such, its effect on the universe today is negligible.
Remarkably enough, however, in its spectrum at high frequencies 
where amplification by the stiff-SFDM-dominated era was greatest, 
which can overlap the range of GW laser interferometer experiments, 
the amplitude can be significant enough to be detectable.
We demonstrate this here,
in light of the SGWB energy density spectrum predicted in \S\ref{sec:Evolution}
and the cosmological constraints on the SFDM particle parameters derived in \S\ref{sec:constraint},
by analyzing the detectability of the amplified inflationary SGWB at the present 
 by current and future laser interferometer experiments, 
as a unique signature of the $\Lambda$SFDM model.

\begin{figure*}[t]
\begin{minipage}{\linewidth}
     \centering
     \hspace*{-0.1in}
     \includegraphics[width=1\linewidth]{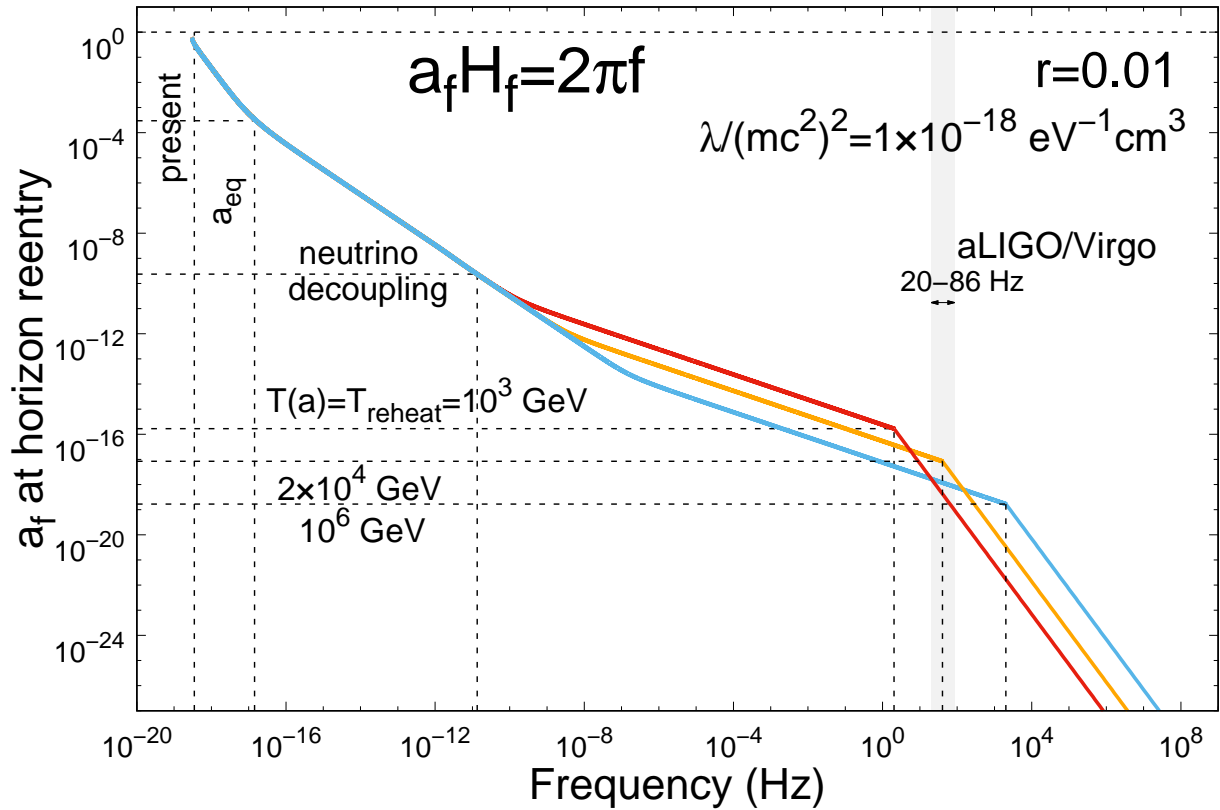}
    \end{minipage}
  \caption{Scale factor $a_f$ at horizon reentry ($a_fH_f=2\pi f$) 
  vs. the comoving frequency $f$ for each mode, plotted for three example $\Lambda$SFDM models
  with different $\tre$.
  $T(a)$ refers to the photon temperature at this scale factor.
  }
 \label{fig:afvsf}
\end{figure*}

The expansion history of the $\Lambda$SFDM universe described in \S\ref{sec:EvolALL} 
is imprinted in the present-day energy density spectrum of the SGWB from inflation, 
$\ogw(f)$, defined as follows:
\beq
	\ogw(f)\equiv\ogw(k=2\pi f/c, a=1).
\eeq 
For each mode whose (comoving) frequency is $f$, 
there corresponds an epoch of horizon reentry at $a_f\equiv a_{k=2\pi f/c}$, 
which determines the outcome of the cosmic evolution of its $\Omega_{\rm GW}(f)$ to the present-day, 
as described in \S\ref{sec:EvolGW}. For different SFDM model parameters 
and values of $\tre$ and $r$, there is a different mapping between $f$ and $a_f$.
This is illustrated in Fig. \ref{fig:afvsf}. 
In general, as long as the Hubble radius increases with time, 
as it does from the end of inflation to the end of matter-domination 
when the cosmological constant begins to dominate afterward, 
$a_f$ increases as $f$ decreases.
For $a_f > a_{\rm eq}$ (or $f< f_{\rm eq} \sim10^{-17}$ Hz, 
the dependence of $a_f$ on $f$ is universal, 
since the expansion history of $\Lambda$SFDM is the same as that of $\Lambda$CDM.
The maximum $a_f$ in all example models corresponds to the moment 
when modes begin to exit, instead of reentering the horizon, 
once $w=-1/3$ for the EOS of the background universe.
From this moment on, all modes that are still outside the horizon 
will never reenter the horizon, as the cosmological-constant-dominated era begins. 
On the other hand, toward the high-frequency end, 
manifest distinctions arise for $a_f$ earlier than the end of the stiff-SFDM-dominated era, 
among the three example $\Lambda$SFDM models.
We note that, since the dependence of $f$ on $a_f$ is different in $\Lambda$SFDM 
from its dependence in $\Lambda$CDM, for $a_f<a_{\rm eq}$, 
so will the dependence of $f$ on the photon temperature $T(a)$ at $a=a_f$ be different.
For $\Lambda$CDM, we can write $f\approx10^{-4}\rm{Hz}\frac{T(a_f)}{10^3\rm{GeV}}$, 
for $a_{\rm reheat}<a_f<a_{\rm eq}$ \cite{2012JCAP...06..027B}, 
while this is not true for $\Lambda$SFDM.
For example, as seen in Fig. \ref{fig:afvsf}, for the example model in which $\tre=2\times10^4$ GeV, 
$f\approx40$ Hz at $T(a_f)=\tre$.
 
We begin by presenting the present-day SGWB energy density spectra, $\ogw(f)$, 
for three example $\Lambda$SFDM models, to guide our discussion. 
In \S\ref{sec:GWspectratoday} below, we will use these models to explain the generic features 
of the inflationary SGWB spectra in relation to that in $\Lambda$CDM 
and in relation to the current and future GW detection experiments.
In \S\ref{sec:maxSNR}, we will revisit these illustrative models as we quantify 
the detectability of the SGWB in $\Lambda$SFDM as a function of the SFDM particle parameters, 
for given values of $r$ and $\tre$.
In the upper plots of Figs. \ref{fig:GWspectra1} -- \ref{fig:GWspectra3}, 
we show the present-day SGWB energy density spectra 
of the same example $\Lambda$SFDM models shown in Fig. \ref{fig:afvsf}. 
For all three models shown in this section, the values of $\lambda/(mc^2)^2$ are fixed 
according to Eq. (\ref{eq:lgtom2}) such that the corresponding core size of an SFDM halo 
is $\sim 0.8~{\rm kpc}$ due to the repulsive self-interaction of SFDM. 
We also hold the value of $r=0.01$ fixed for all these models, for the purpose of comparison.
For $\tre$, in contrast to the choices in \S\ref{sec:EvolALL}, 
we here choose $\tre=10^3,~2\times10^4,$ and $10^6$ GeV, 
such that the span of corresponding $f_{\rm reheat}$, the frequency of the mode 
that reenters the horizon at the end of reheating at $a_{\rm reheat}$, 
is nearly centered on the LIGO sensitive frequency band ($20-86$ Hz) \cite{2017PhRvL.118l1101A}.
\footnote{
This LIGO sensitive band is defined as the range which includes $99\%$ of the
signal for a \emph{flat} spectrum.  This range is
dependent on the assumed shape of the SGWB spectrum, however.
For power-law spectra $\Omega_{\rm GW}(f)\propto f^\beta$, 
the range is different for different values of $\beta$.  
For $\beta = 2/3$ (or 3), for example, this range shifts to $20-98$ (or 305) Hz, 
respectively.}
The SFDM particle mass $m$ is different for each value of $\tre$, as labeled. 
The particular choice of $m$ values will be described below in \S\ref{sec:maxSNR};
all $\Lambda$SFDM models shown in Figs. \ref{fig:GWspectra1} -- \ref{fig:GWspectra3} 
satisfy all of the cosmological constraints described above in \S\ref{sec:constraint}.
We will first describe the shape of $\ogw(f)$ for the example models 
and the respective detectabilities of their SGWBs, 
with special emphasis on the Advanced LIGO/Virgo experiment.

\subsection{Generic features of the present-day energy density spectrum $\ogw(f)$ of the inflationary SGWB and its detectability}
\label{sec:GWspectratoday}

\begin{figure*}[t]
\begin{minipage}{\linewidth}
     \centering
     \hspace*{-0.5in}
     \includegraphics[scale=1.4]{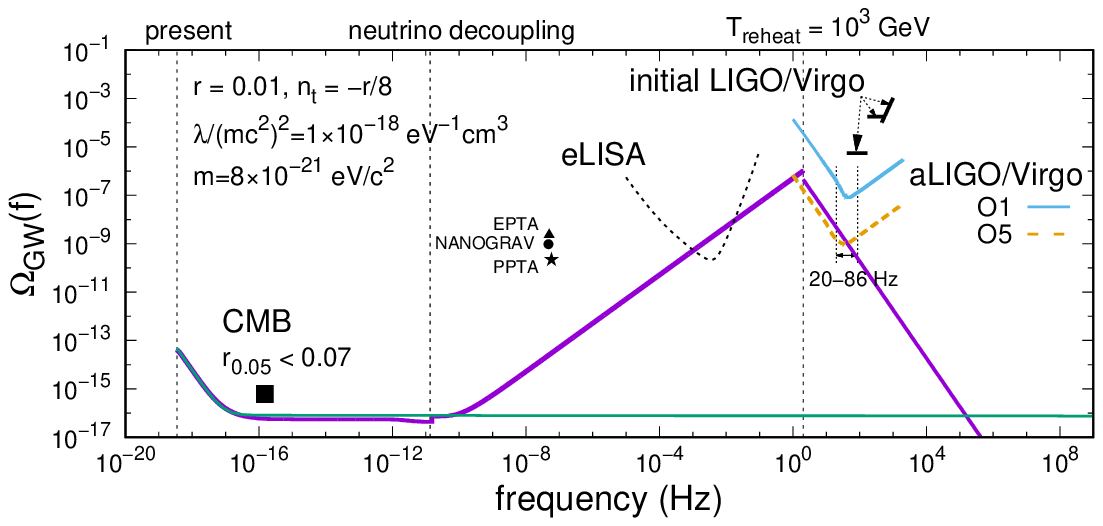}
      \includegraphics[scale=0.7]{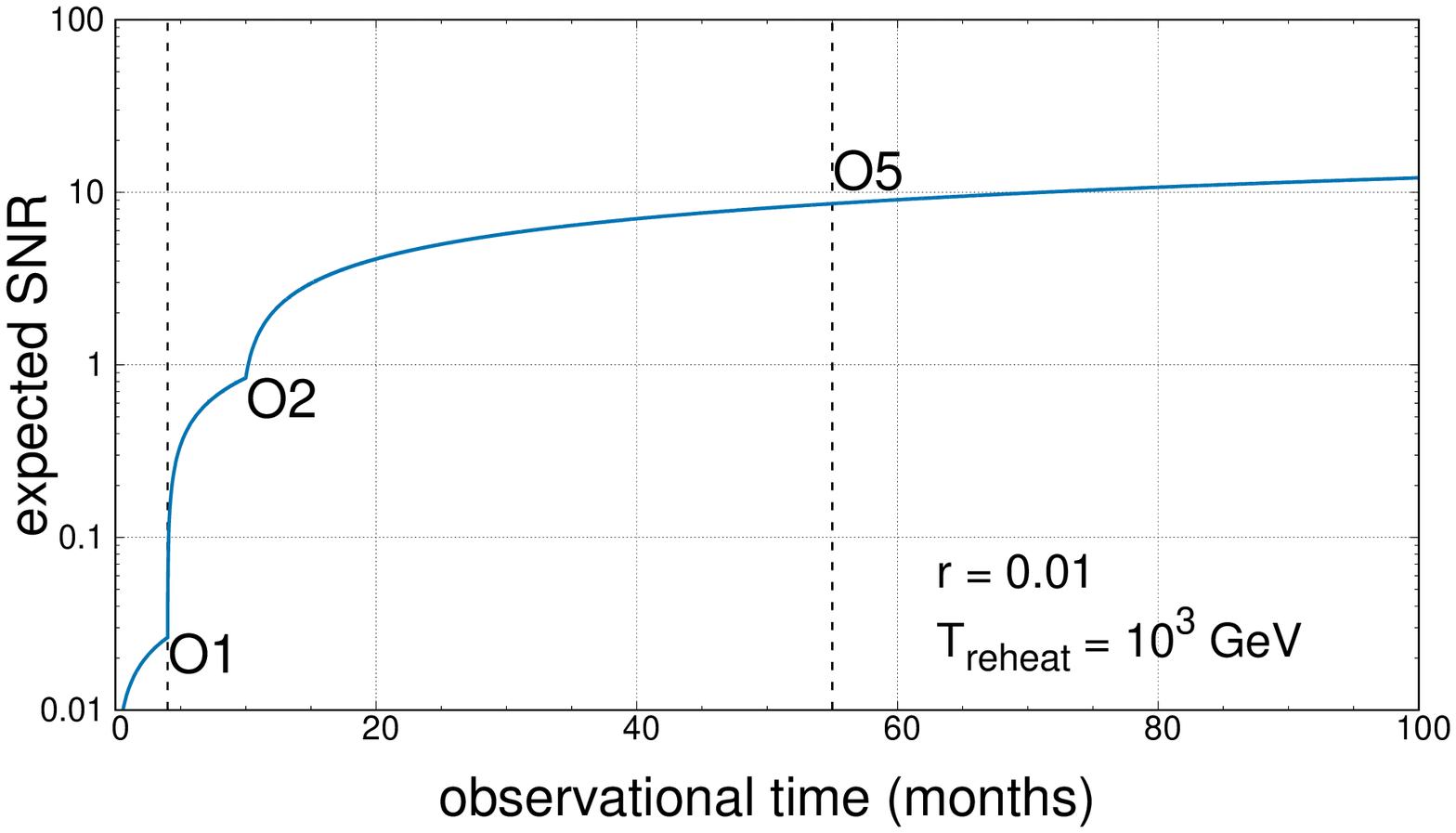}
    \end{minipage}
  \caption{\tx{Upper plot:} Present-day energy density spectrum of the SGWB from inflation.
 The purple curve shows the prediction from one example $\Lambda$SFDM model
 in which reheating ends at $\tre=10^3$ GeV.
 The green curve shows the prediction of the standard $\Lambda$CDM model. $r=0.01$ for both cases.
 [Note: the $e^-e^+$ annihilation results in the little kink of the purple curve
 right before neutrino decoupling. This effect had not been taken into account 
 in the green $\Lambda$CDM curve.]
 The blue solid curve and the yellow dashed curve indicate
 the 1$\sigma$-sensitivity curves of aLIGO/Virgo 
for the two observing runs O1 (with data now analyzed) 
and O5 (theoretical forecast for 2-year run), respectively,
 integrated over frequency for the inflationary SGWB energy density spectra in $\Lambda$SFDM.
 Upper limits from various experiments are shown, including the joint CMB analysis,
 PTA experiments and the (initial) LIGO/Virgo, all at 95\% confidence. 
 (The current upper limit from the aLIGO O1 run is shown in Fig. \ref{fig:GWzoomin}.)
 The LISA sensitivity curve is the predicted strain noise function of eLISA in \cite{2012CQGra..29l4016A} 
 and may be revised in the final design of the upcoming LISA mission.
  \tx{Lower plot:} The expected SNR of the inflationary SGWB 
  predicted by the same $\Lambda$SFDM model 
  \tx{vs.} the cumulative observation time of aLIGO/Virgo.
  The dashed vertical lines indicate the observation time by the end of O1 and O5 runs, respectively.
}

 \label{fig:GWspectra1}
\end{figure*}

As derived in \S\ref{sec:eval_ogw}, the generic energy density spectrum $\ogw(f)$ of 
the primordial SGWB from inflation, predicted by $\Lambda$SFDM models, 
must be approximately piece-wise power laws, 
the power indices of which are determined by the EOS parameters $w=p/\rho$ of the universe
throughout all eras in the expansion history. 
In particular, if we neglect the very weak dependence on the primordial tensor index $n_t$
(i.e. set  $n_t = 0$),
then $\ogw(f)\propto f^{-2}$ for modes which reenter the horizon during the matter-dominated era, 
after $z_{\rm eq}$.
For modes which reenter the horizon earlier than $z_{\rm eq}$, during the radiation-dominated era, 
$\Omega_{\rm GW}(f)\propto f^0$.
These two power laws actually apply to the $\ogw(f)$ predicted by $\Lambda$CDM, 
as well, as indicated by the green curve in the upper plot of, e.g., Fig. \ref{fig:GWspectra1}. 
There, $\ogw(f)$ exhibits a long plateau ($\propto f^0$) 
over a frequency range which covers the bands of most GW experiments at the present, 
e.g., aLIGO/Virgo and the LISA mission.
The amplitude of this plateau depends on the value of $r$ alone, independent of $f$ 
(see, e.g., \cite{2016PhRvX...6a1035L}). 
For $r=0.01$ shown here, this amplitude is $\sim10^{-16}$, 
more than six orders of magnitude below the sensitivities of current GW detectors, 
which is the main reason why the SGWB from inflation (in $\Lambda$CDM) 
was not expected to be detectable 
by current major GW detection experiments listed in \S\ref{sec:intro_inf}.

\begin{figure*}[t]
\begin{minipage}{\linewidth}
     \centering
     \hspace*{-0.5in}
     \includegraphics[scale=1.4]{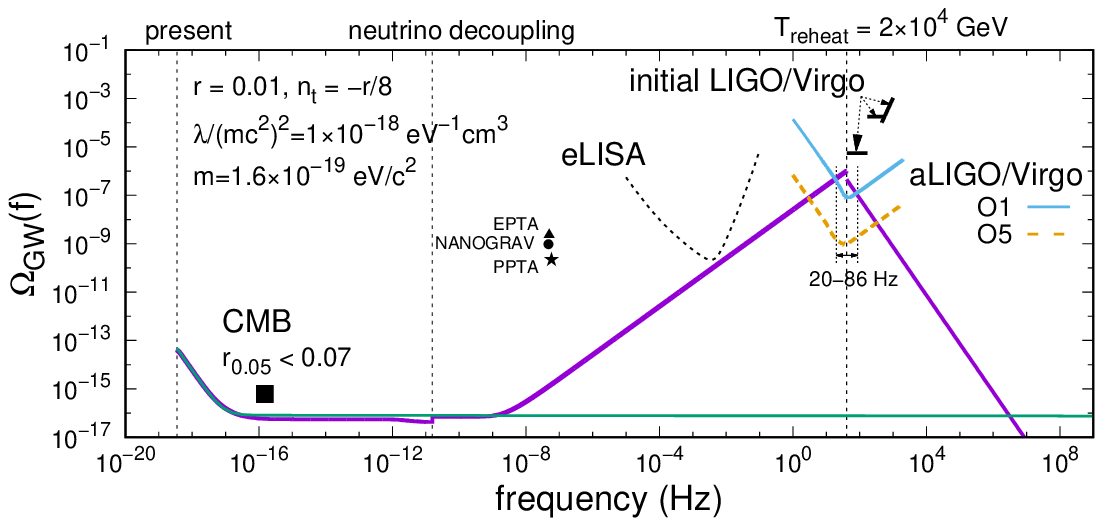}
      \includegraphics[scale=0.7]{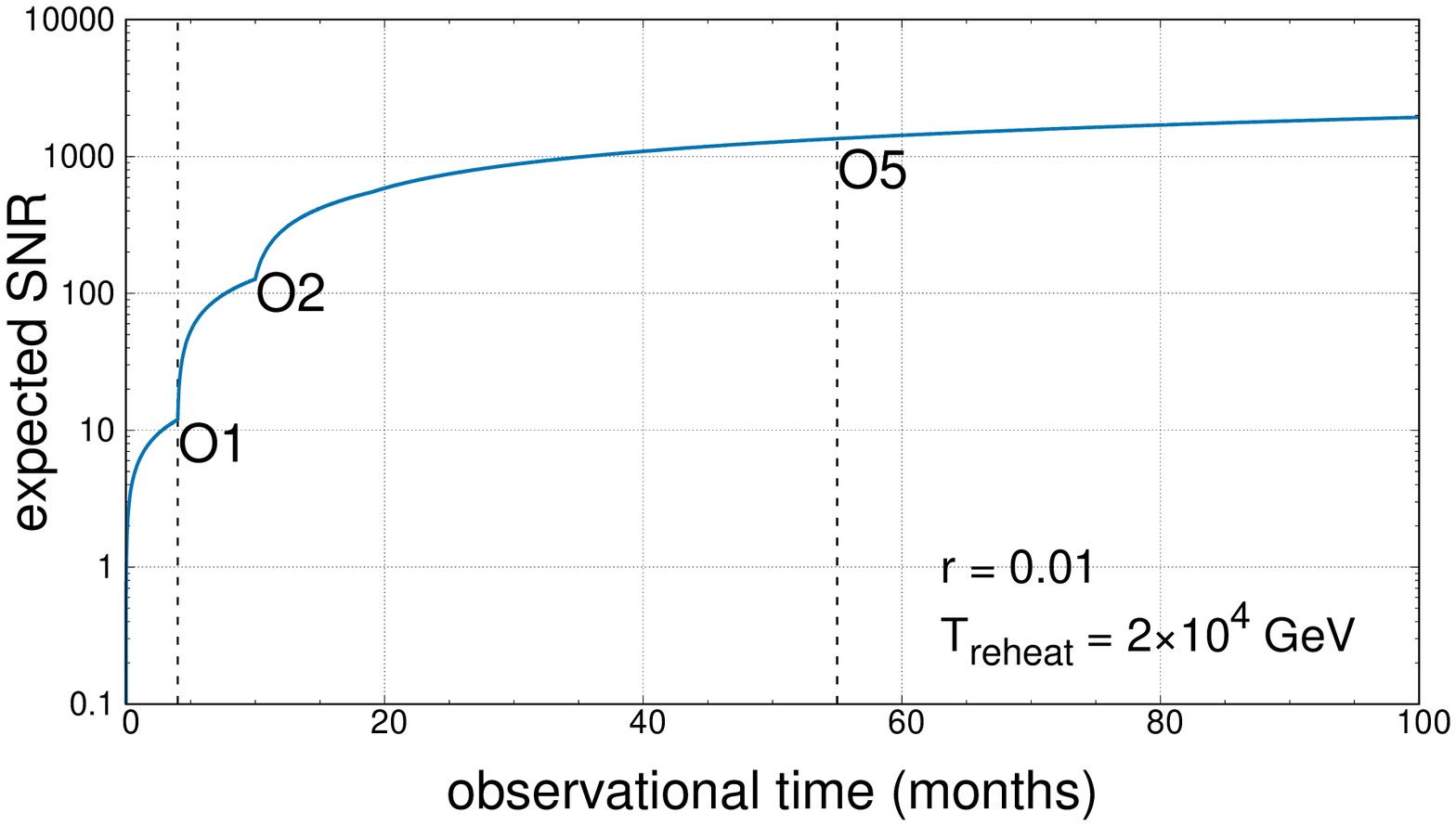}
    \end{minipage}
  \caption{
  Caption same as in Fig. \ref{fig:GWspectra1}, except for a $\Lambda$SFDM model 
  with $\tre=2\times10^4$ GeV.
  }
 \label{fig:GWspectra2}
\end{figure*}

\begin{figure*}[t]
\begin{minipage}{\linewidth}
     \centering
     \hspace*{-0.5in}
     \includegraphics[scale=1.4]{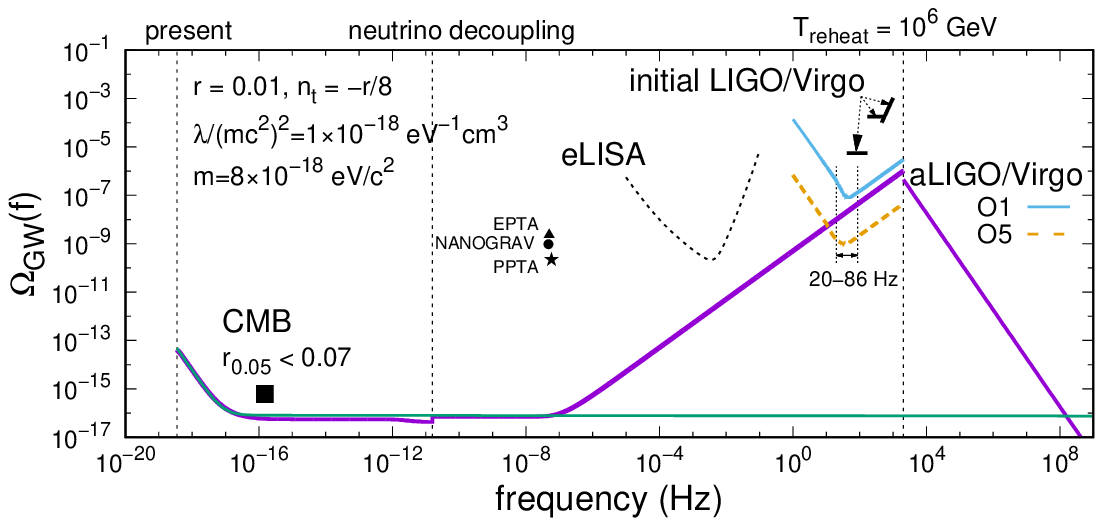}
      \includegraphics[scale=0.7]{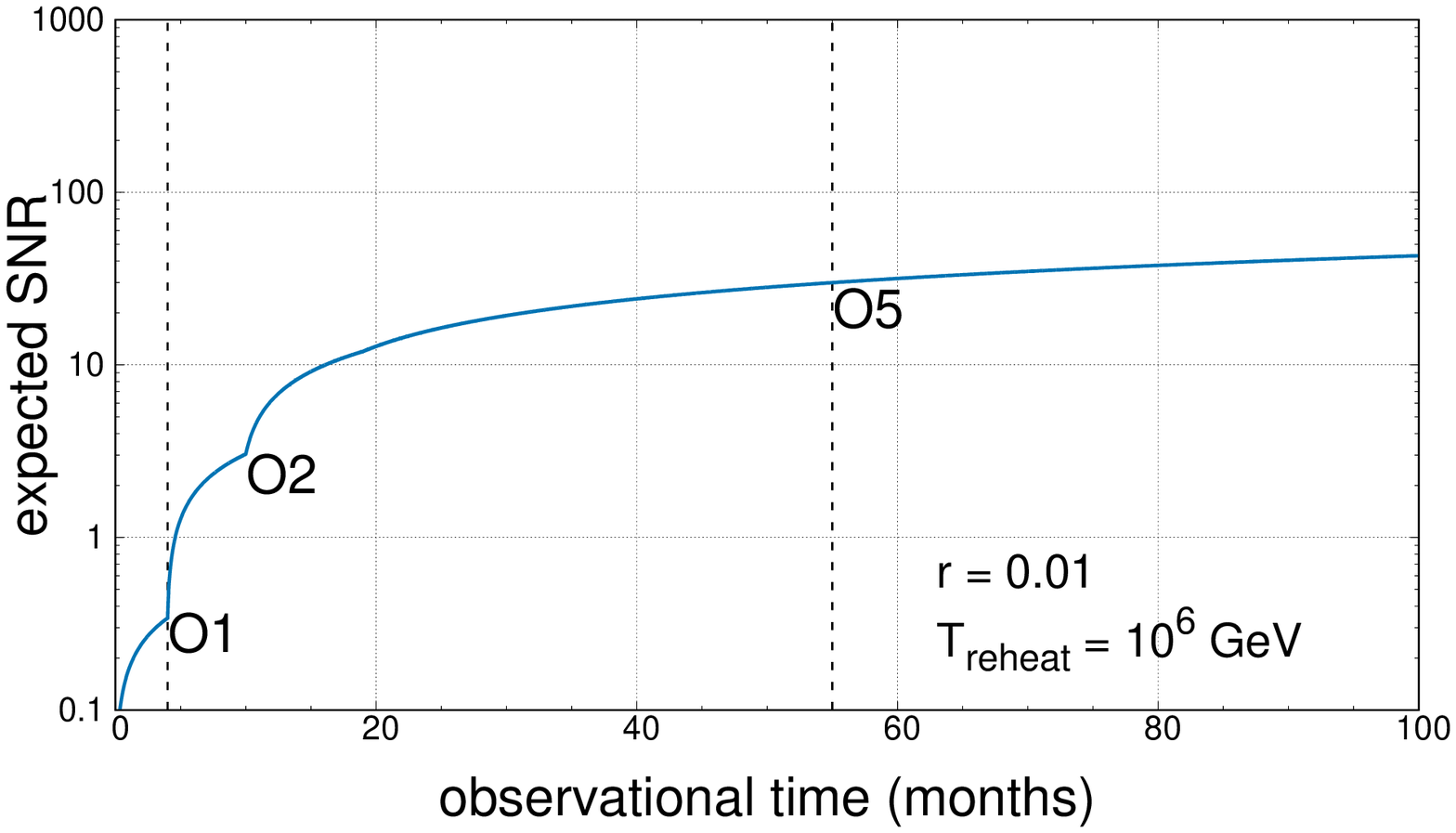}
    \end{minipage}
  \caption{
  Caption same as in Fig. \ref{fig:GWspectra1}, except for a $\Lambda$SFDM model 
  with $\tre=10^6$ GeV.
  }

 \label{fig:GWspectra3}
\end{figure*}

However, we will now show that the SGWB from inflation, 
predicted by the $\Lambda$SFDM model, has the potential to be detectable 
by current GW experiments like aLIGO/Virgo, 
i.e., can reach their detection sensitivities for a wide range of model parameters,
due to the amplification of the SGWB during the stiff-SFDM-dominated era.
Its present-day energy density spectra $\ogw(f)$ are indicated by purple curves 
in Figs. \ref{fig:GWspectra1} -- \ref{fig:GWspectra3}. 
These show that the SGWB spectrum for $\Lambda$SFDM departs dramatically
at high frequencies from that of standard $\Lambda$CDM:
$\ogw(f)\propto f^1$ for modes reentering the horizon during the stiff-SFDM-dominated era, 
while $\ogw(f)\propto f^{-2}$ for modes reentering even earlier, during the reheating era
(corresponding to even higher frequencies). 
Therefore, $\ogw(f)$ in $\Lambda$SFDM models, amplified by the stiff era, 
has a characteristic triangle-shaped feature at high frequencies, peaked at $f_{\rm reheat}$.
\footnote{
In Figs. \ref{fig:GWspectra1} -- \ref{fig:GWspectra3}, the discontinuity in $\ogw(f)$ 
at $f_{\rm reheat}$ is due to the fact that we assume an instantaneous change of 
the EOS at the end of reheating, from $w=0$ to $w=1$, in our $\Lambda$SFDM model. 
We will adopt a more realistic model for reheating in the future, in which $w$ changes smoothly. 
}
The baseline of this triangle sits on the plateau 
which corresponds to modes that reenter the horizon during RD,
and highly overlaps the long plateau for the $\Lambda$CDM model mentioned above, 
as shown in Figs. \ref{fig:GWspectra1} -- \ref{fig:GWspectra3}.

In Figs. \ref{fig:GWspectra1} -- \ref{fig:GWspectra3}, 
we display a comprehensive collection of previous constraints on 
the cosmological SGWB, from various types of observations. 
Specifically, ranging from lowest to highest frequencies, they are from
the BICEP2/Keck Array CMB polarization experiment \cite{2016PhRvL.116c1302B}, 
pulsar timing array (PTA) experiments (NANOGrav \cite{2016ApJ...821...13A},
PPTA \cite{2015Sci...349.1522S}, EPTA \cite{2015MNRAS.453.2576L}), 
and the (initial, pre-2015) LIGO experiment \cite{2014PhRvL.113w1101A,2015PhRvD..91b2003A}.
All three example models shown here satisfy all these constraints on the SGWB. 
In fact, these constraints are weaker than the ones from $z_{\rm eq}$ and BBN 
discussed in \S\ref{sec:constraint}. 
Therefore, we do not utilize them to constrain the SFDM particle parameters. 
However, the frontier laser interferometer experiments, aLIGO/Virgo
\cite{2016PhRvL.116m1102A,2017PhRvL.118l1101A}, in operation today,  
and LISA \cite{2012JCAP...06..027B} (currently in its Pathfinder stage) in 
the future, are capable of placing much better and more useful constraints 
on the inflationary SGWB, or quite possibly even  \emph{detecting} it.  In fact, as we shall
show below, the new data from the O1 run of aLIGO, recently reported in 
\cite{2017PhRvL.118l1101A}, should already be sensitive enough to detect 
the strongest possible signals predicted by our $\Lambda$SFDM model
for a limited range of model parameters. This will allow us here   
to place the first meaningful constraints on the model which are
based upon this new observable.
With regard to the future, the curve in Figs. \ref{fig:GWspectra1} -- \ref{fig:GWspectra3} 
labeled ``eLISA'' is the expected strain noise function of eLISA in \cite{2012CQGra..29l4016A} 
and may be revised in the final design of the upcoming LISA mission.
If $\ogw(f)$ is higher than (i.e., intersects) this noise sensitivity curve, 
it is possible for LISA to detect this SGWB.

In this paper, we concentrate mostly on the detectability of the SGWB from inflation 
by the ongoing aLIGO/Virgo experiment, whose O1 run has so far detected several 
GW signals from binary black hole merger events, as reported in \cite{2016PhRvL.116f1102A}. 
This same experiment can also detect a stochastic background 
(i.e. either of diffuse origin or from unresolved sources), but that
requires a different strategy for analyzing the data, by 
considering the correlation of the strains measured by different detectors.  
That is one of the major, additional goals of the experiment \cite{2016PhRvL.116m1102A},
in fact.  As mentioned above, the first results of analyzing the O1 data run 
to search for the SGWB  were presented just recently in \cite{2017PhRvL.118l1101A}.
Although the present-day SGWB, were it detected by aLIGO/Virgo, 
could have a variety of origins other than the primordial SGWB from inflation, 
such as unresolved black hole mergers \cite{2016PhRvL.116m1102A}, 
we will only consider the inflationary SGWB in this paper, 
which has a unique spectral shape in $\ogw(f)$ as predicted by $\Lambda$SFDM, 
and probe its detectability characterized by the SNR. 

As shown by \cite{1999PhRvD..59j2001A}, the SNR of any fixed SGWB today, 
for a certain laser interferometer experiment, 
is proportional to the square root of the accumulated observation time. 
Therefore, we can in principle enhance the detectability of a reasonably-motivated SGWB 
to a required level of significance, provided enough observation time.
The solid and dashed curves in the upper plots of Figs. \ref{fig:GWspectra1} -- \ref{fig:GWspectra3}, 
labeled by ``aLIGO/Virgo'', indicate the ``integrated 1$\sigma$ sensitivity curves'' for
detecting the inflationary SGWB predicted by $\Lambda$SFDM, 
for the two observing runs O1 (with data now analyzed) and O5 (theoretical forecast), respectively. 
The calculation of these curves is described in Appendix \ref{app:SNR_O1}, 
based upon a modification of the ``power-law integrated (PI) sensitivity curves'' 
developed by \cite{2013PhRvD..88l4032T} 
as a handy tool for visualizing the sensitivity of GW detectors for $\ogw(f)$ spectra which are
assumed to be pure power-laws (for which we are grateful to Joseph Romano for letting us 
modify his code).  The latter, for example, includes the case thought to describe
the background from unresolved binary black hole mergers (i.e. power-law index 2/3, up to a 
turn-over frequency).   The power-law assumption underlying the PI curves
is an approximation that reflects the fact
that the frequency band of greatest sensitivity of aLIGO/Virgo is narrow.  However, in our case,
the strong triangle feature of the spectrum is not amenable to approximation as a single power-law,
so we have used the actual non-power-law shape of the SGWB for our $\Lambda$SFDM model 
in producing the curves in Figs. \ref{fig:GWspectra1} -- \ref{fig:GWspectra3}, instead.
In our case, the way to interpret the integrated sensitivity curves is the following: 
if the predicted $\ogw(f)$ for the inflationary SGWB from a given set of $\Lambda$SFDM
model parameters touches the curve for the O1(O5) run at any $f$, 
this SGWB will be detected with 1$\sigma$ significance (SNR $=1$) by the O1(O5) run,
respectively.
\footnote{
We note that this interpretation of the integrated sensitivity curves differs from that used
to interpret the PI curves for pure power-law spectra.   For the latter spectra, the interpretation is
as follows:  for each point on the PI sensitivity curve, a spectrum which is tangent to the curve
at that point (for which the power-law index is given by the slope of the tangent to the curve at that
point) would be detected at 
1$\sigma$ significance (SNR $=1$), or a confidence level of $68\%$.  Hence, a single sensitivity curve
encodes the detectability for a range of spectral indices at once.  
}
The dashed curve (O5) is much lower than the solid curve (O1) (i.e. can detect a smaller $\ogw(f)$), 
which reflects the fact that the design sensitivity of O5 is higher and the integration time is longer
than those of O1. 
These curves are calculated by integrating $\ogw(f)$ 
over frequency, convolved with the LIGO strain sensitivity, 
which is concentrated in the $20-86$ Hz band (see footnote 13, page 33).

For the SGWB from standard inflation, which is enhanced in $\Lambda$SFDM, 
its predicted $\ogw(f)$ has a triangle-shaped feature as described above, 
by which $\ogw(f)$ can possibly reach the aLIGO/Virgo sensitivity curves, 
an impossible task for the corresponding inflationary SGWB in $\Lambda$CDM  
(see Figs. \ref{fig:GWspectra1} -- \ref{fig:GWspectra3} and \cite{2016PhRvX...6a1035L}). 
To see this in more detail, there are three cases for our predicted $\ogw(f)$, 
with regard to the position of the peak of the triangle in $\ogw(f)$, at $f_{\rm reheat}$, 
relative to the narrow frequency band of peak sensitivity of aLIGO/Virgo, $20-86$ Hz.
These cases can be expressed as:
\begin{enumerate}[leftmargin=5\parindent]
\item[Case 1.] \quad$f_{\rm reheat}< 20$ Hz,
\item[Case 2.] \quad20 Hz $<f_{\rm reheat}< 86$ Hz,
\item[Case 3.] \quad$f_{\rm reheat}> 86$ Hz.
\end{enumerate}

We choose the values of $\tre=10^3,~2\times10^4,~10^6$ GeV,  
in the three example models shown here, 
such that each of them fits one of the above three cases, respectively. 
Intuitively, one should expect that among these three models, for which
the peak amplitudes $\ogw(f_{\rm reheat})$ are all approximately equal,
the maximally detectable model, i.e., the one with the highest SNR for a given observation time, 
must be the one with $\tre=2\times10^4$ GeV that fits Case 2, 
where $f_{\rm reheat}$ lies inside the peak sensitivity band of LIGO.
Indeed, this is shown to be true by the SNR plots of Figs. \ref{fig:GWspectra1} -- \ref{fig:GWspectra3}, 
for the SNR from the SGWB predicted by $\Lambda$SFDM  
vs. the accumulated observation time of aLIGO/Virgo.
[Note: These plots can be compared to the right panel of Fig. 1 in \cite{2016PhRvL.116m1102A},
except that the latter are based upon assuming a spectrum appropriate for a model
of the background from unresolved binary black hole mergers and the SNR there is based upon the 
theoretical forecast for all observing runs O1 through O5, 
while in Figs. \ref{fig:GWspectra1} -- \ref{fig:GWspectra3} 
we use the noise characteristics of the actual O1 data, as we describe in Appendix \ref{app:SNR_O1}, 
and only use the theoretical forecast for O2 through O5.] 
From Fig. \ref{fig:GWspectra2} we see that if $\tre=2\times10^4$ GeV and $r=0.01$, 
the expected SNR should already have achieved a value greater than $10$, 
by the end of the recent O1 run, for this SFDM parameter choice
$(\lambda/(mc^2)^2,~m)$ = ($10^{-18}$ eV$^{-1}$cm$^3$, $1.6\times10^{-19}$ eV$/c^2$).
Consequently, a nondetection of the SGWB by aLIGO O1 would rule out this example case.
On the other hand, consider the case in Fig. \ref{fig:GWspectra3}, instead, 
where the values of $r$ and $\lambda/(mc^2)^2$ are the same 
but $\tre=10^6$ GeV and $m = 8\times10^{-18}$ eV$/c^2$.
While its expected SNR is less than $1$ for O1,  
even this case will reach an SNR $\sim30$ by the end of O5 in 2022.
Apparently, a wider range of SFDM parameters and reheat temperatures 
than that to which aLIGO O1 is sensitive will be accessible by the end of the
aLIGO/Virgo O5 run.  

{\bf
This shows that the $\Lambda$SFDM model promises to be detectable 
via its predicted SGWB from inflation, or else will be seriously constrained 
with regard to its particle parameters, over the course of the ongoing aLIGO/Virgo
experiment.
}
 We will quantify this in more detail below in  
\S\ref{sec:maxSNR}.  There, we shall go beyond the three example cases above,
by considering the expected SNR for a range of cases and include an analysis of the data 
accumulated in the O1 run to determine which of these cases are either consistent with the data
or else already excluded by it.

\subsection{Implications from $\Lambda$SFDM models marginally satisfying cosmological constraints}
\label{sec:maxSNR}

The total $\Omega_{\rm GW}(a=1)$ at the present, integrated over all frequencies, 
is equal to the total area underneath the spectrum curve $\ogw(f)$. 
As we confirm, this total area is dominated by the area of the triangle 
for a wide range of $\Lambda$SFDM model parameters, 
including all three example models shown here.
For these models, $r$ and $\lambda/(mc^2)^2$ are fixed 
as described at the beginning of \S\ref{sec:SGWBspectra}, 
but $\tre$ is different in each model, and for each $\tre$, 
the SFDM particle mass $m$ is adjusted to be the marginal value 
such that all the example models satisfy the current cosmological constraints 
derived in \S\ref{sec:constraint}. 
For other given values of $r$ and $\lambda/(mc^2)^2$, 
the marginally-allowed $\Lambda$SFDM models can be defined similarly, 
as follows.  For fixed values of $r$ and $\lambda/(mc^2)^2$, there is a
family of marginally-allowed cases for different values of $\tre$.   For
each value of $\tre$, the value of $m$ is adjusted to match the lower bound
of the vertical shaded region of allowed mass values in Fig. \ref{fig:psmultiple}.
This value of $m$ serves to maximum the predicted value of $\ogw(f)$ amongst
the allowed cases for which the other parameters are the same.
In this subsection, we will study the detectability of these marginal $\Lambda$SFDM models.

First, among all $\Lambda$SFDM models which satisfy the cosmological constraints, 
these marginal ones have the highest detectability. 
This can be shown by decreasing $m$ but fixing all other parameters in an allowed model, 
until $m$ reaches its lower bound. 
During this procedure, SFDM ends its stiff phase later, 
while the beginning of the stiff phase does not change for fixed $\tre$,
so the duration of the stiff-SFDM-dominated era becomes prolonged.
As a result, the SGWB experiences more boost, 
and $\ogw(f)$ has higher amplitudes in the triangle, 
i.e., for modes which reenter the horizon by the end of the stiff era. 
This leads to a larger SNR of $\ogw(f)$ measured by aLIGO/Virgo.
Therefore, the marginally-allowed $\Lambda$SFDM models,
which includes the three example models here, 
are motivated by the fact that they maximize the detectability of the predicted SGWB 
today and, hence, are the best starting point for comparison with the data.

In the upper plots of Figs. \ref{fig:GWspectra1} -- \ref{fig:GWspectra3} for these models, 
as $T_{\rm reheat}$ changes from $10^3$ GeV to $10^6$ GeV, 
the peak frequency of the spectrum at $f_{\rm reheat}$ shifts from low to high, 
passing through the $20-86$ Hz sensitive band of aLIGO/Virgo, 
whereas the peak amplitude $\ogw(f_{\rm reheat})$, 
and thus the area of the triangle in $\ogw(f)$, remains almost the same.
This is not surprising, because for these marginally-allowed models 
with the same values of $r$ and $\lambda/(mc^2)^2$, 
they must produce approximately the same amount of $\rho_{\rm GW}$ of the SGWB, 
at epochs which correspond to the cosmological constraints.
Therefore, the corresponding total $\ogw$ at the present is nearly the same for all marginal models, 
dominated by the area of the triangle in $\ogw(f)$, as mentioned above.
Interestingly, we find that $f_{\rm reheat}$ is nearly proportional to $T_{\rm reheat}$ 
among these marginally-allowed models. 
We provide an analytical explanation for this relation in Appendix \ref{app:fproptoT}.

\begin{figure*}[t]
     \centering
     \hspace*{-0.2in}
     \includegraphics[width=1.0\linewidth]{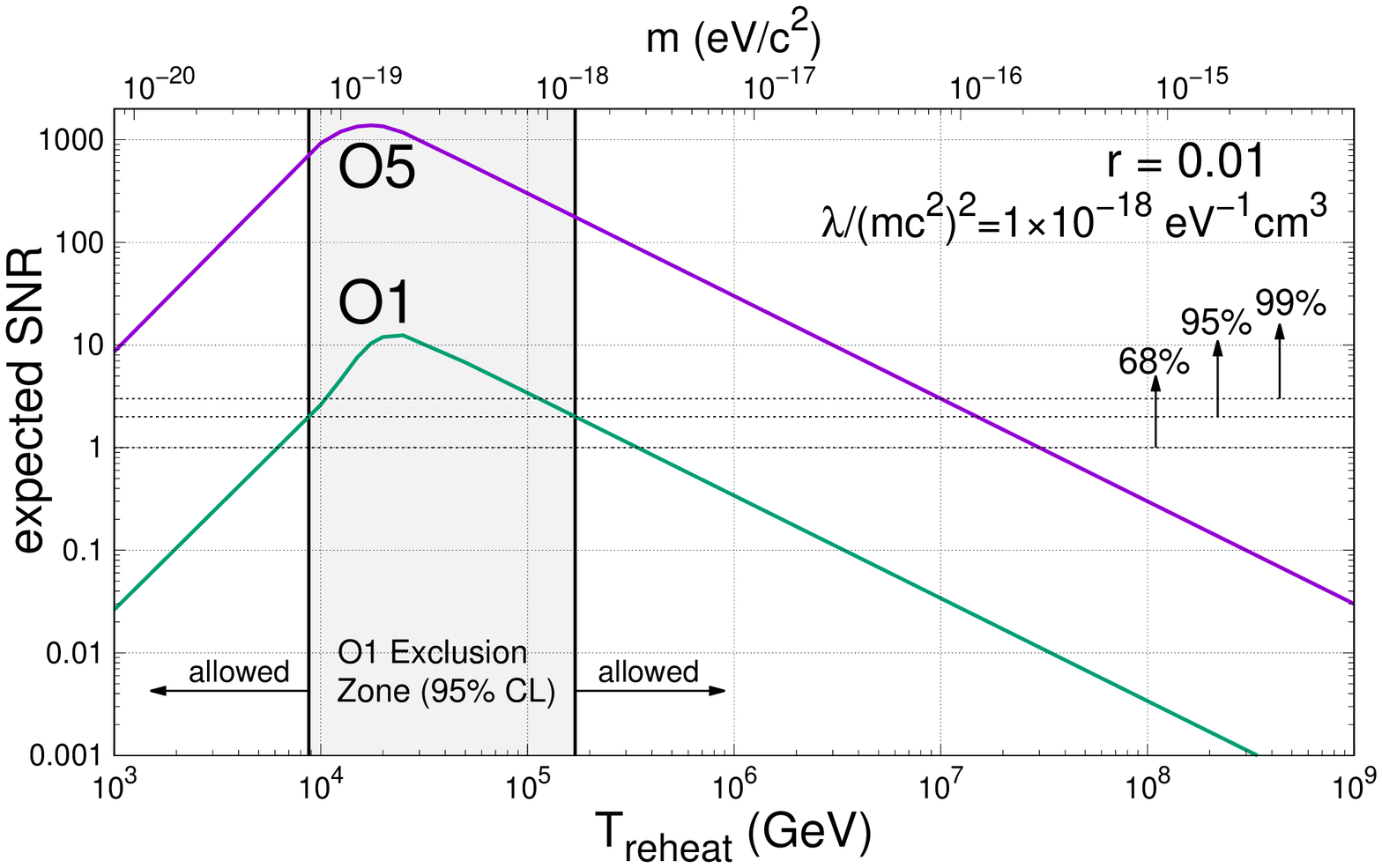}
  \caption{Dependence of the expected SNR on $\tre$, 
  for the inflationary SGWB measured by aLIGO/Virgo by the end of the O1 and O5 runs, respectively. 
  Both curves are predictions of $\Lambda$SFDM models which assume $r=0.01$ and 
  $\lambda/(mc^2)^2=1\times10^{-18}$ eV$^{-1}$cm$^3$, and, for each value of $\tre$, 
  the particle mass $m$ has the value which 
  marginally satisfies current cosmological constraints for that case. 
  The maximally detectable model corresponds 
  to a value of $\tre$ between $10^4$ and $10^5$ GeV.
  }
 \label{fig:SNRdist}
\end{figure*}

Although the total present-day $\Omega_{\rm GW}$ is almost constant for all marginal models, 
the detectability of their predicted SGWB, by aLIGO/Virgo, 
is highly distinguishable from one model to another 
(see lower plots of Figs. \ref{fig:GWspectra1} -- \ref{fig:GWspectra3}). 
As we discussed in \S\ref{sec:GWspectratoday}, 
this is apparently due to the narrowness of the LIGO sensitive frequency band 
and the strong dependence on $f_{\rm reheat}$ of the overlap 
between this band and the peak of the SGWB spectrum.
The maximally detectable case, with the largest expected SNR, 
is when $f_{\rm reheat}$ lies inside this window.
Therefore, among the marginally-allowed $\Lambda$SFDM models 
with fixed $r$ and $\lambda/(mc^2)^2$, 
we can maximize the predicted SGWB signal by tuning $\tre$ 
(and with it, the corresponding marginally-allowed value of $m$) 
so as to center $f_{\rm reheat}$ on the LIGO sensitive band.
Among the three example models, 
the most detectable is that with $\tre=2\times10^4$ GeV.
We can plot the dependence of the expected SNR on $\tre$
for a given family of marginally-allowed $\Lambda$SFDM models 
with constant $r$ and $\lambda/(mc^2)^2$, and locate the value of  $\tre$ which
corresponds to the maximally detectable model.  For illustrative purposes, we choose one set
of values, $r=0.01$ and $\lambda/(mc^2)^2=1\times10^{-18}$ eV$^{-1}$cm$^3$,
and plot this dependence of the SNR on $\tre$ in
Fig. \ref{fig:SNRdist}.

As seen in Fig. \ref{fig:SNRdist}, for both the aLIGO/Virgo O1 and O5 runs, 
the expected SNR has a peak between $\tre=10^4$ GeV and $10^5$ GeV, 
which corresponds to the most sensitive (lowest strain noise) frequency range of the experiment.
For $r=0.01$, which is still 7 times below the current upper bound from CMB polarization measurements,
the maximally detectable $\Lambda$SFDM model has an expected SNR $>10$ 
even for the completed O1 run.  After 2 years of the future O5 run, the same model can 
reach an SNR greater than 1000.    These SNRs would increase or decrease if we assumed 
values of $r$ which are larger or smaller than this, respectively.  This establishes, however,
that the $\Lambda$SFDM model is capable of producing a detectable signal for the SGWB
for some range of model parameters.  

Now that the O1 run is finished, we can use the data to compare with these predictions.
As of yet, no detection of the SGWB has been reported for the O1 run \cite{2017PhRvL.118l1101A}.
The significance of a null detection depends upon the assumed spectrum of $\ogw(f)$, so the
confidence level (C.L.) of this null detection is model-dependent.   
\cite{2017PhRvL.118l1101A} has analyzed the data for power-law spectra, 
$\ogw(f) = \Omega_\beta(f/f_{\rm ref})^\beta$, 
where $f_{\rm ref} = 25$ Hz, and reported 95\% C.L. upper limits for $\Omega_\beta$ as
a function of $\beta$.  In their Table 1, for example, they report this upper limit for   
$\beta = 0$ as $\Omega_\beta \leq 1.7\times 10^{-7} $.  
The analysis in \cite{2017PhRvL.118l1101A}
did not extend to models like ours in which the spectrum is not a pure power-law, 
but is rather a broken power-law which defines the triangle feature we have described above,
which peaks at a frequency determined by the value of $\tre$.   However, as shown in
Fig. \ref{fig:SNRdist}, when the value of $\tre$ is chosen so that $f_{\rm reheat}$ is 
outside the range of the LIGO sensitive band (i.e. Case 1 when $f_{\rm reheat} < 20$ Hz,
Case 3 when $f_{\rm reheat} > 86$ Hz), the SNR is roughly the same as it would
be for the power-law spectrum with the same slope as the corresponding side of the
triangle feature (i.e. $\beta = -2$ or $1$ for Cases 1 and 3, respectively).   These
spectral indices for Cases 1 and 3 are within the range of the spectral indices
for which the analysis in \cite{2017PhRvL.118l1101A} reports a null detection.
This indicates that if the data were analyzed for our predicted spectrum in Cases 1
and 3, a null detection would also be reported.  

This fact allows us to place a meaningful constraint on a range of $\Lambda$SFDM model
parameters.  For the illustrative cases shown in Fig. \ref{fig:SNRdist}, for example, 
the signal predicted by those marginally-allowed model parameters can be so strong
that some of those model parameters can already be excluded at some C.L. by the null
detection reported in \cite{2017PhRvL.118l1101A}.  As shown
by the horizontal lines drawn across the SNR plot in  Fig. \ref{fig:SNRdist} for
the O1 run, each corresponding to a different constant value of SNR, 
there are two points on each horizontal line which intersect the curve of SNR vs. $\tre$, 
one on each side of the peak SNR.   For all values of  $\tre$ between these two points,
the SNR is larger than it is at the two points.   A null detection that implies a
95\% C.L. upper limit (SNR $=2$) means that the $\Lambda$SFDM model parameters
for the marginally-allowed cases with those values of $\tre$ and all the points in between 
are inconsistent with the data at the 95\% C.L.   For the illustrative cases plotted in 
 Fig. \ref{fig:SNRdist}, the marginally-allowed $\Lambda$SFDM models 
for which $r=0.01$ and $\tre$ ranges between about $8.75\times10^3$ and $1.7\times10^5$ GeV 
(for which the corresponding masses
are in the range between $7\times10^{-20}$ and $1.36\times10^{-18}$ eV$/c^2$), 
are excluded at the 95\% C.L., based upon the O1 data. 
{\bf
This provides a new kind of cosmological constraint on the $\Lambda$SFDM model.
}

On the other hand, as aLIGO/Virgo improves its sensitivity and accumulates more data 
over time from O1 to O5, 
the expected SNR of the SGWB predicted by any given $\Lambda$SFDM model 
will also increase.
As a result, by the end of the final observing run O5, 
$\Lambda$SFDM models with a wider range of parameters, 
which were not yet detectable by the O1 run, will become accessible.
In particular, the marginally-allowed models with 
$r=0.01$ and $\lambda/(mc^2)^2=1\times10^{-18}$ eV$^{-1}$cm$^3$ 
predict an inflationary SGWB signal with an SNR $>3$, by the end of O5, 
for those in which $\tre$ lies approximately inside $(6\times10^2\rm{~GeV},~10^7\rm{~GeV})$, 
and the corresponding SFDM mass range is about
$(5\times10^{-21}{\rm~eV}/c^2,~8\times10^{-17}{\rm~eV}/c^2)$, as shown in Fig. \ref{fig:SNRdist}.
Table \ref{table:tab2} summarizes the results for the ranges of $\tre$ and the corresponding $m$ 
for these marginally-allowed $\Lambda$SFDM models which are detectable by 
aLIGO/Virgo at $2\sigma$ or $3\sigma$ confidence levels 
by the end of its O1 and O5 runs, respectively.
{\bf
These results demonstrate that, in the future, the $\Lambda$SFDM model has the great 
potential of having its signature imprint on the primordial SGWB from inflation
detected by the Advanced LIGO/Virgo experiment.
} 

\begin{table*}[t]
\vspace{0.1in}
{\renewcommand{\arraystretch}{1.5}
\renewcommand{\tabcolsep}{.1cm}
\hspace{-0.3in}\vspace{0.1in}
\begin{tabular}[t]{cc|cc|cc}
\hline
LIGO run & Epoch & $\tre/\rm{GeV~(SNR}>2)$ & $m/(\rm{eV}/c^2)~(\rm{SNR}>2)$ & $\tre/\rm{GeV~(SNR}>3)$ & $m/(\rm{eV}/c^2)~(\rm{SNR}>3)$\\
\hline
O1 & 2015-2016 & $(8.75\times10^3,~1.7\times10^5)$ & $(7\times10^{-20},~1.36\times10^{-18})$ & $(1.05\times10^4,~1.125\times10^5)$ & $(8.4\times10^{-20},~9\times10^{-19})$\\
\hline
O5 &  2020-2022 & $(5\times10^2,~1.5\times10^7)$ & $(4\times10^{-21},~10^{-16})$ & $(6\times10^2,~10^7)$ & $(5\times10^{-21},~8\times10^{-17})$\\
\hline
\end{tabular}
}\caption{LIGO-detectable parameter ranges of $\tre$ and $m$ 
for $\Lambda$SFDM models with $r=0.01$ and 
$\lambda/(mc^2)^2=1\times10^{-18}$ eV$^{-1}$cm$^3$
that marginally satisfy the cosmological constraints,
by the end of the O1 and O5 observing runs of aLIGO/Virgo, respectively.
The detectable ranges for this illustrative family of models 
correspond to $2\sigma$ and $3\sigma$ detections, respectively. 
We note that the O1 run is now completed with a null detection, so the ranges for O1 can be
interpreted as excluded at 95\% and 99\% confidence, respectively.
} 
\label{table:tab2}
\end{table*}
In conclusion, the $\Lambda$SFDM model shows a great prospect 
of detectability by the Advanced LIGO/Virgo experiment, 
thanks to its unique prediction of the present-day energy density spectrum 
of the primordial SGWB from inflation.

\section{Discussion}
\label{sec:discussion}

\subsection{What happens to $\Lambda$SFDM if $N_{\rm eff,BBN}\approx N_{\rm eff,standard}$?}\label{sec:lowerneff}
\vspace{-0.1in}
In \S\ref{sec:constraint}, we apply the cosmological observables, 
$z_{\rm eq}$ and $N_{\rm eff,BBN}$, to constrain the SFDM particle parameters, 
through constraining the background expansion history of the $\Lambda$SFDM universe.
These constraints result in the allowed range of the parameters $(\lambda/(mc^2)^2, m)$, 
expressed as shaded region in the two-dimensional parameter space, 
as a function of the tensor-to-scalar ratio $r$ and the reheat temperature $\tre$,
(see Figs. \ref{fig:pssingle} and \ref{fig:psmultiple}).
We adopt conservative thresholds, the $1\sigma$ confidence intervals from measurements,
 for both the $z_{\rm eq}$ and $N_{\rm eff,BBN}$ constraints. 
 These thresholds lead to the shapes of the allowed regions as thin stripes, for all cases.
 In particular, since the $-1\sigma$ confidence limit of the measured value of $N_{\rm eff,BBN}$ 
 in Eq. (\ref{equation:neff}) is greater than the standard value, $N_{\rm eff,standard}=3.046$,
 all of the allowed $\Lambda$SFDM models can explain a higher value of $N_{\rm eff}$ 
 at BBN than at recombination, as mildly suggested by current measurements, 
 mentioned in \S\ref{sec:constraintBBN}.

However, if we adopt a more relaxed threshold, e.g., the $2\sigma$ confidence interval, 
particularly for the $N_{\rm eff,BBN}$ constraint, 
we will allow a much broader range of $\Lambda$SFDM models 
which satisfy these cosmological constraints.
In fact, the $2\sigma$ confidence interval of the measured value of $N_{\rm eff,BBN}$ 
contains the standard value $N_{\rm eff,standard}$ (see \cite{2015PhRvD..91h3505N}).
Therefore, there would be then no lower bound from the BBN constraint 
on the value of $\Delta N_{\rm eff,BBN}$ predicted by the $\Lambda$SFDM model. 
Only an upper bound on $\Delta N_{\rm eff,BBN}$ would be left, 
translated to a lower bound on $m$ for any allowed value of $\lambda/(mc^2)^2$.
As a result, the allowed ranges of $(\lambda/(mc^2)^2, m)$ 
illustrated in Figs. \ref{fig:pssingle} and \ref{fig:psmultiple} would amount to the whole ``quadrants'' 
above the solid and dashed-dotted curves (for the $z_{\rm eq}$ constraint), 
free from the dashed curves. The quadrant regions, 
as opposed to the stripe-shaped shaded regions in Figs. \ref{fig:pssingle} and \ref{fig:psmultiple}, 
allow the $\lambda\to0$ limit, in which SFDM is (nearly) non-self-interacting.
This implies that while the non-self-interacting SFDM model is mildly disfavored 
by the $1\sigma$ confidence interval from current measurements of $N_{\rm eff,BBN}$,
it is consistent with the $2\sigma$ limits.

Furthermore, should the measured value of $N_{\rm eff,BBN}$ decrease in the future 
to the extent of strongly favoring $N_{\rm eff,standard}$, 
the allowed ranges of SFDM particle parameters can be adjusted accordingly. 
In that case, the allowed regions in the parameter space would be like the quadrants described above.
\vspace{-0.3in}


\subsection{SGWB from inflation versus that from unresolved binary black hole mergers?} 
\begin{figure*}[t]
     \centering
     \hspace*{-0.3in}
     \includegraphics[width=1.05\linewidth]{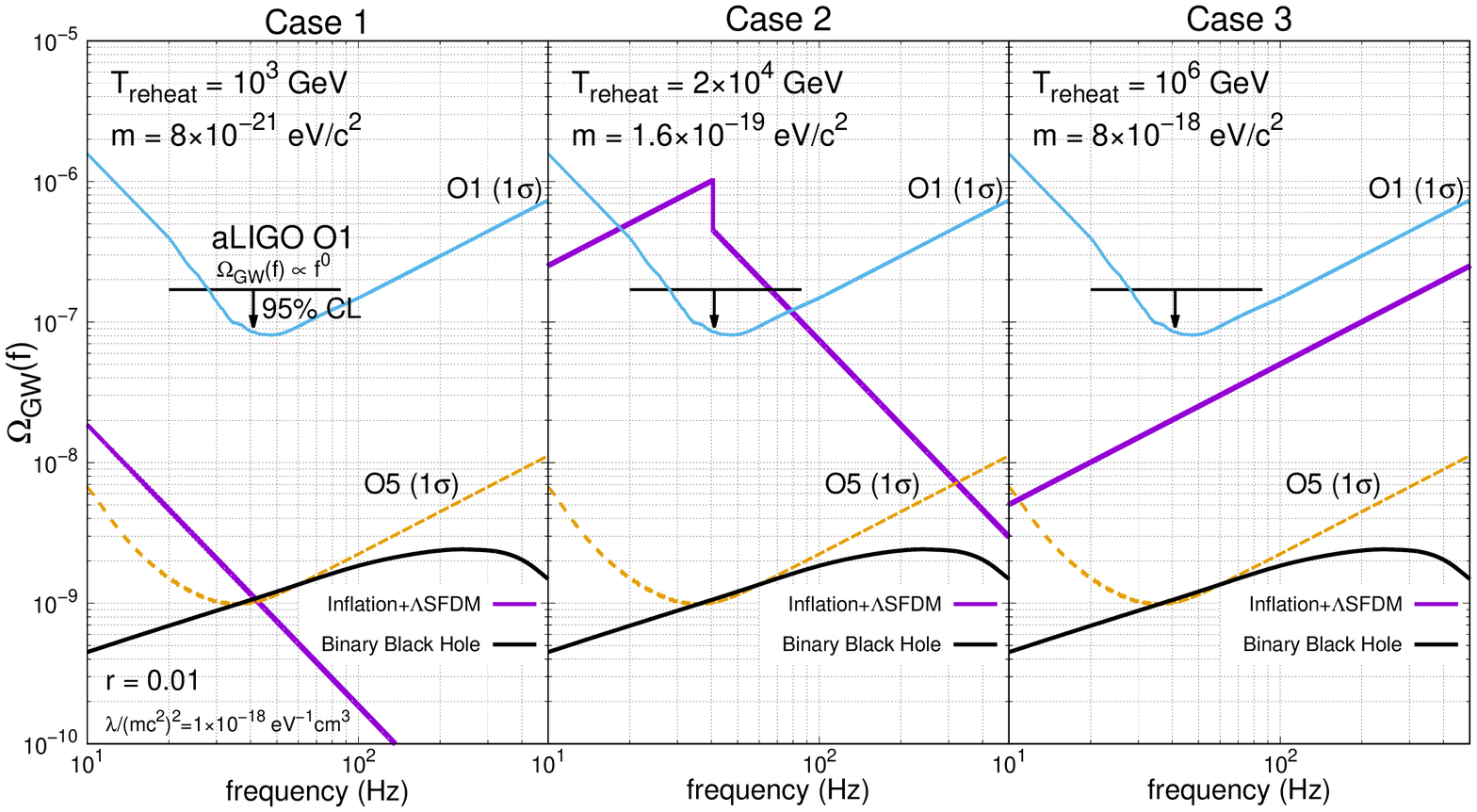}
  \caption{Present-day energy density spectra of the SGWB from inflation (purple curves) 
  for the same illustrative $\Lambda$SFDM models 
  as in Figs. \ref{fig:GWspectra1} -- \ref{fig:GWspectra3} 
  but zoomed in around the LIGO sensitive frequency band ($20-86$ Hz), 
  compared with the predicted energy spectrum of the SGWB
  from unresolved binary black hole merger events (black curves), 
  for which the merger rate is constrained by data from the completed aLIGO O1 run 
  \cite{2017PhRvL.118l1101A}.
  All the three $\Lambda$SFDM cases marginally satisfy the cosmological constraints 
  described in \S\ref{sec:constraint}, 
  for which $r=0.01$ and $\lambda/(mc^2)^2=1\times10^{-18}$ eV$^{-1}$cm$^3$, 
  but the values of $\tre$ and $m$ vary. 
  The new 95\% C.L. upper limit for an SGWB with a flat power spectrum, 
  from the recent O1 run data \cite{2017PhRvL.118l1101A}, is also shown here.
  }
 \label{fig:GWzoomin}
\end{figure*}

\vspace{-0.1in}
Since LIGO has a narrow sensitive frequency band ($20-86$ Hz), 
for any potential SGWB signal,
it is conventional to assume a power law for its present-day energy density spectrum, 
$\Omega_{\rm GW}(f)$, inside this band,
and convolve this power-law spectrum with observational data 
to test this potential signal or to put an upper bound on it.
This assumption is applicable to the SGWB from unresolved binary black hole mergers, 
since theoretical modeling suggests a power-law spectrum for such a signal, 
$\Omega_{\rm GW}(f)\propto f^{2/3}$, within the LIGO band 
\cite{2016PhRvL.116m1102A, 2017PhRvL.118l1101A}.
However, the $\Omega_{\rm GW}(f)$ of the inflationary SGWB predicted by the $\Lambda$SFDM model 
has a unique triangle-like spectral shape as described in \S\ref{sec:SGWBspectra}, 
for which the power-law based detection analysis may be invalid.
In addition, by tuning the model parameters, 
the inflationary SGWB in $\Lambda$SFDM can achieve an amplitude within the LIGO band
which is comparable to or much greater than that from known astrophysical sources.
In Fig. \ref{fig:GWzoomin},  for example, 
we compare the current predictions for the SGWB from unresolved binary 
black hole mergers with the inflationary SGWB predictions of the $\Lambda$SFDM model
for the three illustrative, marginally-allowed cases in Figs. \ref{fig:GWspectra1} -- \ref{fig:GWspectra3}.
The SNR of the SGWB from unresolved binary black hole mergers 
is currently predicted to be less than 10 at 90\% C.L. by the end of O5 (in 2022) 
\cite{2016PhRvL.116m1102A}, 
while in $\Lambda$SFDM, if, for example, we assume values of $r = 0.01$ and 
$\lambda/(mc^2)^2 = 1\times 10^{-18}$ eV$^{-1}$cm$^{3}$,
the expected SNR of the inflationary SGWB for the family of marginally-allowed cases
ranges from $\sim 3$ to $> 1000$ by then, for $6\times10^2<\tre~(\rm{GeV})<10^7$ 
(see Fig. \ref{fig:SNRdist} and Table \ref{table:tab2}).
Therefore, it will be important for aLIGO/Virgo and future GW detection experiments
to consider the SGWB from inflation predicted by the $\Lambda$SFDM model
and develop a means of distinguishing this potential SGWB signal 
from that sourced by binary black hole mergers, e.g., via their different spectral shapes.
For that reason, it will be interesting to consider the possibility of
simultaneous detection of the SGWB in different frequency bands, as should be
possible in the future with, for example, the LISA space-based mission, as
we shall discuss below.

\vspace{-0.2in}
\subsection{Future detectability of the SGWB from inflation in $\Lambda$SFDM with LISA?}
\vspace{-0.1in}
We have briefly mentioned the prospective constraints on the present-day SGWB
from the space laser interferometer mission LISA, currently in its Pathfinder stage, 
in \S\ref{sec:GWspectratoday}.
LISA can potentially probe SGWB signals 
from various cosmological and astrophysical sources, in the milli-Hertz frequency range.
According to our examples in Figs. \ref{fig:GWspectra1} -- \ref{fig:GWspectra3}, 
the SGWB from inflation in the $\Lambda$SFDM model is predicted to lie above
the LISA design sensitivity curve for Case 1 values of $\tre$.
Therefore, the synergy between LISA and LIGO will prospectively provide crucial information 
about the spectral shape of $\Omega_{\rm GW}(f)$ 
over frequencies which span the bands of these two experiments, 
and thus the strongest constraint on the inflationary SGWB predicted by $\Lambda$SFDM, 
in terms of its triangle-shaped $\Omega_{\rm GW}(f)$ at high frequencies. 
In other words, if this signal is detected by both experiments 
and consistent with a $\Lambda$SFDM model, 
it will be a ``smoking-gun'' for SFDM and will determine its particle parameters 
as well as $T_{\rm reheat}$ to a good accuracy. 
On the contrary, if both experiments suggest null detection of any SGWB signal, 
it would place stringent constraints on the SFDM particle parameters and $\tre$, 
in the context of the $\Lambda$SFDM model.
\vspace{-0.1in}

\section{Summary and conclusion}
\label{sec:conclusion}
\vspace{-0.1in}

We extended our analysis in Paper I of the cosmological evolution of a universe 
in which dark matter is comprised of ultralight self-interacting bosonic particles 
which form a Bose-Einstein condensate, described by a classical complex scalar field (SFDM)
with a global $U(1)$ symmetry. In this case, the comoving particle density, 
or charge density, conserved after particle production during reheating, 
is large enough to account for all the dark matter, 
a form of asymmetric dark matter. Here we connect the evolution of the SFDM to its origin
in the context of standard inflation, including the tensor modes 
and their associated stochastic gravitational-wave background (SGWB), 
and self-consistently account for the effect 
of the evolution of the background universe and the SGWB on each other.

Unlike standard CDM, which is always non-relativistic 
once it decouples from the thermal bath, 
SFDM has an evolving equation of state (EOS). 
As we had shown previously, there are four eras 
in the evolution of a homogeneous $\Lambda$SFDM universe:
the familiar radiation-dominated, matter-dominated and
$\Lambda$-dominated eras common to standard $\Lambda$CDM as well, 
but also an earlier era dominated by SFDM with a stiff equation of state.
In this paper, we embedded this model self-consistently into the standard inflationary paradigm 
by postulating that inflation is followed by an extended epoch of reheating (with matter-like EOS), 
from which SFDM emerges, as well as the particles of the Standard Model. 
We assumed that most of the energy density of the inflaton field 
goes into the creation of the DM bosons, 
which quickly condense into their ground state, thereafter giving rise to SFDM in its stiff phase. 
The subdominant energy density of standard model particles constitutes the radiation component. 
We adopted an instant transition, where the end of reheating at $T = T_{\rm{reheat}}$
is followed by the stiff-SFDM-dominated era of $\Lambda$SFDM. 

Standard inflation predicts a stochastic background of gravitational waves, 
mainly encoded in a finite value of the tensor-to-scalar ratio $r$, 
which is related to the energy scale of inflation. 
We have shown that this SGWB is amplified during the stiff-SFDM-dominated epoch 
compared with what it would be in a $\Lambda$CDM universe. 
SFDM in its relativistic phases (first stiff, then radiation-like EOS) 
and this amplified SGWB from inflation both add to 
the number of relativistic degrees of freedom, $N_{\rm{eff}}$, 
in the early universe before and around Big Bang nucleosynthesis (BBN), 
and possibly up to the time of matter-radiation equality at $z_{\rm{eq}}$. 
It is necessary to ensure that the stiff-SFDM-dominated era ends no later than when BBN begins. 
Moreover, since the combined energy density of SFDM plus the amplified SGWB 
must preserve $z_{\rm{eq}}$ from CMB measurements, 
SFDM should be nonrelativistic by the time of $z_{\rm{eq}}$. 
The constraints derived in Paper I on the SFDM particle parameters, 
boson mass $m$ and two-particle self-interaction $\lambda > 0$,
required to make the SFDM compatible with these observables had to be modified here
to account for the presence of the SGWB.
Since many cosmological observables are dependent upon the ratio $\lambda/(mc^2)^2$, 
rather than $\lambda$, we express our results for the constraints on SFDM
in terms of the parameter pair ($\lambda/(mc^2)^2,~m$). 

We considered two values for $r$, $r=0.01$ and $r=0.1$, focusing on the former in particular, 
which is still seven times below current upper bounds from CMB measurements. 
We chose several values of the reheat temperatures, 
spanning a wide range from $1$ to $10^9$ GeV, 
to probe the range of impacts of SFDM on the inflationary SGWB. 
To this end, we solved the fully-coupled Klein-Gordon and Einstein field equations 
for the time-dependence of different $\Lambda$SFDM models, 
self-consistently accounting for their amplification of the SGWB from inflation. 
We studied the back-reaction of the energy density of the enhanced SGWB 
on the expansion history of the universe, 
which in turn affected the SGWB, requiring us to develop an elaborate numerical methodology. 
We incorporated important additional effects, 
like the effect of electron-positron annihilation on the thermal history, 
as well as the damping of tensor modes due to the free streaming of neutrinos.

The amplification of the SGWB from inflation in $\Lambda$SFDM 
makes possible the prospective detection of the latter, 
using current and upcoming gravitational wave observatories. 
In fact, we calculated the present-day gravitational wave energy spectra, 
$\ogw(f)$, and found that detection of the SGWB at high frequencies 
is within reach of the Advanced LIGO/Virgo experiment and possibly LISA in the future. 
We have shown that, for SFDM particle parameters that satisfy the above cosmological constraints, 
the amplified SGWB is currently detectable by aLIGO for a broad range of reheat temperatures, 
for values of the tensor-to-scalar ratio currently allowed by CMB polarization measurements. 
Using the actual noise characteristics of the aLIGO O1 run 
(kindly provided us by the LIGO Scientific Collaboration), 
we determined the expected SNR 
for the inflationary SGWB in $\Lambda$SFDM for a range of model parameters.
The null detection of the SGWB by the aLIGO O1 run, 
recently reported by \cite{2017PhRvL.118l1101A}, has already provided 
a new kind of cosmological constraint on SFDM as illustrated by the case in 
Fig. \ref{fig:GWspectra3} and the middle panel of Fig. \ref{fig:GWzoomin}, 
with an excluded range of cases shown in Fig. \ref{fig:SNRdist}. 
A wider range of SFDM parameters and reheat temperatures 
will be accessible to the aLIGO O5 run, 
potentially detecting this unique signature of the SFDM model.

\vspace{0.1in}
$\bullet$ {\bf Cosmological constraints on SFDM particle parameters}
\vspace{0.05in}

In \S\ref{sec:constraint}, we described in detail 
how observational constraints on $N_{\rm eff,BBN}$ and $z_{\rm eq}$ constrain 
the allowed range of SFDM particle parameters $(\lambda/(mc^2)^2,~m)$ 
for given values of $r$ and $\tre$. Details aside, a rough summary of those results 
can be described as follows. For $\lambda/(mc^2)^2$, we found
\begin{equation}\label{eq:lambdam2rough}
     10^{-18}~\rm{eV}^{-1}~\rm{cm}^3\lesssim\frac{\lambda}{(mc^2)^2} \lesssim 4 \times 10^{-17}~\rm{eV}^{-1}~\rm{cm}^3,
\end{equation} 
For $m$, we found 
\begin{equation}
m\gtrsim5\times10^{-21}\times\f{\tre}{10^3~{\rm GeV}}\sqrt{\f{r}{0.01}}{~\rm eV}/c^2,
\end{equation}
for $r\gtrsim0.01$ and $\tre\gtrsim10^3$ GeV, and $m\gtrsim5\times10^{-21}{~\rm eV}/c^2$ 
for $r\gtrsim0.01$ and $\tre<10^3$ GeV. 
As discussed in \S\ref{sec:lowerneff}, 
if we relax the $N_{\rm eff,BBN}$ constraint, such that $\Delta N_{\rm eff,BBN}=0$ is allowed, 
then the lower limit in Eq. (\ref{eq:lambdam2rough}) goes away, 
and even $\lambda\to0$ is allowed.

\vspace{0.1in}
$\bullet$ {\bf Detectability of the SGWB from inflation in $\Lambda$SFDM}
\vspace{0.05in}

As described in \S\ref{sec:SGWBspectra}, 
the detectability of the SGWB $\ogw(f)$ amplified in $\Lambda$SFDM 
depends upon the SFDM particle parameters, $\tre$, and $r$.
For fixed $r$ and $\tre$, the maximum predicted signal corresponds to 
the pairs of $(\lambda/(mc^2)^2,~m)$ which marginally satisfy the cosmological constraints 
and maximize the duration of the stiff era.
For each of the allowed value of $\lambda/(mc^2)^2$, 
the minimum allowed value of $m$ maximizes this duration.
For LIGO, the overall maximum predicted signal (at fixed $r$) 
corresponds to this maximum when $\tre$ is chosen 
so that $f_{\rm reheat}$ lies inside the LIGO sensitive frequency band. 
For $r=0.01$, this corresponds to $\tre\simeq2\times10^4$ GeV, 
for which we predict an SNR $\sim10$ for the recent aLIGO/Virgo O1 run.
The null detection in the O1 data recently reported \cite{2017PhRvL.118l1101A},
therefore, excludes this particular maximally-detectable case.

In the future, we will be able to compare the $\Lambda$SFDM model predictions 
to this O1 data for the \emph{full} range of model parameters allowed by 
the cosmological constraints described in \S\ref{sec:constraint} 
to determine what subset of these allowed parameters also satisfy 
this \emph{new} cosmological constraint from direct measurement of the SGWB today.
While that is beyond the scope of the present paper, we have, however, 
made such a determination for a representative family of marginally-allowed cases, 
for $r=0.01$ and $\lambda/(mc^2)^2=10^{-18}~\rm{eV}^{-1}~\rm{cm}^3$, as follows.
Null detection by the O1 run now excludes at 95\% confidence. the range 
$8.75\times10^3\lesssim\tre~(\rm{GeV})\lesssim 1.7\times10^5$, for which the corresponding masses
are in the range $7\times10^{-20}\lesssim m~(\rm{eV}/c^2)\lesssim1.36\times10^{-18}$.

A wider range of $\Lambda$SFDM model parameters will be accessible to aLIGO/Virgo as time goes on.
For $r=0.01$ and $\lambda/(mc^2)^2=10^{-18}~\rm{eV}^{-1}~\rm{cm}^3$, for example, 
a $3\sigma$ detection of the inflationary SGWB is predicted for the O5 run 
if $6\times10^2\lesssim\tre~(\rm{GeV})\lesssim10^7$.
For these $\tre$ ranges, the ranges of particle masses in the marginally-allowed models
correspond to $5\times10^{-21}\lesssim m~(\rm{eV}/c^2)\lesssim8\times10^{-17}$ (O5).

For parameters in these ranges, our predicted SNR for aLIGO/Virgo 
for the SGWB from inflation in $\Lambda$SFDM can exceed 
current predictions of the background from unresolved binary black hole mergers 
in \cite{2017PhRvL.118l1101A}, as shown in Fig. \ref{fig:GWzoomin}. 
It will be important, therefore, to consider this inflationary SGWB in $\Lambda$SFDM 
in interpreting the current and future GW detection results.

We have also shown here that, for a range of values of $\tre$ and allowed values of $r$, 
the inflationary SGWB in $\Lambda$SFDM may also be detectable by LISA.
In that case, the difference in spectral shape 
between the primordial and black-hole merger GW backgrounds 
may allow them to be distinguished.


\vspace{-0.1in}
\begin{acknowledgments}
\vspace{-0.1in}
During the work on this paper, we benefitted from discussions with many people 
to whom we express our gratitude: Joel Meyers, Dustin Lorshbough, Aditya Aravind, 
Thomas Crawford, Can Kilic, Raphael Flauger, Takeshi Kobayashi, and Eiichiro Komatsu.
We thank Paul Lasky especially for sharing his code 
for plotting current experimental constraints on the present-day SGWB background.
We are grateful to members of the LIGO Scientific Collaboration for helpful discussion, 
including Tania Regimbau and David Reitze. 
We give special thanks to Joseph Romano for sharing his code 
for computing aLIGO/Virgo sensitivity and the signal-to-noise ratios 
for a stochastic background of GWs with a power-law spectra.
We modified this code to compute these for the non-power-law
spectrum of the SGWB from inflation in our $\Lambda$SFDM model.
We are grateful to Andrew Matas and the LIGO Scientific Collaboration 
for providing us with the noise characteristics for the O1 run and 
to Joseph Romano, again, for helping us to interpret it.

This work was supported in part by U.S. NSF grant AST-1009799, NASA grant NNX11AE09G, 
NASA/JPL grant RSA Nos.1492788 and 1515294, 
and supercomputer resources from NSF XSEDE grant TG-AST090005 
and the Texas Advanced Computing Center (TACC) at the University of Texas at Austin. 
TRD was supported by the U.S. Department of Energy under grants 
DE-FG02-95ER40899 and DE-SC0007859.
TRD acknowledges the support by the Austrian Science Fund FWF 
through a Lise Meitner fellowship (M2008-N36).

\end{acknowledgments}

\appendix

\section{Gravitational Waves in a FLRW universe}
\label{ap:GW}

\subsection{Effective stress-energy tensor of gravitational waves}
\label{ap:GWtmn}

It is instructive to show how it is that tensor perturbations associated with gravitational waves 
also contribute an effective mean stress-energy to the background curvature of the universe, 
which is spatially homogeneous on large scales.

For a FLRW universe of which the metric is defined in Eq. (\ref{eq:metric}), 
only allowing tensor perturbations, 
let us evaluate the left-hand side of the Einstein field equations (\ref{eq:EFE}), 
\begin{equation}\label{eq:expansion}
    R^\mu_{~\nu}-\frac{1}{2}R=\left(R^{\mu~(0)}_{~\nu}-\frac{1}{2}R^{(0)}\right) +
    \left(R^{\mu~(2)}_{~\nu}-\frac{1}{2}R^{(2)}\right),
\end{equation}
where we have expanded the left-hand side in perturbations $h_{ij}$, up to the second order.
On the right-hand side of the expansion above, the zeroth-order term contributed by
the unperturbed FLRW metric $\bar g_{\mu\nu}$ is familiar, of which the nonzero components are
\begin{equation}
    R^{0~(0)}_{~0}-\frac{1}{2}R^{(0)}=\frac{3\dot{a}^2}{c^2a^2},
\end{equation}
\begin{equation}
    R^{i~(0)}_{~i}-\frac{1}{2}R^{(0)}=\frac{-a\ddot{a}+\dot{a}^2}{c^2a^2}.
\end{equation}
The first-order term in the expansion vanishes. The second-order term (of the order $O(h^2)$)
due to tensor perturbations, can be moved to the right-hand side of
the Einstein field equations (\ref{eq:EFE}), and hence viewed as an effective contribution
to the total stress-energy tensor $T^{\mu}_{~\nu}$. That is to say,
$T_{\mu\nu,~\rm{GW}}$ purely results from the spatial metric perturbations,
rather than the stress-energy tensor of an intrinsic cosmic component.

The stress-energy carried by GWs can not be localized within a wavelength \cite{1973grav.book.....M}.
Instead, it is only meaningful to interpret the effective $T_{\mu\nu,~\rm{GW}}$ as a macroscopic average
over several wavelengths. With this understanding, we see that the stress-energy of GWs
indeed contributes to the curvature of the homogeneous background universe.
In other words, it back-reacts to the zeroth-order term in Eq. (\ref{eq:expansion})
once moved to the right-hand side. 
Let us, for simplicity, focus on subhorizon modes. 
We can then explicitly write down the stress-energy tensor of GWs,
\begin{IEEEeqnarray}{rCl}\label{eq:isaacson}
    T_{\mu\nu,~\rm{GW}} & \equiv & -\frac{c^4}{8\pi G}\left(\langle R^{~(2)}_{\mu\nu}\rangle-\frac{1}{2}\bar g_{\mu\nu}\langle R^{(2)}\rangle\right)\nonumber\\
    & = & \frac{c^4}{32\pi G}\langle (a^2h_{ij})_{;\mu}(\f{1}{a^2}h^{ij})_{;\nu}\rangle,
\end{IEEEeqnarray}
where the brackets $\langle\cdot\rangle$ denote the spatial average over several wavelengths, 
and the semicolon denotes the covariant derivative with respect to the background metric 
$\bar g_{\mu\nu}$.
This was first derived by Isaacson in \cite{1968PhRv..166.1263I, 1968PhRv..166.1272I}.
Therefore, $T_{\mu\nu,~\rm{GW}}$ is also known as the Isaacson tensor.

Particularly, the time-time component of $T_{\mu\nu,~\rm{GW}}$ defines the energy density of GWs, 
\begin{IEEEeqnarray}{rCl}\label{eq:rhogwdef}
	\rho_{\rm GW} & \equiv & T^0_{~0,~\rm{GW}}= \frac{c^4}{32\pi G}\langle (a^2h_{ij})_{;0}(\f{1}{a^2}h^{ij})^{;0}\rangle\nonumber\\
	& = &  \frac{c^2}{32\pi G}\langle\p_th_{ij}\p_th^{ij}\rangle.
\end{IEEEeqnarray}
Remember that $h^{ij}=h_{ij}$ 
(see also \cite{1999PhRvD..59j2001A, 2006PhRvD..73l3515W, 2012JCAP...06..027B}).

\subsection{Fourier decomposition of $h_{ij}$}
\label{ap:GWfourier}

It is customary to move into $\mathbf{k}$-space by Fourier transforming the tensor perturbations, 
\beq\label{eq:fourier}
	h_{ij}(\mathbf{x},t) = \sum_P \int\f{\ud^3\mathbf{k}}{(2\pi)^3}h_{\mathbf{k}}^P(t) {\rm e}^{i\mathbf{k}\cdot\mathbf{x}}\mathbf{\epsilon}_{ij}^P(\mathbf{k}),
\eeq
where $\mathbf{k}$ is the comoving wave vector, 
and $\mathbf{\epsilon}_{ij}^P(\mathbf{k})$ are the spin-2 polarization tensors 
for the ``plus'' and ``cross'' polarization states, $P=+$ or $\times$, 
with respect to the wave vector $\mathbf{k}$. 
Both $\mathbf{\epsilon}_{ij}^+(\mathbf{k})$ and $\mathbf{\epsilon}_{ij}^\times(\mathbf{k})$ 
are symmetric, traceless ($\sum_{i}\mathbf{\epsilon}_{ii}^P(\mathbf{k})=0$), 
and perpendicular to the direction in which the plane wave propagates (transverse), 
$\mathbf{\epsilon}_{ij}^P(\mathbf{k})\cdot\mathbf{k}=0$. 
Also, $\mathbf{\epsilon}_{ij}^P(\mathbf{-k})=\mathbf{\epsilon}_{ij}^P(\mathbf{k})$. 
They follow such normalization convention,
\beq
	\sum_{i,j}\mathbf{\epsilon}_{ij}^P(\mathbf{k})\mathbf{\epsilon}_{ij}^{P'}(\mathbf{k})=2\delta_{PP'},
\eeq
where $\delta_{PP'}$ is the Kronecker delta.
In three-dimensional space with Cartesian coordinates, 
if $\mathbf{k}$ goes along the $z$-direction, the explicit form of $\mathbf{\epsilon}_{ij}^P$ 
can be written as
\begin{IEEEeqnarray}{rCl}
	\mathbf{\epsilon}_{ij}^+ & = & \mathbf{e}_x\otimes\mathbf{e}_x-\mathbf{e}_y\otimes\mathbf{e}_y=\left(\begin{array}{ccc}
	1 & 0 & 0 \\
	0 & -1 & 0 \\
	0 & 0 & 0
	\end{array}\right),\nonumber\\
	\mathbf{\epsilon}_{ij}^\times & = & \mathbf{e}_x\otimes\mathbf{e}_y+\mathbf{e}_y\otimes\mathbf{e}_x=\left(\begin{array}{ccc}
	0 & 1 & 0 \\
	1 & 0 & 0 \\
	0 & 0 & 0
	\end{array}\right),
\end{IEEEeqnarray}
where $\mathbf{e}_x$ and $\mathbf{e}_y$ are unit polarization vectors in the $xy$ plane, 
both perpendicular to $\mathbf{k}$.

\section{Thin-horizon approximation: analytical solution and asymptotic behavior 
of tensor modes}
\label{app:thinhor}

\begin{figure*}[t]
\begin{minipage}{0.5\linewidth}
     \centering
     \includegraphics[width=8cm]{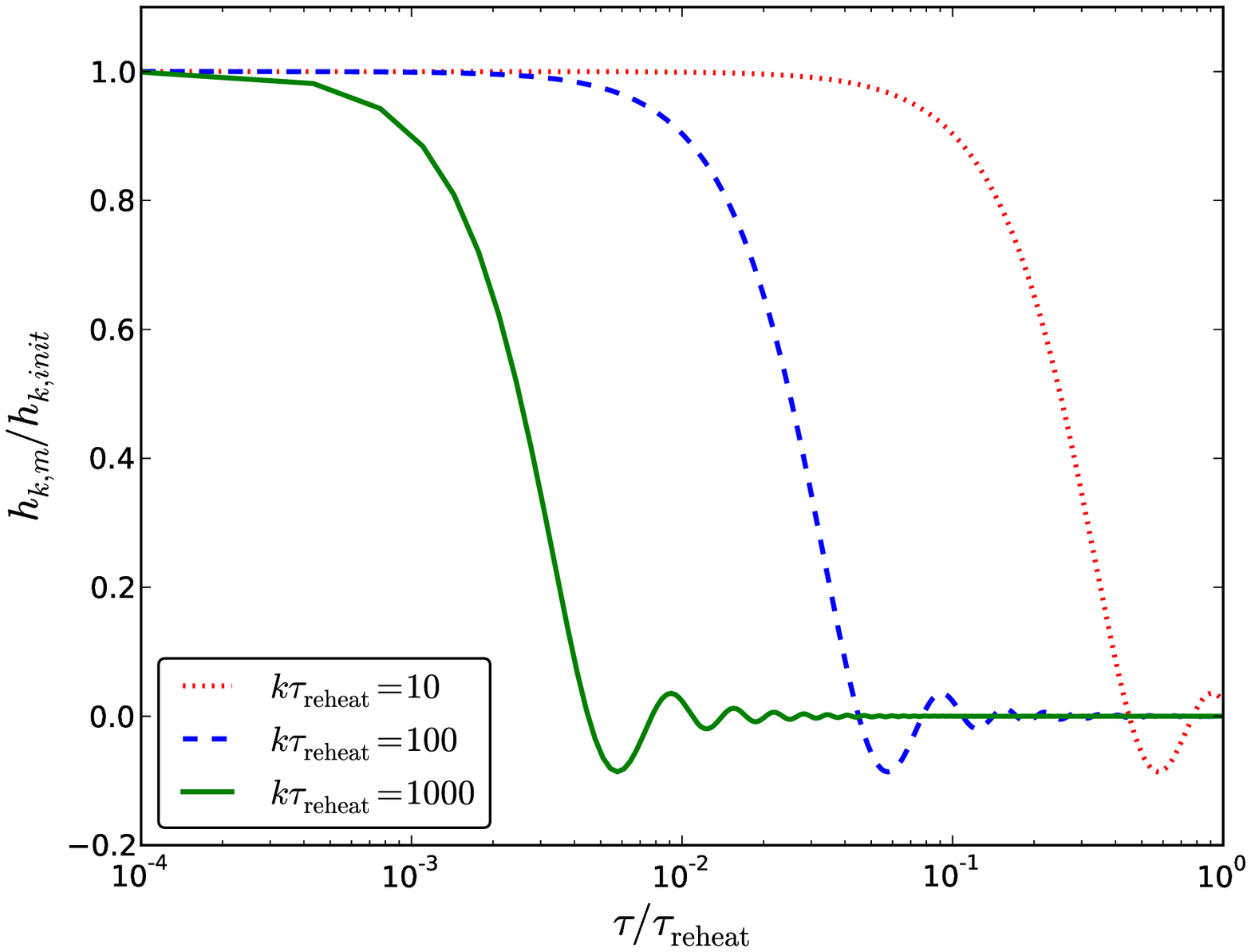}
    \end{minipage}%
    \begin{minipage}{0.5\linewidth}
      \centering
      \includegraphics[width=8cm]{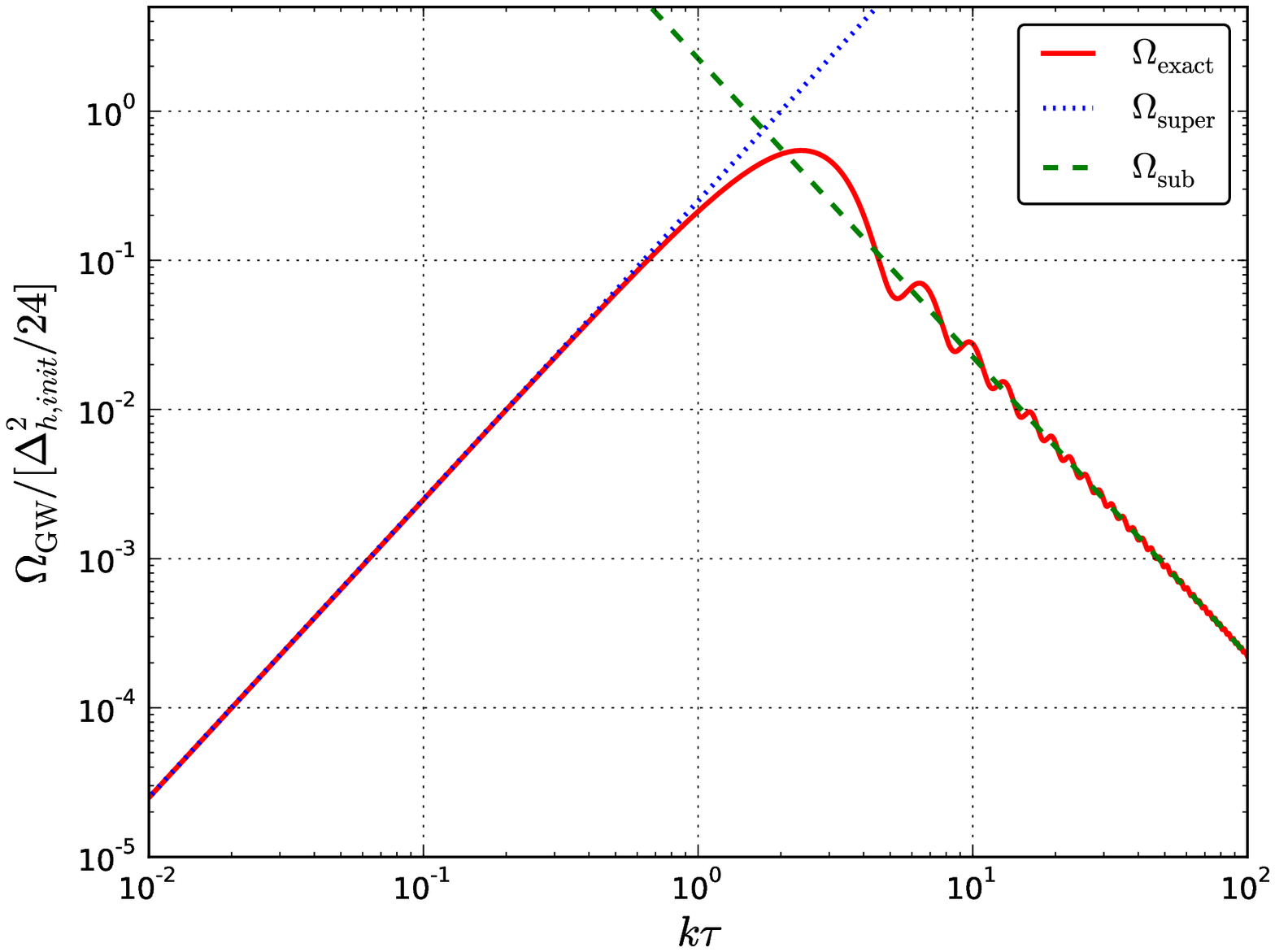}
    \end{minipage}
 \caption{\tx{Left-hand plot:} Tensor perturbations for different $k$-modes, 
 as they reenter the horizon during reheating (with $w=0$) at different times. 
 At $\tau/\tau_{\rm reheat} = 1$, the reheating era gives rise to the stiff era. 
 The tensor modes (strains) are normalized over their initial amplitude $\hini$, 
 for each $k$.
 \tx{Right-hand plot:} The exact solution for $\ogw(k, \tau)$ as a function of $k\tau$ (solid curve), 
 as well as the respective asymptotic expressions (superhorizon in dot, subhorizon in dash), 
 for a reheating era with $w=0$. $\ogw$ is normalized over $\Delta_{h,~\rm{init}}^2/24$.}
 \label{appogw}
\end{figure*}

In this appendix, we show that the thin-horizon approximation is valid for tensor modes 
which reenter the horizon during an era with constant $w$ for the EOS of the background universe, 
the case of most interest to us throughout the $\Lambda$SFDM expansion history.
For this purpose, we show an example of how well the exact analytical solution 
matches the asymptotic sub- and superhorizon evolution, in their respective regime of validity 
(we draw this example from other work in progress, Rindler-Daller, Shapiro, Li, in prep.). 
Fig. \ref{appogw} shows plots of the evolution of $h_k(\tau)$ and $\ogw(k,\tau)$ 
in the case of a matter-like ($w=0$) EOS of the background universe.
We confirm that the range in $k\tau$ around horizon crossing is rather narrow, 
justifying the thin-horizon approximation in which the horizon crossing is deemed 
to occur suddenly at $k=aH(a)/c$ for each $k$.

\section{Marginally-allowed $\Lambda$SFDM models with given $r$ and $\lambda/(mc^2)^2$: $f_{\rm reheat}\propto T_{\rm reheat}$}
\label{app:fproptoT}

It can be analytically shown that for $\Lambda$SFDM models 
which marginally satisfy the cosmological constraints, 
with given values of $r$ and $\lambda/(mc^2)^2$,
$f_{\rm reheat}$ is nearly proportional to $T_{\rm reheat}$.

First, let us express $f_{\rm reheat}$ as follows:
\begin{equation}
	f_{\rm reheat} = \frac{f_{\rm reheat}}{f_{\rm sr}}\frac{f_{\rm sr}}{f_{\rm r,late}}f_{\rm r,late},
\end{equation}
where $f_{\rm sr}$ is the frequency of the mode that reenters the horizon 
(approximately) at $a_{\rm sr}=2\pi f_{\rm sr}/H_{\rm sr}$, 
the transition between the stiff-SFDM-dominated era and the radiation-dominated era, 
and $f_{\rm r,late}$ is the frequency of a mode that reenters later in the radiation-dominated era 
at $a_{\rm r,late}=2\pi f_{\rm r,late}/H_{\rm r,late}$. 
Both $f_{\rm r,late}$ and $a_{\rm r,late}$ are required to be the same for all models, 
the feasibility of which is guaranteed by the fact that 
these marginally-allowed $\Lambda$SFDM models 
share a uniform expansion history in the radiation-dominated era, 
since the values of $r$ and $\lambda/(mc^2)^2$ are fixed.

Since the area of the triangle in $\ogw(f)$ is almost constant, 
the triangle itself must be almost identical for all marginally-allowed $\Lambda$SFDM models, 
as the slopes of the two ``sides'' of the triangle are fixed by the power-law indices. 
In other words, they can be approximated by the same triangle which slides on a fixed plateau, 
whose height is determined by the value of $r$ alone.
Thus, the ratio $f_{\rm reheat}/f_{\rm sr}$ must be the same for all marginally-allowed models, 
since the x-axis in the $\ogw(f)$ plots is logarithmic. 
This implies that the number of e-foldings between $a_{\rm reheat}$ and $a_{\rm sr}$ 
must be the same as well, shown by the following equation:
\begin{IEEEeqnarray}{rCl}
	\frac{f_{\rm reheat}}{f_{\rm sr}} & = & \frac{a_{\rm reheat}H_{\rm reheat}}{a_{\rm sr}H_{\rm sr}}
	 =  \left(\frac{a_{\rm reheat}}{a_{\rm sr}}\right)\left(\frac{a_{\rm reheat}}{a_{\rm sr}}\right)^{-3}\nonumber\\
	& = & \left(\frac{a_{\rm reheat}}{a_{\rm sr}}\right)^{-2}.
\end{IEEEeqnarray}
Also,
\begin{IEEEeqnarray}{rCl}
	\frac{f_{\rm sr}}{f_{\rm r,late}} & = & \frac{a_{\rm sr}H_{\rm sr}}{a_{\rm r,late}H_{\rm r,late}}=
	\left(\frac{a_{\rm sr}}{a_{\rm r,late}}\right)\left(\frac{a_{\rm sr}}{a_{\rm r,late}}\right)^{-2}\nonumber\\
	& = &  \left(\frac{a_{\rm sr}}{a_{\rm reheat}}\right)^{-1}\left(\frac{a_{\rm reheat}}{a_{\rm r,late}}\right)^{-1}.
\end{IEEEeqnarray}
Combining the above two equations yields
\begin{IEEEeqnarray}{rCl}
	f_{\rm reheat} & = & f_{\rm r,late}\left(\frac{a_{\rm reheat}}{a_{\rm sr}}\right)^{-2}\left(\frac{a_{\rm sr}}{a_{\rm reheat}}\right)^{-1}\left(\frac{a_{\rm reheat}}{a_{\rm r,late}}\right)^{-1}\nonumber\\
	& = & f_{\rm r,late}\left(\frac{a_{\rm reheat}}{a_{\rm sr}}\right)^{-1}\left(\frac{a_{\rm reheat}}{a_{\rm r,late}}\right)^{-1}\nonumber\\
	& \propto & a_{\rm reheat}^{-1}.
\end{IEEEeqnarray}
Since $T_{\rm reheat}\propto a_{\rm reheat}^{-1}$ to a very good accuracy, ignoring the details in the thermal history of the universe, we conclude that $f_{\rm reheat}\propto T_{\rm reheat}$, 
for all marginally-allowed $\Lambda$SFDM models with fixed $r$ and $\lambda/(mc^2)^2$.

\section{Calculating the expected SNR and the integrated sensitivity curve for a given SGWB signal for aLIGO/Virgo with the noise characteristics from the completed O1 run}
\label{app:SNR_O1}

\subsection{Expected signal-to-noise ratio}

According to \cite{1999PhRvD..59j2001A}, 
a potential SGWB signal can be detected by cross-correlating the strain outputs 
of two laser interferometric GW detectors, e.g., the Advanced LIGO/Virgo experiment 
\cite{2016PhRvL.116m1102A, 2017PhRvL.118l1101A}. 
For this study, the expected SNR for a generic SGWB 
whose spectrum is $\Omega_{\rm GW}(f)$ today
can be derived as \cite{1999PhRvD..59j2001A}
\footnote{
For simplicity, the treatment described in this Appendix applies 
for a cross-correlation study with only two detectors, 
as is the case for the completed aLIGO O1 run. However, 
for a network of detectors (i.e., no less than three detectors, 
as is the case for the full aLIGO/Virgo experiment after Virgo comes online), 
this treatment can be easily generalized to combine the SNR values from each pair of detectors, 
as shown by \cite{2013PhRvD..88l4032T}.
}
\begin{equation}\label{eq:snr}
	{\rm SNR}=\frac{3H_0^2}{10\pi^2}\sqrt{2T}\left[\int_0^\infty\ud f\left(\frac{\gamma^2(f)}{f^6P_1(f)P_2(f)}\right)\Omega_{\rm GW}^2(f)\right]^{1/2},
\end{equation}
where $P_1(f)$ and $P_2(f)$ are the one-sided strain noise power spectral densities 
of the two detectors; $\gamma(f)$ is the normalized isotropic overlap reduction function
\cite{1992PhRvD..46.5250C, 1993PhRvD..48.2389F}; 
and $T$ is the accumulated coincident observation time. $T=29.85$ days for the O1 run. 
Eq. (\ref{eq:snr}) is consistent with Eq. (21) in \cite{2013PhRvD..88l4032T} 
and can be rearranged into
\begin{equation}\label{eq:snr2}
	{\rm SNR}=\sqrt{2T}\left[\int_0^\infty\ud f\frac{\ogw^2(f)}{\Omega_{\rm eff}^2(f)}\right]^{1/2},
\end{equation}
where $\Omega_{\rm eff}(f)$ is defined as 
\begin{equation}
	\Omega_{\rm eff}(f)\equiv\frac{10\pi^2}{3H_0^2}\left(\frac{\gamma^2(f)}{f^6P_1(f)P_2(f)}\right)^{-1/2}.
\end{equation}
Therefore, calculating the SNR for a given SGWB 
amounts to determining the function $\Omega_{\rm eff}(f)$ 
which basically reflects the noise characteristics of a given observing run.
For future observing runs, e.g. aLIGO/Virgo O5, one can estimate the noise characteristics 
and provide a theoretical prediction of $\Omega_{\rm eff}(f)$, 
as shown by \cite{2013PhRvD..88l4032T}, to calculate the expected SNR
for a given SGWB signal using Eq. (\ref{eq:snr2}).
However, since the aLIGO O1 run is already completed, it is reasonable 
to replace the theoretical function of $\Omega_{\rm eff}(f)$ 
with the actual noise characteristics from the O1 run data
\cite{2017PhRvL.118l1101A}, as explained below.

It is shown, for example, in \cite{2015PhRvD..91b2003A} 
(a cross-correlation analysis for two colocated LIGO detectors with data from the initial LIGO S5 run)
that $\Omega_{\rm eff}(f)$ can be related to the expectation value of the variance of 
the frequency-dependent estimator for the amplitude of a flat SGWB signal ($\ogw(f)=\Omega_0$), 
as follows:
\begin{equation}\label{eq:sigma_est}
	\Omega^2_{\rm eff}(f) = (2 T\Delta f)\cdot \sigma^2_{\hat\Omega_0}(f),
\end{equation}
where $\Delta f$ is the width of the frequency bin, and $\sigma^2_{\hat\Omega_0}(f)$ 
is the variance of the estimator $\hat\Omega_0(f)$ in each frequency bin. 
A detailed derivation of the equation above can be found in \cite{1999PhRvD..59j2001A}.
While the function $\Omega_{\rm eff}(f)$ depends \emph{only} 
upon the noise characteristics of the experiment of interest,
independent from the spectral form of the SGWB signal,
Eq. (\ref{eq:sigma_est}) implies that one can use $\Omega_{\rm eff}(f)$ 
to construct frequency-dependent variance estimators for power-law SGWB spectra, 
and, particularly, $\Omega_{\rm eff}(f)$ is encoded in the variance estimator, 
$\sigma^2_{\hat\Omega_0}(f)$, for the flat spectrum.
We communicated with the LIGO Scientific Collaboration with regard to 
the noise characteristics (encoded in) $\sigma^2_{\hat\Omega_0}(f)$
in the recently reported SGWB analysis with O1 data 
(plotted in Fig. 1 of \cite{2017PhRvL.118l1101A}), 
which they kindly provided us for the entire frequency range ($20-1726$ Hz) 
to which the aLIGO O1 run is sensitive. 
The width of the frequency bin in their analysis is $\Delta f=0.031$ Hz.
We are thus able to calculate the expected SNR for the inflationary SGWB 
predicted in our $\Lambda$SFDM model as follows:
\begin{equation}\label{eq:SNR_O1}
	{\rm SNR}=\left(\sum_{f=f_{\rm min}}^{f_{\rm max}} \frac{\ogw^2(f)}{ \sigma^2_{\hat\Omega_0}(f)}\right)^{1/2},
\end{equation}
where the summation is over the frequency bins of $\sigma^2_{\hat\Omega_0}(f)$. 
We use Eq. (\ref{eq:SNR_O1}) to calculate the expected SNR for any given SGWB signal 
for the completed aLIGO O1 run.

\subsection{Integrated Sensitivity Curves}

The construction of the frequency-integrated sensitivity curves 
for the inflationary SGWB spectrum predicted in $\Lambda$SFDM
is analogous to the procedure developed in \cite{2013PhRvD..88l4032T} 
where they constructed the sensitivity curves for arbitrary power-law spectra, 
$\ogw(f)\propto f^\beta$.
However, as described in \S\ref{sec:GWspectratoday}, the SGWB from inflation
predicted in our model has a triangle-shaped feature with fixed slopes, 
which can be parametrized as the following broken power-law spectrum:
\begin{widetext}
\begin{numcases}{\hspace{-0.2in}\ogw(f)=}
    \ogw(f_{\rm reheat})\left(\f{f}{f_{\rm reheat}}\right), & $\hspace{0.3in}~f\lesssim f_{\rm reheat},$\label{eq:flfpeak}\\[1em]
    \f{9\pi}{64}\cdot\ogw(f_{\rm reheat})\left(\f{f}{f_{\rm reheat}}\right)^{-2}, &  $\hspace{0.3in} ~f>f_{\rm reheat}.$ \label{eq:fgfpeak}
\end{numcases}
\end{widetext}
As explained in \S\ref{sec:GWspectratoday}, at $f=f_{\rm reheat}$, 
the SGWB spectrum has the maximum value, $\ogw(f_{\rm reheat})$, 
which corresponds to the peak of the triangle.
Therefore, to construct a sensitivity curve with a fixed value of SNR, e.g. SNR $=1$, 
we can carry out the following procedure.
\begin{enumerate}
\item We choose a sample of values of $f_{\rm reheat}$ over a frequency range
which includes the available range of the noise characteristics, $(f_{\rm min},~f_{\rm max})$. 
For each value of $f_{\rm reheat}$, we calculate the corresponding value of $\ogw(f_{\rm reheat})$ 
which yields that fixed SNR, using Eq. (\ref{eq:snr2}) or (\ref{eq:SNR_O1}).

\item For each pairs of values for $f_{\rm reheat}$ and $\ogw(f_{\rm reheat})$ in the sample, 
plot the spectrum $\ogw(f)$ using Eqs. (\ref{eq:flfpeak}) and (\ref{eq:fgfpeak}).
The envelope of these spectra yields the integrated sensitivity curve 
for the inflationary SGWB in $\Lambda$SFDM with the fixed SNR.

\end{enumerate}

The interpretation of these integrated sensitivity curves is as follows 
(repeating the description in \S\ref{sec:GWspectratoday}):
for the curve with SNR $=1$, for example,
if the predicted $\ogw(f)$ for the inflationary SGWB from a given set of $\Lambda$SFDM
model parameters touches the curve for the O1(O5) run at any $f$, 
this SGWB will be detected with 1$\sigma$ significance (SNR $=1$) by the O1(O5) run,
respectively.
These curves are plotted in Figs. \ref{fig:GWspectra1} -- \ref{fig:GWspectra3}.
The O1 curve uses the actual noise characteristics from data
while the O5 curve is based on the theoretical prediction of $\Omega_{\rm eff}(f)$ 
provided by \cite{2013PhRvD..88l4032T}.

\bibliography{SFDM_cosmology}

\end{document}